\documentclass[12pt,a4]{article}

\usepackage{lmodern,sansmath}
\usepackage{textcomp}
\usepackage{microtype}
\usepackage{amsmath, amssymb, amsthm, dsfont, bm}         
\usepackage{amsfonts}         

\usepackage[round]{natbib}
\usepackage[colorlinks=true,linkcolor=red,citecolor=blue]{hyperref}

\usepackage{color}

\usepackage{booktabs}
\usepackage{graphicx}

\usepackage{caption}
\usepackage{subcaption}

\usepackage{rotating}

\newcommand{\dt}{\delta t}
\newcommand{\dtsd}{\dfrac{\delta t}{2}}

\newcommand{\bv}{ {\bf v} }
\newcommand{\four}[1][]{\mathcal F_{#1}}
\newcommand{\para}{/\!\!/}

\newcommand{\Ekini}{E_{\rm kin}^{\rm i}}
\newcommand{\Ekinc}{E_{\rm kin}^{\rm c}}
\newcommand{\Eint}{E_{\rm int}}
\newcommand{\Eq}{E_{\rm q}}

\newcommand{\calD}{{\cal D}}

\newcommand{\ie}{{\em i.\thinspace{}e. }}
\newcommand{\eg}{{\em e.\thinspace{}g. }}

\newcommand{\ds}{\displaystyle}
\newcommand{\pl}{\partial }
\newcommand{\bfx}{\vec{x}}

\def\M{{\mathcal M}}

\def\vec#1{\mathbf{#1}}

\newcommand{\hb}{\hbar}

\def\aho{a_{\hbox{\footnotesize h}}}

\def\Imag{{\mathcal{I}mag}}

\newcommand{\ext}{{\rm ext}}
\newcommand{\trap}{{\rm trap}}
\newcommand{\scal}{{\rm ref}}

\newcommand{\adim}[1]{\tilde{#1}}

\makeatletter
\renewcommand*{\eqref}[1]{%
	\hyperref[{#1}]{\textup{\tagform@{\ref*{#1}}}}%
}
\makeatother

\begin{document}

\title{\vspace{-1cm} Quantum turbulence simulations using the Gross-Pitaevskii equation: high-performance computing and new numerical benchmarks}

\author{Michikazu Kobayashi$^1$,  Philippe Parnaudeau$^2$, Francky Luddens$^3$,\\
	Corentin  Lothod{\'e}$^3$, Luminita Danaila$^4$, Marc Brachet$^5$\\ and Ionut Danaila$^{3, *}$
	\\ \\
	$^1$Department of Physics, Kyoto University, Japan\\
	\\ 
	$^2$Institut Pprime, CNRS,\\
	Universit{\'e} de Poitiers - ISAE-ENSMA - UPR 3346,\\
	86962 Futuroscope Chasseneuil Cedex, France\\
	\\ 
	$^3$Universit{\'e} de Rouen Normandie,\\ Laboratoire de Math{\'e}matiques Rapha{\"e}l Salem,\\  
	F-76801 Saint-{\'E}tienne-du-Rouvray, France \\
	\\ 
	$^4$Universit{\'e} de Rouen Normandie,\\ CORIA,  CNRS UMR 6614,\\ 
	F-76801 Saint-{\'E}tienne-du-Rouvray, France\\
	\\ 
	$^5$Laboratoire de Physique de l'\'Ecole Normale Sup\'erieure,\\ 
	ENS, Universit\'e PSL, CNRS, Sorbonne Universit\'e,\\ Universit\'e de Paris, 
	F-75005 Paris, France\\
	\\ 
	$^*$ Corresponding author: ionut.danaila@univ-rouen.fr
 }

\date{\today}
\maketitle

\begin{abstract}
This paper is concerned with the numerical investigation of Quantum Turbulence (QT) described by the Gross-Pitaevskii (GP) equation.  Numerical simulations are performed using a parallel (MPI-OpenMP) code based on a pseudo-spectral spatial discretization and  second order splitting for the time integration.
We start by revisiting (in the framework of high-performance/high-accuracy computations) well-known GP-QT settings, based on the analogy with classical vortical flows: Taylor-Green (TG) vortices and Arnold-Beltrami-Childress (ABC) flow.  Two new settings are suggested to build the initial condition for the QT simulation. They are based on the direct manipulation of the wave function by  generating a smoothed random phase (SRP) field, or seeding random vortex rings (RVR) pairs.  
The new initial conditions have the advantage to be simpler to implement than the TG and ABC approaches, while generating statistically equivalent QT fields.
Each of these four GP-QT settings is described in detail by defining corresponding benchmarks that could be used to validate/calibrate new GP codes. We offer a comprehensive description of the numerical and physical parameters of each benchmark. We  analyze the results in detail and present values, spectra and structure functions of main quantities of interest (energy, helicity, etc.)  that are useful to describe the turbulent flow.  Some general features of QT are identified, despite the variety of initial states.

\end{abstract}

\begin{flushleft}
\end{flushleft}

\section{Introduction}

The study of quantum fluids, realized in superfluid helium and atomic Bose-Einstein condensates (BEC), has become a central topic in various fields of physics, such as low temperature physics, fluid dynamics of inviscid flows, quantum physics, statistical physics, cosmology, etc. One of the striking features of quantum fluids is the nucleation of vortices with quantized (fixed) circulation, when an external forcing is applied (rotation, stirring, etc). The observation of quantized vortices, as a signature of the superfluid (zero-viscosity) nature of these flow systems, was extensively explored in different experimental settings of superfluid helium or BEC.  Configurations with a large number of quantized vortices tangled in space can evolve to Quantum Turbulence (QT), generally referred to as {\em vortex tangle turbulence}. While QT in superfluid helium has been largely studied in the last two decades (see dedicated volumes \cite{QT-review-2002-vinen,QT-book-2001-barenghi,QT-book-2008-barenghi,QT-book-2009-tsubota}), only recent experimental and theoretical studies \citep{QT-2010-seman,QT-2011-seman,QT-experiment-2014-Kwon,QT-experiment-2016-Navon} reported different possible routes to QT in BEC. 

The focus of this paper is on the numerical study of QT in superfluid helium. A large body of literature in this field  is based on the macroscopic description of the superfluid using the Gross-Pitaevskii (GP) mean-field equation. The GP equation  is a  nonlinear Schr{\"o}dinger (NLS) equation with cubic nonlinearity.
It describes, in the theoretical limit of absolute zero temperature, the time evolution of the single-particle complex wave function $\psi$ (identical for all particles). Quantized vortices appear then as topological line defects, resulting from  the $U(1)$ symmetry breaking of the phase shift of $\psi$. QT in superfluid helium based on the GP  model (denoted hereafter as GP-QT) implies complicated space and time interactions between a large number of quantized vortices. For a comprehensive description of different models of QT, see recent reviews by \cite{QT-book-2009-tsubota,QT-review-2012-brachet,QT-review-2014-Barenghi-PNAS,QT-review-2017-tsubota-num}.

There are several challenges when setting a numerical simulation to investigate GP-QT: \\
{\em (i)}  generate a physically and mathematically sound initial state with many quantized vortices that finally evolve to a statistically steady state of QT,
{\em (ii)} use accurate numerical methods that preserve the invariants of the GP equation when long time-integration is necessary, {\em (iii)} design numerical codes affording large grid resolutions, necessary to accurately capture the dynamics of vortices and {\em (iv)}  compute appropriate (statistical) diagnostic tools to analyze the superfluid flow evolution.

We use in this contribution a modern parallel (MPI-OpenMP) numerical code satisfying the requirements {\em (ii)}  and {\em (iii)}. The code is called GPS (Gross-Pitaevskii Simulator) \citep{HPC-Parnaudeau-2015} and is based on a Fourier-spectral  space discretization and up-to-date numerical methods: a semi-implicit backward-Euler scheme with Krylov preconditioning for the stationary GP equation \citep{BEC-CPC-2014-antoine-duboscq} and various schemes (Strang splitting, relaxation, Crank-Nicolson) for the real-time GP equation \citep{BEC-review-2013-antoine-besse-bao}. The GPS code offers a solid framework to address in detail challenges {\em (i)} and {\em (iv)}, for which we review previous models and bring new contributions.

As in classical turbulence (CT), the numerical and physical accuracy of the initial condition is crucial in computing properties of numerically generated QT.
Using the  hydrodynamic analogy for the GP model (through the Madelung transform, as explained below), pioneering numerical simulations of GP-QT \citep{Nore1993154,Nore97a,Abid2003509} suggested initial conditions and statistical analysis tools inspired from CT. A velocity field, derived from  the well-known  classical flow with Taylor-Green (TG) vortices, was imposed to the superfluid flow. An initial wave function, with nodal lines corresponding to vortex lines of the velocity, was thus generated. This initial wave function was  then used in the Advective Real Ginzburg-Landau equation (ARGLE), equivalent to the imaginary-time GP equation with Galilean transformation (see also below), to reduce the acoustic emission of the initial field. The result of the ARGLE procedure was finally used as initial field for the time-dependent GP simulation.  A similar approach was more recently used by replacing the TG vortices with the   Arnold-Beltrami-Childress (ABC) classical vortex flow \citep{Clark16,Clark17}.  If this approach is well suited to control the hydrodynamic characteristics of the initial superfluid flow (Mach number, helicity), it  involves supplemental technicalities and computations through the ARGLE procedure. We suggest in this paper two new approaches to generate the initial condition for the GP-QT simulations, based on the direct manipulation of the wave function. The ARGLE procedure is thus avoided. The first method prescribes a smoothed random-phase (SRP) for the wave function, while the second one  generates random vortex rings (RVR). The two new methods, which are simply to implement, are shown to develop QT fields with similar statistical properties as those obtained using the TG or ABC classical initial conditions. Nevertheless, the dynamics of the superfluid flow is different.
Compared to TG and ABC cases, in the SRP case the initial field is vortex free and dominated by the compressible kinetic energy; vortices nucleate progressively and do not display long vortex lines. For the RVR flow, evolution is opposite to that observed for the SRP case: in the early stages of the time evolution the incompressible kinetic energy is dominant; the compressible kinetic energy then starts to increase due to sound emissions through vortex reconnections. However, like in well-documented TG and ABC cases, a Kolmogorov-like scaling of the incompressible kinetic energy spectrum is obtained for the new SRP and RVR cases.

Concerning the analysis of the QT field, we present classical diagnostic tools (inspired from CT), as the energy decomposition and associated spectra \citep{Nore1993154,Nore97a} and also new ones, as the second-order structure function, not reported in the previously cited studies. This supplements the statistical description of the superfluid flow. We also carefully investigate the influence of numerical parameters (as the grid resolution of a vortex, the maximum resolved wave-number and the computed local Mach number) on the characteristic of the QT. This topic is generally very briefly addressed in physical papers on GP-QT.

Starting from the observation that in previously published studies of GP-QT, the focus was mainly given to the physics of turbulence, this paper is also intended to define in detail numerical benchmarks in the framework of parallel computing. We start by revisiting classical GP-QT settings (based on TG and ABC flows). 
New results obtained with our high-performance/high-accuracy parallel code are compared with available data in the literature. We then present the new numerical benchmarks, based on random phase fields or random vortex rings generation. The new benchmarks offer a new perspective in comparing the classical  CT based approaches to more GP-oriented models.  For all benchmarks, we offer a comprehensive description of the numerical and physical parameters and give checkpoint values for validating each step of the simulation. This could be useful to verify new numerical methodologies or tune/validate new modern GP numerical codes.

The organization of the paper is as follows. Section  \ref{sec:Model} introduces the GP mean field equation, its stationary version and the hydrodynamic analogy, derived using the Madelung transform. The main conservation laws are emphasized and  the Bogoliubov-de Gennes (BdG) linearized system for the response of the system to small oscillations is derived.   Section \ref{sec:QTscales} introduces the main characteristics of the QT flow that are used to set numerical simulations: the healing length, the sound velocity, the Bogoliubov dispersion relation,  the compressible and incompressible kinetic energy and the helicity. The notion of quantum vortex and its representation on a computational grid are also discussed. The numerical method used to solve the GP equation and the associated GPS code are described in Section \ref{sec:NUM}. An extended subsection is devoted to the derivation of dimensionless equations, by setting a unified framework covering different existing approaches in the literature. The particular numerical methods used in this study  to advance the GP wave function in imaginary-time (ARGLE) or real-time (GP) are also described. Section \ref{sec:prepar} presents in detail four different approaches to generate the initial field for the simulation of decaying GP-QT: Taylor-Green (TG), Arnold-Beltrami-Childress (ABC), smoothed random phase (SRP) and  random vortex rings (RVR).  Each method is associated to a benchmark. The results obtained for the four benchmarks are discussed in Section \ref{sec:Results}. We  present values, spectra and structure functions of main quantities of interest (energy, helicity, etc.)  that could be useful to benchmark numerical codes simulating QT with the  GP model. Finally, 
the main features of the benchmarks and their possible extensions are summarized in Section \ref{sec:Conclusion}.



\section{Mathematical and physical formulation}\label{sec:Model}

We present in this section the GP model used for the numerical simulation of QT. The equations in this part apply to both BEC and superfluid helium QT. We first introduce  the GP mean field equation and its stationary version. The hydrodynamic analogy, important for relating QT to CT, is then derived using the Madelung transform. The main conservation laws are emphasized as important checkpoints for numerical simulations. Finally, we derive the Bogoliubov-de Gennes (BdG) linearized system for the response of the system to small oscillations. The BdG formalism will be used in the next section to derive the dispersion relation for the QT flow.

\subsection{The Gross-Pitaevskii model}\label{sec:GPEModel}

In the zero-temperature limit, the superfluid system of weakly interacting bosons of mass $m$, is described by the Gross-Pitaevskii mean field equation \citep{BEC-book-2003-pita}:
\begin{equation}
i \hbar \frac{\partial }{\partial t} \psi(\vec{x},t) =
\left( -\frac{\hbar^2}{2 m} \nabla^2 + {V}_{\trap}(\vec{x}) +  g\,|\psi(\vec{x},t)|^2 \right) \psi(\vec{x},t),
\label{eq-gpe}
\end{equation}
where $V_{\trap}$ is the external trapping potential and $g$ the non-linear interaction coefficient
\begin{equation}
g=\frac{4\pi \hbar^2 a_s}{m},
\label{eq-g}
\end{equation}
with $a_s$ the s-wave scattering length for the binary collisions within the system.  

The complex wave function $\psi$ is generally represented as (Madelung transformation):
\begin{equation}
\psi =\sqrt{n(\vec{x},t)}\, e^{i \theta (\vec{x},t)}, \quad n(\vec{x},t) = |\psi(\vec{x},t)|^2 = \psi(\vec{x},t) \psi^{\ast}(\vec{x},t),
\label{eq-psi}
\end{equation}
with $n$ the atomic density and $\theta$ the phase of the order parameter. We denote by $\psi^\ast$ the complex conjugate of the wave function.
Inserting \eqref{eq-psi} in \eqref{eq-gpe} we obtain by  separating  the imaginary and real parts, the time-evolution equation for the atomic density:
 \begin{equation}
\frac{\pl n}{\pl t} +  \frac{\hbar}{m}\left(\nabla n \cdot \nabla \theta + n \nabla^2 \theta  \right) =  0, \quad \mbox{or}\quad
\frac{\pl n}{\pl t} +  \frac{\hbar}{m}\nabla \cdot (n \nabla \theta) =  0,
\label{eq-gpe-density}
\end{equation}
and the equation for the phase:
 \begin{equation}
\hbar \frac{\pl \theta}{\pl t} + \frac{\hbar^2}{2 m} (\nabla \theta)^2 + V_{\trap} + g n - \frac{\hbar^2}{2 m} \frac{1}{\sqrt{n}} \nabla^2 (\sqrt{n}) = 0.
\label{eq-gpe-phase}
\end{equation}
The last term in the left-hand side of \eqref{eq-gpe-phase} is called {\em quantum pressure} and is a direct consequence of the Heisenberg uncertainty principle \citep{BEC-book-2003-pita}. It depends on the gradient of density, suggesting that quantum effects are important in non-uniform gases and, for uniform systems, close to vortex cores.
The system of equations \eqref{eq-gpe-density}-\eqref{eq-gpe-phase} is equivalent to the original GP equation \eqref{eq-gpe} and will be used in the next section to derive a hydrodynamic analogy.

It is important to define the integral quantities conserved by the GP equation \eqref{eq-gpe}. First, after multiplying \eqref{eq-gpe} by $\psi^{\ast}$, integrating in space 	and taking the imaginary part, we obtain that the number of atoms ${N}$ in the system is conserved ($\pl{N}/\pl t =0$), with
\begin{equation}
{N}=\int |\psi|^2  \, d{\vec x}= \int n \, d{\vec x}.
\label{eq-N}
\end{equation}
Second, after multiplying \eqref{eq-gpe} by $\pl \psi^{\ast}/\pl t$, integrating in space 	and taking the real part, we obtain that 
	the energy of the system is also conserved:
\begin{equation}
\label{eq-energy}
{E}(\psi) = \int \left(\frac{\hb^2}{2m} |\nabla \psi|^2 + V_{\trap}\, |\psi|^2 +{\frac{1}{2} g} |\psi|^4 \right)\, d{\vec x}.
\end{equation}
One can also obtain that  $\pl {E}/\pl t =0$ by directly differentiating \eqref{eq-energy}.

Finally, it is useful to introduce stationary solutions of the GP equation. These are obtained by considering that the wave function evolves in time as:
 \begin{equation}
 \psi (\vec{x},t) = \Psi(\vec{x})\exp(-i \mu t / \hbar),
 \label{eq-psi-stat}
 \end{equation}
 with $\mu$ the chemical potential. Note that $|\psi|=|\Psi|$ and the number of atoms in \eqref{eq-N} is conserved by the stationary field.
The time-evolution GP equation \eqref{eq-gpe} then reduces to the stationary (time-independent) GP equation:
\begin{equation}
-\frac{\hb^2}{2m} \nabla^2 \Psi + V_{\trap} \Psi + g |\Psi|^2 \Psi = \mu \Psi.
\label{eq-gpe-stat}
\end{equation}
The chemical potential $\mu$ is fixed by the normalization condition \eqref{eq-N} and expressed from \eqref{eq-psi-stat} and \eqref{eq-gpe-stat} as:
\begin{equation}
{\mu}
= \frac{1}{ {N}}\,\int \left(\frac{\hb^2}{2m} |\nabla \Psi|^2 + V_{\trap}\, |\Psi|^2 + g |\Psi|^4 \right)\, d{\vec x}
=  \frac{1}{ {N}} \left({E}(\Psi) + \int  \frac{1}{2} g |\Psi|^4 \, d{\vec x}\right).
\label{eq-mu}
\end{equation}

\subsection{Hydrodynamic analogy}\label{sec:hydro}

The hydrodynamic analogy of the GP equation \eqref{eq-gpe} is obtained by relating the wave function $\psi$ to a superfluid flow of mass density 
\begin{equation}
\rho (\vec{x},t)  = m\, n(\vec{x},t) = m\, |\psi(\vec{x},t)|^2,
\label{eq-rho}
\end{equation}
and velocity
\begin{equation}
\vec{v} (\vec{x},t)  = \frac{\hbar}{m}\, \nabla \theta (\vec{x},t)  =  \frac{\hbar}{\rho}\, \frac{ \psi^\ast \nabla \psi - \psi \nabla \psi^\ast}{2 i}. 
\label{eq-vel}
\end{equation}
For a flow of non-vanishing density ($\rho \neq 0$), we infer from \eqref{eq-vel} that the superfluid is irrotational:
\begin{equation}
\nabla \times \vec{v} =0.
\label{eq-vel-rot}
\end{equation}
The velocity $\vec{v}$ is usually related to the current density \citep{BEC-book-2003-pita}:
 \begin{equation}
\vec{j} (\vec{x},t) = n(\vec{x},t)\, \vec{v} (\vec{x},t) = \frac{1}{m}\,  \rho (\vec{x},t)\, \vec{v} (\vec{x},t).
 \label{eq-j}
 \end{equation}
Taking the divergence of \eqref{eq-j} and using \eqref{eq-vel} we obtain the expression:
 \begin{equation}
\nabla\cdot \vec{j}  =  \frac{\hbar}{m}\, \frac{\psi^\ast \nabla^2 \psi - \psi \nabla^2 \psi^\ast}{2 i} =  \frac{\hbar}{m}\left(\nabla n \cdot \nabla \theta + n \nabla^2 \theta  \right)  
= \frac{\hbar}{m}\nabla \cdot (n \nabla \theta).
\label{eq-div-j}
\end{equation}
We infer from \eqref{eq-div-j} and   \eqref{eq-gpe-density} that the equation for the time evolution of the density could be now written as:
 \begin{equation}
\frac{\pl n}{\pl t} + \nabla\cdot \vec{j}  =  0 \Longleftrightarrow 
\frac{\pl \rho}{\pl t} + \nabla\cdot (\rho \vec{v})  =  0.
\label{eq-cont}
\end{equation}
Equation \eqref{eq-cont} is the continuity equation of the superflow, expressing the conservation of the number of particles  ${N}$ given by \eqref{eq-N}.

To complete the hydrodynamic equations, we take the gradient of the equation  \eqref{eq-gpe-phase} and use \eqref{eq-vel} to obtain the equation for the time evolution of the velocity:
 \begin{equation}
 \frac{\pl \vec{v}}{\pl t} + \frac{1}{2} \nabla (\vec{v}^2) =  - \frac{1}{m} \nabla\left(gn + V_\trap\right) + \frac{\hbar^2}{2 m^2} \nabla\left(\frac{1}{\sqrt{\rho}} \nabla^2 (\sqrt{\rho})\right).
\label{eq-mom-vel}
\end{equation}
Using the identity 
$$\frac{1}{2} \nabla (\vec{v}^2)  = (\vec{v} \cdot \nabla)\vec{v} + \vec{v} \times (\nabla \times \vec{v}),$$
and \eqref{eq-vel-rot}, we infer from \eqref{eq-mom-vel} that, after neglecting the last (quantum pressure) term and the trapping potential ($V_\trap = 0$):
\begin{equation}
\frac{\pl \vec{v}}{\pl t} + (\vec{v} \cdot \nabla)\vec{v} =  -\frac{1}{\rho}\nabla\left(\frac{g \rho^2}{2 m^2}\right). 
\label{eq-mom-vel-E}
\end{equation}

Equations \eqref{eq-cont} and \eqref{eq-mom-vel-E} are similar to the Euler equations describing the evolution of a compressible, barotropic and inviscid classical flow with pressure: 
\begin{equation}
\label{eq-pressure}
P = \frac{g \rho^2}{2 m^2}.
\end{equation}
Notice that \eqref{eq-mom-vel} is exactly the non-conservative form  for the time evolution of the momentum $\rho \vec{v}= m \vec{j}$.  As a consequence,
equations \eqref{eq-cont} and \eqref{eq-mom-vel-E} show that the total momentum of the superfluid,
 \begin{equation}
\vec{p} = \int (\rho \vec{v}) \, d{\vec x}= m \int \vec{j} \, d{\vec x} = {\hbar} \int \Imag(\psi^\ast \nabla \psi) \, d{\vec x},
\label{eq-mom-total}
\end{equation}
is also conserved. Note that the momentum conservation can be directly derived from \eqref{eq-gpe}. We recall that the energy \eqref{eq-energy} is conserved; it can be rewritten as:
\begin{equation}
\label{eq-energy-rho}
{E} = \int \left(\frac{1}{2} \rho \vec{v}^2 + \frac{\hb^2}{2m^2} |\nabla \sqrt{\rho}|^2 + \frac{\rho}{m} V_\trap + {\frac{g \rho^2}{2 m^2}}  \right)\, d{\vec x}.
\end{equation}
To summarize, the three main integral invariants of the GP equation \eqref{eq-gpe} are: the number of particles ${N}$ \eqref{eq-N}, the energy  ${E}$, in the form \eqref{eq-energy} or \eqref{eq-energy-rho}, and, when $V_\trap = 0$, the momentum $\vec{p}$ \eqref{eq-mom-total}.

\subsection{The Thomas-Fermi limit}

The Thomas-Fermi regime is characterized by strong interactions (the kinetic energy is negligible compared to the interaction energy). 
If the kinetic energy (first term) is neglected in \eqref{eq-gpe-stat}, the stationary solution $\Psi$ takes a simple form, that can be expressed from the  equation:
 \begin{equation}
V_\trap(\vec{x}) + g |\Psi|^2 = \mu.
\label{eq-TF}
\end{equation}
The Thomas-Fermi limit \eqref{eq-TF} can  also be derived from \eqref{eq-mom-vel}   by considering $\vec{v}=0$ and neglecting the quantum pressure term. 
For a harmonic trapping potential with $V_{\trap} = m \omega^2 \Vec{x}^2 / 2$, with a reference trapping frequency $\omega$, 
the Thomas-Fermi regime is attained when $N a_s/\aho \gg 1$, with $\aho=\sqrt{{\hb}/{(m \omega)}}$ the harmonic oscillator length. In the absence of a trapping potential, we obtain the Bogoliubov relation $\mu = g n$ \citep{BEC-book-2003-pita}.

\subsection{Small amplitude oscillations}\label{sec:BdG}

Since  the GP equation (\ref{eq-gpe}) sustains wave solutions, it is interesting to estimate the response of the system to small perturbations. The Bogoliubov-de Gennes model is based on the linearisation of  (\ref{eq-gpe}) assuming that:
\begin{equation}
\psi({\vec{x}},t)= \left[ \Psi({\vec{x}}) + \delta \psi(\vec{x},t) \right] \exp(-{i \mu t/\hb}) ,
\label{eq-BdG-psi}
\end{equation}
where  $\Psi({\vec{x}})$ is a stationary solution satisfying \eqref{eq-gpe-stat}, and $\delta \psi$ a perturbation of small amplitude.  After discarding the terms of order $\delta \psi^2$, we obtain the following linearized equation:
\begin{equation}
\label{eq-BdG-delta-psi}
\mu \delta \psi + i \hb \frac{\partial \delta\psi}{\partial t} = -\frac{\hb^2}{2m} \nabla^2 {\delta \psi} + V_{\trap} {\delta \psi} 
+ g \Psi^2  \delta \psi^\ast + 2 g |\Psi|^2 \delta \psi.
\end{equation}
We consider the following form of the perturbation:
\begin{equation}
\delta \psi = 
\left(a({\bf x}) e^{-i \omega t} + {b^\ast}({\bf x}) e^{i {\omega}^\ast t} \right),
\label{eq-BdG-deltapsi}
\end{equation}
where $a, b$ are complex functions (small amplitudes) and $\omega$ is a complex frequency. From (\ref{eq-BdG-delta-psi}) we separate the terms in $e^{-i \omega t}$ and $e^{i \omega^\ast t}$  and obtain the following Bogoliubov-de Gennes equations:
\begin{align}
\label{eq-BdG-aplitude1}
\left(
-\frac{\hb^2}{2m} \nabla^2 + V_{\trap} - \mu + 2 g |\Psi|^2 \right) a + g \Psi^2 b  &= \phantom{-} \hb \omega\, a,\\
\label{eq-BdG-aplitude2}
g (\Psi^\ast)^2 a  + \left(-\frac{\hb^2}{2m} \nabla^2 + V_{\trap} -\mu  + 2 g |\Psi|^2 \right) b  &= - \hb \omega\, b.
\end{align}
Solving the system of differential equations \eqref{eq-BdG-aplitude1}-\eqref{eq-BdG-aplitude2} provides the eigenfrequencies $\omega$ and the modes amplitudes $a$ and $b$. Note from \eqref{eq-BdG-deltapsi} that if $(a, b, \omega)$ is a solution of this system, then $(b^\ast, a^\ast, -\omega^\ast)$ is also solution (representing the same physical oscillation). Moreover, due to the Hamiltonian nature of the system \eqref{eq-BdG-aplitude1}-\eqref{eq-BdG-aplitude2}, $\omega^\ast$ is also an eigenfrequency. Consequently, modes corresponding to eigenfrequencies with a non-zero imaginary part are amplified in time. Such modes are associated to the dynamic instability of the system.

\section{Description of the quantum turbulent flow}\label{sec:QTscales}
Since the focus of this study is the QT in superfluid helium, we set to zero the trapping potential  ($V_{\trap}=0$) in the remainder of the paper. In this section, we introduce the main characteristics of the QT flow that have to be resolved in numerical simulations. We first analyze the constant density uniform superflow, which is the background of the QT field. Important characteristic scales for properly designing numerical simulations are introduced: the healing length and the sound velocity. The response of this uniform flow to small perturbations is also presented by deriving the Bogoliubov dispersion relation. This relation is an important guide in setting the numerical resolution  in spectral methods. It also defines the domain of the validity of the GP model in describing superfluid helium QT. We then introduce the notion of quantum vortex and its representation on a computational grid. Finally, we present the main integral quantities that will be used to analyze the QT field: the compressible and incompressible kinetic energy and the helicity.

\subsection{Uniform flow and characteristic scales of the system}

An elementary solution to the stationary GP equation \eqref{eq-gpe-stat}, in absence of the trapping potential, is a flow with constant density $\rho=\rho_0$.  From  \eqref{eq-gpe-stat} and \eqref{eq-psi-stat}, we infer that the corresponding wave functions and chemical potential are related by
 \begin{equation}
\mu_0 = g |\Psi_0|^2 = g |\psi_0|^2 = g n_0 = g\, \frac{\rho_0 }{m}.
\label{eq-rho0-pot}
\end{equation}
The solution $\Psi_0$ could be taken as real and represents a first approximation of a quantum flow developing in a container of volume $V$, far from the walls. The number of atoms in the container and the energy of the system follow from  \eqref{eq-N} and \eqref{eq-energy}, respectively:
 \begin{equation}
N_0 = |\Psi_0|^2\, V = n_0 V =\frac{\rho_0 }{m}\, V.
\label{eq-rho0-N0}
\end{equation}
 \begin{equation}
E_0 = \frac{1}{2} g |\Psi_0|^4\, V = \frac{g N_0^2}{2 V}. 
\label{eq-rho0-E0}
\end{equation}
It is now possible to determine the constitutive relation of the ideal barotropic fluid:
 \begin{equation}
P_0 = - \frac{\pl E_0}{\pl V} =  \frac{g N_0^2}{2 V^2} = \frac{1}{2}  g n_0^2 = \frac{1}{2} \frac{g \rho_0^2}{m^2},
\label{eq-rho0-P0}
\end{equation}
which was put into evidence in the momentum equation \eqref{eq-mom-vel} and also defined in \eqref{eq-pressure}. Note that the pressure of the superflow does not vanish at zero temperature, as in an ideal gas \citep{BEC-book-2003-pita}. We also infer from \eqref{eq-rho0-P0} that the flow is compressible, with the sound velocity $c$ defined as in classical hydrodynamics:
 \begin{equation}
c = \sqrt{\frac{\pl P_0}{\pl \rho_0}} = \frac{\sqrt{g \rho_0}}{m} =  \sqrt{\frac{g n_0}{m}}.
\label{eq-rho0-c}
\end{equation}
Combining \eqref{eq-rho0-pot} and \eqref{eq-rho0-c} we obtain the equality:
 \begin{equation}
\mu_0  = m c^2.
\label{eq-rho0-pot2}
\end{equation}

The sound velocity gives a characteristic velocity of the system. To have a complete space-time description of the system we need to introduce a characteristic length scale. The healing length indicates the distance over which density variations take place in the system. It can de derived in several ways. If in the stationary GP equation \eqref{eq-gpe-stat} we assume that the first (kinetic-energy) term  balances the non-linear interaction term over the length $\xi$, we can use the following approximation in the stationary limit:
\begin{equation}
 \frac{\hbar^2}{2 m} \frac{\Psi}{\xi^2} \approx  g\,|\Psi|^2 \Psi \Longrightarrow \xi = \frac{\hbar}{\sqrt{2 m g |\Psi|^2}}.
\label{eq-rho0-xi-1}
\end{equation}
For a constant density flow, using \eqref{eq-rho0-pot} and \eqref{eq-rho0-pot2}, the expression of the healing length becomes:
\begin{equation}
\xi = \frac{\hbar}{\sqrt{2 m g n_0}} = \frac{\hbar}{\sqrt{2 m \mu_0}}  = \frac{1}{\sqrt{2}} \frac{\hbar}{m c}.
\label{eq-rho0-xi-2}
\end{equation}
The same expression for the healing length could be obtained from the momentum equation \eqref{eq-mom-vel} by imposing that the pressure term balances the quantum pressure term over the healing length. It results that the quantum pressure is negligible for distances $R \gg  \xi$, which is exactly the domain of validity of the hydrodynamic analogy with the Euler equations.

\subsection{Dispersion relation and Bogoliubov excitation regimes}\label{sec:dispersion}

The Bogoliubov-de Gennes linearized system derived in \S\ref{sec:BdG} is greatly simplified in the case of the flow of uniform density described by \eqref{eq-rho0-pot}, with real wave function $\Psi_0=\sqrt{n_0}$.
 The perturbations are then taken as plane waves:
\begin{equation}
\label{eq-BdG-aplitude_ab}
a(\vec{x}) = u e ^{i \vec{k}\cdot \vec{x}}, \quad b(\vec{x}) = v e ^{i \vec{k}\cdot \vec{x}},
\end{equation}
with $\vec{k}$ the wave number vector.  The BdG system \eqref{eq-BdG-aplitude1}-\eqref{eq-BdG-aplitude2} becomes for this case:
\begin{align}
\label{eq-BdG-aplitude1_0}
\left(\frac{\hb^2}{2m} \vec{k}^2 + g n_0 - \hb \omega \right) u +  (g n_0) v  &= 0,\\
\label{eq-BdG-aplitude2_0}
 (g n_0) u  +\left(\frac{\hb^2}{2m} \vec{k}^2 + g n_0 + \hb \omega \right)  v   &=  0,
\end{align}
with a non-trivial solution if
\begin{equation}
\label{eq-BdG-dispersion-1}
(\hb \omega)^2  =  \left(\frac{\hb^2}{2m} \vec{k}^2\right)^2  + (g n_0) \frac{\hb^2}{m} \vec{k}^2.
\end{equation}
Using \eqref{eq-rho0-c} to express the sound velocity and \eqref{eq-rho0-xi-2} for the healing length, the Bogoliubov dispersion relation \eqref{eq-BdG-dispersion-1} becomes:
\begin{equation}
\label{eq-BdG-dispersion-2}
\omega  =  (c k) \sqrt{1 + \frac{\xi^2 k^2}{2}}.
\end{equation}
This dispersion relation is linear for $(k \xi \ll 1)$ and the excitations in this regime are called {\em phonons} (sound waves). 
Going back to the momentum equation \eqref{eq-mom-vel}, we can easily see that in the phonons regime, the quantum pressure is negligible in front of the hydrodynamic pressure. Consequently, the validity of the hydrodynamic analogy is limited to the phonons excitations, a regime where the quantum pressure could be neglected. 
The transition between the phonon regime and the particle regime takes place at $(k \xi \sim 1)$ and the dispersion relation of the free particle is recovered for ($k \xi \gg 1$), when  $\omega \to c \xi k^2/\sqrt{2}$ (or $\omega \to \hbar k^2 / 2 m$).
The information on the separation of different regimes is quite useful for numerical simulations, since the quantity $(k \xi)$ must be carefully assessed  to represent correctly  the density waves on the the computational grid.  Note that in helium II, due to strong interactions between particles, the dispersion relation has a different shape, with a linear regime followed by a quadratic regime with a maxon (local maximum) and a roton (local minimum) \citep{QT-book-2016-barenghi}. The excitations in the quadratic region near the minimum of the dispersion curve are called rotons. Consequently, using the GP equation allows us to capture only the phonons regime of excitations in a quantum flow.
Note that GP equation can be modified by including non-local terms to model any dispersion curve \cite{Berloff2014-nonzero}.

\subsection{Quantized vortices}\label{sec:qvortices}

The definition \eqref{eq-vel} of the superflow velocity becomes singular along lines with vanishing density ($\rho=0$). The lines along which both real and imaginary part the order parameter are zero define topological defects, known as {\em quantum vortices}. The hydrodynamic analogy through the Madelung transformation becomes singular when vortices are present in the superflow. A detailed review of mathematical problems related to the Madelung transformation in presence of quantum vortices is offered in \cite{BEC-math-2013-carles-NL}. Another intriguing consequence of the hydrodynamic analogy appears when calculating the circulation $\Gamma$ along a regular path ${\mathcal C}$ surrounding a simply connected domain $S$:
\begin{equation}
\label{eq-vortex-gamma}
\Gamma = \oint_{\mathcal{C}} \vec{v}  \cdot d\vec{l}.
\end{equation}
From \eqref{eq-vel-rot} we infer by applying the Stokes theorem that $\Gamma=0$ if $S$ is a simply connected domain without vortices.  When a vortex line crosses the domain $S$, the curl of the velocity expressed by  \eqref{eq-vel-rot}  becomes a Dirac function and the circulation $\Gamma$ takes non zero values. 
It is important to note, however,  that vortex solutions are not singular solutions of the GP equation \eqref{eq-gpe}. Indeed, a straight-line vortex solution $\Psi_{v}$ is represented using cylindrical coordinates $(r, \varphi, z)$ by:
\begin{equation}
\label{eq-vortex-psi}
\Psi_{v} = |\Psi_{v}(r)|\, e^{i \kappa \varphi},
\end{equation}
where $\kappa$ is necessarily an integer to  ensure that the wave function is single valued. Using \eqref{eq-vel} we infer that the velocity around the vortex line  is tangential and singular for $r=0$:
\begin{equation}
\label{eq-vortex-vel}
\vec{v}_{v} = \frac{\hbar}{m}\frac{1}{r} \frac{\pl (\kappa \varphi)}{\pl \varphi}\, \vec{e}_\varphi= \frac{\hbar}{m}\frac{\kappa}{r}\, \vec{e}_\varphi.
\end{equation}
The circulation around this vortex line follows from \eqref{eq-vortex-gamma}:
\begin{equation}
\label{eq-vortex-gammav}
\Gamma_{v} = (2 \pi) \kappa  \frac{\hbar}{m} = \kappa  \frac{h}{m}.
\end{equation}
This quantification of the vortex line circulation is the outstanding difference between quantum and classical hydrodynamics. The integer $\kappa$ is usually referred to as winding number or charge of the vortex.

The vortex solution \eqref{eq-vortex-psi} can be further developed by assuming that:
\begin{equation}
\label{eq-vortex-psi-mod}
|\Psi_{v}(r)|  = \sqrt{n_0} \, f(\eta),\quad \eta=\frac{r}{\xi},
\end{equation}
with $n_0$ the far-field background solution with constant density and $\xi$ the healing length. Injecting \eqref{eq-vortex-psi} with \eqref{eq-vortex-psi-mod} in the stationary GP equation \eqref{eq-gpe-stat} and using definitions \eqref{eq-rho0-pot} and \eqref{eq-rho0-xi-2}, we obtain the following ordinary differential equation for $f$:
\begin{equation}
\label{eq-vortex-feta}
- \frac{1}{\eta} \frac{d}{d \eta}\left(\eta \frac{d f}{d \eta}\right) + f \left(\frac{\kappa^2}{\eta^2}-1 \right) + f^3 =0,
\end{equation}
with limit conditions: $f \rightarrow 0$ for $\eta \rightarrow 0$ and $f \rightarrow 1$ for $\eta \rightarrow \infty$. The asymptotic behaviour of this solution near the origin ($r=0$) is well-known \citep{BEC-math-1990-neu}:
\begin{equation}
\label{eq-vortex-as}
f(\eta) \sim  \eta^{|\kappa|} + {\cal O} (\eta^{|\kappa|+2}), \quad \eta \rightarrow 0,
\end{equation}
suggesting that the vortex core, \ie the region near the vortex line where the density is varying in a significant way, is of the order of the healing length $\xi$.

Using \eqref{eq-vortex-vel},  the kinetic energy ($\int \rho \vec{v}_{v}^2 d\vec{x}$) of the vortex solution, which is the main contribution to the total energy \eqref{eq-energy-rho}, results to be proportional to $\kappa^2$  \citep[\eg][]{QT-book-2016-barenghi}.
This implies that a multiply quantized vortex with $\kappa > 1$ is energetically unstable and split into $\kappa$-singly quantized vortices in GP-QT.
As shown in \cite{Takeuchi2018}, the multiply quantized vortex is also dynamically unstable, since the complex frequency $\omega$ obtained in the Bogoliubov-de Gennes equations \eqref{eq-BdG-aplitude1}-\eqref{eq-BdG-aplitude2} has a non-zero imaginary part.


\subsection{Energy decomposition}\label{sec:energ-decomp}

The energy is the first integral quantity that will be used to characterize the QT flow field. The accuracy of numerical simulations in conserving this quantity is an important checkpoint to validate the numerical scheme and the grid resolution. As in CT, energy spectra will be used to identify different (Kolmogorov) regimes/ranges in the structure of the turbulent field. Starting from the observation that the QT field can be viewed as a background uniform flow to which a large number of quantum vortices are superimposed, the total energy of the system in QT studies \citep[\eg][]{Nore97a,Nore97b} is generally computed using the form:
\begin{equation}\label{eq-energy-mod}
{E}_T(\psi) = \int \left(\frac{\hb^2}{2m} |\nabla \psi|^2 
+{\frac{1}{2} g} \left(|\psi|^2 -  |\Psi_0|^2\right)^2 \right)\, d{\vec x},
\end{equation}
where $ |\Psi_0|^2=n_0$ is the atomic density of the uniform flow. This expression is strictly equivalent to the form \eqref{eq-energy} of the energy because of the conservation of the number of atoms \eqref{eq-N}. The corresponding GP equation, equivalent to \eqref{eq-gpe} is then:
\begin{equation}
i \hbar \frac{\partial }{\partial t} \psi(\vec{x},t) = \frac{\delta E_T}{\delta \psi^\ast} =
\left( -\frac{\hbar^2}{2 m} \nabla^2  +  g\,\left(|\psi(\vec{x},t)|^2 - |\Psi_0|^2\right) \right) \psi(\vec{x},t),
\label{eq-gpe-modif}
\end{equation}
Using the hydrodynamic analogy presented in \S\ref{sec:hydro}, the total energy \eqref{eq-energy-mod} can be also presented as:
\begin{equation}
\label{eq-energy-mod-rho}
{E}_T(\rho, \vec{v}) = \int \left(\frac{1}{2} \rho \vec{v}^2 + \frac{\hb^2}{2m^2} |\nabla \sqrt{\rho}|^2 + {\frac{1}{2 m^2} g} \left(\rho -  \rho_0\right)^2  \right)\, d{\vec x}.
\end{equation}
Again \eqref{eq-energy-mod-rho} is strictly equivalent to \eqref{eq-energy-rho}. The three terms is \eqref{eq-energy-mod-rho} correspond to
\citep{Nore97a,Nore97b}: \\
-- the kinetic energy
\begin{equation}
\label{eq-energy-Ekin}
E_{\rm kin} = \int \frac{|\sqrt{\rho} {\bf v}|^2}{2}\, d\vec{x},
\end{equation}
-- the so-called quantum energy (expressed using \eqref{eq-rho0-xi-2})
\begin{equation}
\label{eq-energy-Eq}
E_{\rm q}= \int \frac{\hb^2}{2m^2} |\nabla \sqrt{\rho}|^2\, d\vec{x} =
\int  c^2 \xi^2 |\nabla \sqrt{\rho}|^2 \, d\vec{x}.
\end{equation}
-- and the internal energy (expressed using  \eqref{eq-rho0-c}):
\begin{equation}
\label{eq:def-eint-mod}
E_{\rm int} = \int  {\frac{1}{2 m^2} g} \left(\rho -  \rho_0\right)^2   \, d\vec{x} = \int \frac{c^2 (\rho-\rho_0)^2}{2 \rho_0}  \, d\vec{x}.
\end{equation}
The kinetic energy $E_{\rm kin}$ can be further decomposed \citep{Nore97a,Nore97b} as the sum of a {\em compressible} part $E_{\rm  kin}^{\rm c}$
and an  {\em incompressible} part $E_{\rm kin}^{\rm i}$:
\begin{equation}
\label{eq:def-ekin-decomp}
E_{\rm kin}^{\rm c} = \int \frac{|(\sqrt{\rho} {\bf v})^{\rm c}|^2}{2}\, d{\bf x}, \qquad
E_{\rm kin}^{\rm i} = \int  \frac{|(\sqrt{\rho} {\bf v})^{\rm i}|^2}{2}\, d{\bf x},
\end{equation}
owing to the Helmholtz decomposition:
\begin{align}
(\sqrt \rho
{\bf v})=(\sqrt \rho {\bf v})^{\rm c}+ (\sqrt \rho {\bf v})^{\rm i}, \quad \mbox{with}\,\,
\nabla \times (\sqrt{\rho} {\bf v})^{\rm c}=0, \quad \mbox{and}\,\, \nabla \cdot (\sqrt \rho {\bf v})^{\rm i}=0.
\end{align}

\subsection{Spectra and structure functions}\label{sec:spectra}

Spectra of the different components of the energy and structure functions of velocity will be used to analyze the QT field, as in CT. The energy spectra  are computed using the following expressions resulting after applying Parseval's theorem for the Fourier transform:
\begin{align}
\begin{split}
E_{\rm kin}^i(k) &= \frac{1}{2 (2 \pi)^3} \int_{|{\bf k}| = k}  |\mathcal{F}_{\bf k}(\sqrt{\rho} {\bf v})^i|^2 \,d\Omega_{\bf k}, \\
E_{\rm kin}^c(k) &= \frac{1}{2 (2 \pi)^3} \int_{|{\bf k}| = k} |\mathcal{F}_{\bf k}(\sqrt{\rho} {\bf v})^c|^2 \,d\Omega_{\bf k}, \\
E_{\rm int}(k) &= \frac{c^2}{2 \rho_0 (2 \pi)^3} \int_{|{\bf k}| = k}  |\mathcal{F}_{\bf k}(\rho-\rho_0)|^2 \,d\Omega_{\bf k}, \\
E_{\rm q}(k) &= \frac{c^2 \xi^2}{(2 \pi)^3} \int_{|{\bf k}| = k} |\mathcal{F}_{\bf k}(\nabla \sqrt{\rho})|^2 \,d\Omega_{\bf k},
\end{split}
\end{align}
where $\mathcal{F}_{\bf k}$ is the Fourier transform
\begin{equation}
\mathcal{F}_{\bf k}(f({\bf x})) = \int  f({\bf x}) e^{- i {\bf k} \cdot {\bf x}} \,d{\bf x}, \quad
\mathcal{F}^{-1}_{\bf x}(g({\bf k})) = \frac{1}{(2 \pi)^3} \int  g({\bf k}) e^{i {\bf k} \cdot {\bf x}} \,d{\bf k},
\end{equation}
and $\Omega_{\bf k}$ is the solid angle in the spectral space. Note that the components of the energy \eqref{eq-energy-Ekin}-\eqref{eq:def-eint-mod} can also be represented and computed in the spectral space following:
\begin{align}
\begin{split}
E_{\rm kin} &= \frac{1}{2 (2 \pi)^3} \int  |\mathcal{F}_{\bf k}(\sqrt{\rho} {\bf v})|^2 \,d{\bf k} = \int  E_{\rm kin}(k)\, dk, \\
E_{\rm int} &= \frac{c^2}{2 \rho_0 (2 \pi)^3} \int  |\mathcal{F}_{\bf k}(\rho-\rho_0)|^2 \,d{\bf k}= \int  E_{\rm int}(k)\, dk, \\
E_{\rm q} &= \frac{c^2 \xi^2}{(2 \pi)^3} \int  |\mathcal{F}_{\bf k}(\nabla \sqrt{\rho})|^2 \,d{\bf k} = \int  E_{\rm q}(k)\, dk.
\end{split}
\end{align}

The structure function for the velocity following the $x$-direction (with unitary vector $\vec{e}_x$) is defined as:
\begin{equation}
S^p_{\para}(r) = \int \left(\left( \vec{v}(\vec{x}+r \vec{e}_x)-\vec{v}(\vec{x}) \right)\cdot \vec{e}_x\right)^p,
\label{strfunc-def}
\end{equation}
where $p$ is the order of the structure function and $r$ the length scale. Similar expressions are used for the structure functions following the $y$ and $z$ directions. Assuming a homogeneous and isotropic distribution of the QT velocity field statistics, averaging over different directions should give the same results. As a verification, for $p=2$ and large length scale $r$, the structure function could be reasonably approximated by:
\begin{equation}
\lim_{r\rightarrow \infty} S^2_{\para}(r) \simeq 2\int |\vec{v}(\vec{x})\cdot \vec{e}_x|^2 = 2\int v_x^2.
\label{strfunc-verif}
\end{equation}

\subsection{Helicity}\label{sec:helicity}

The helicity is another important integral quantity characterizing the QT flow field.
The definition of helicity in a classical flow is:
\begin{equation}
H = \int {\bf v}\cdot
{\bm \omega} \, d\vec{x},
\label{eq-helicity}
\end{equation}
where ${\bm \omega}=\nabla \times {\bf v}$ is the
vorticity.
In a quantum fluid, the vorticity concentrates in  vortex cores as
\begin{equation}
{\bm \omega} ({\bf
	r}) = \frac{h}{m} \int  \frac{d {\bf r}_0}{d s} \delta({\bf r} -
{\bf r}_0(s)) \, d s,
\label{eq-vorticity-quantum-fluid}
\end{equation}
where ${\bf r}_0(s)$ denotes the position of the vortex
line, $\delta$ is the Dirac delta function and $s$ the arclength.
Therefore, only quantized vortices bring a non-zero contribution to the helicity.
We consider in the following the case in which all vortices form closed loops.
Then, the formalism described by \eqref{eq-helicity} and \eqref{eq-vorticity-quantum-fluid} is topologically equivalent to \citep{QT-1969-Moffat1969}:
\begin{equation}
H = \left(\frac{h}{m}\right)^2 \left( \sum_{i \neq j} Lk_{ij} + \sum_i SL_i \right),
\label{eq-quantum-helicity-topological}
\end{equation}
where $Lk_{ij} \in \mathbb{Z}$ denotes the Gauss linking number
\begin{equation}
Lk_{ij} = \frac{1}{4 \pi} \int_{C_i} \int_{C_j} \frac{({\bf X}_i - {\bf X}_j)}{|{\bf X}_i - {\bf X}_j|^3} \cdot (d{\bf X}_i \times d{\bf X}_j),
\label{eq-linking-number}
\end{equation}
between two vortex loops $i$ and $j$.
${\bf X}_i$ and ${\bf X}_j$ denote the position vectors of points on the centerlines $C_i$ and $C_j$ ($i \neq j$) for vortex loops $i$ and $j$, respectively.
$SL_i \in \mathbb{Z}$ denotes the C\u{a}lug\u{a}reanu-White self-linking number of each single vortex loop.
$SL_i$ can be also written as the sum $SL_i = Wr_i + Tw_i$ of the writhe $Wr_i \in \mathbb{R}$ and the total twist $Tw_i \in \mathbb{R}$.
The writhe is defined by
\begin{equation}
Wr_i = \frac{1}{4 \pi} \int_{C_i} \int_{C_i} \frac{({\bf X}_i - {\bf Y}_i)}{|{\bf X}_i - {\bf Y}_i|^3} \cdot (d{\bf X}_i \times d{\bf Y}_i),
\label{eq-writhing-number}
\end{equation}
where ${\bf X}_i$ and ${\bf Y}_i$ denote two distinct points on the same curve $C_i$.
The total twist $Tw_i$ can be further decomposed in terms of normalized total torsion $T_i \in \mathbb{R}$ given by
\begin{equation}
T_i = \frac{1}{2 \pi} \int_{C_i}  \tau(s)\, ds,
\end{equation}
and intrinsic twist $N_i \in \mathbb{Z}$ of the ribbon $R(C_i,C_i^\ast)$, where $\tau(s)$ is the torsion of $C_i$. The second curve $C_i^\ast$ of the ribbon $R(C_i,C_i^\ast)$ is fixed in order for the ribbon to be identified by the points of constant phase that lie on the $\epsilon$-portion of the iso-phase surface.

Within the definition \eqref{eq-quantum-helicity-topological}, the helicity always vanishes due to the identity  \citep{QT-2015-rica}:
\begin{equation}
\sum_{i \neq j} Lk_{ij} + \sum_i Wr_i = - \sum_i Tw_i.
\end{equation}
As a consequence,  the centerline helicity
\begin{align}
\begin{split}
H
&= \left(\frac{h}{m}\right)^2 \left( \sum_{i \neq j} Lk_{ij} + \sum_i Wr_i \right) \\
&= \frac{h^2}{4 \pi m^2} \sum_{i,j} \int_{C_i} \int_{C_j} \frac{({\bf X}_i - {\bf X}_j)}{|{\bf X}_i - {\bf X}_j|^3} \cdot (d{\bf X}_i \times d{\bf X}_j),
\end{split}
\label{eq-centerline-helicity}
\end{align}
has also been introduced as a nonzero helicity in quantum fluid.
Note that the double integral in Eq. \eqref{eq-centerline-helicity} means Eq. \eqref{eq-linking-number} for $i \neq j$ and Eq. \eqref{eq-writhing-number} for $i = j$.
Although the centerline helicity is just the quantification of how knotted vortex lines are, its calculation requires detailed extraction of all centerlines of the quantized vortices. This can be done  by using, for example, the interpolation of the computed wave function at sub-grid scales  \citep{QT-2014-Scheeler2014}.

A new method which yields the same results as the centerline helicity was introduced in \cite{Clark16}.
Because the superfluid velocity ${\bf v}$ diverges at the vortex cores as shown in Eq. \eqref{eq-vortex-vel}, the direct calculation of the helicity looks ill-defined.
However, only the superfluid velocity perpendicular to the quantized vortex has a singularity, while the component parallel to the quantized vortex remains regular.
Using this observation, a  regularized velocity is defined as:
\begin{equation}
{\bf v}_{\rm reg} ={v_\parallel {\bf w}}/{\sqrt{{w}_j{w}_j}},
\end{equation}
where
\begin{equation}
{\bf w} = \frac {\hbar} {im} \bm{\nabla}
\psi^\ast \times \bm{\nabla}  \psi,
\end{equation}
and
\begin{equation}
v_\parallel=\frac {\hbar \,
	{w}_i \left[(\partial_i \partial_j \psi) \partial_j
	\psi^\ast - (\partial_i \partial_j \psi^\ast) \partial_j \psi
	\right]} {2 i m \sqrt{{w}_k{w}_k}(\partial_l \psi)(\partial_l
	\psi^\ast)}.
\end{equation}
The resultant regularized helicity 
\begin{equation}
H_{\rm reg} = \int {\bf
	v}_{\rm reg} \cdot {\bm \omega} \, d\vec{x},
\label{Hreg}
\end{equation}
can be well defined as another formalism for the centerline helicity. 
This expression was proven useful and efficient  in computing the helicity of flows with hundreds of thousands of knots  \citep{Clark16}.

\section{Numerical method and computational code}\label{sec:NUM}

Numerical simulations were performed using the parallel code called GPS (Gross-Pitaevskii Simulator) \citep{HPC-Parnaudeau-2015}. The code is based on a Fourier-spectral  space discretization and recent up-to-date numerical methods: a semi-implicit backward-Euler scheme with Krylov preconditioning for the stationary GP equation \citep{BEC-CPC-2014-antoine-duboscq} and various schemes (Strang splitting, relaxation, Crank-Nicolson) for the real-time GP equation \citep{BEC-review-2013-antoine-besse-bao}. GPS  is written in Fortran 90 and  uses a two-level communication scheme based on MPI across nodes and OpenMP within nodes.  Only one external library, FFTW \citep{HPC-FFTW}, is required for the computation. Initially designed to simulate BEC configurations (with or without rotation), the GPS code was adapted in this study for the simulation of QT flows. We present in this section the main features of the numerical system: the particular scaling used to obtain the GP dimensionless equations, and the particular numerical methods used to prepare the initial state and then to advance in real-time the GP wave function.

\subsection{Scaling and dimensionless equations}\label{sec:scaling}

For the numerical resolution of the GP equation \eqref{eq-gpe}, it is convenient to use a dimensionless form obtained after scaling all physical quantities with the characteristic scales of the QT field introduced in \S\ref{sec:QTscales}. We present below a general formalism that covers the various forms of scaling used in the physical and mathematical GP literature. We start by considering  general reference scales $(L_\scal, v_\scal)$ for length and velocity, respectively. A natural scale for the wave function $\psi$ is $\psi_\scal = \sqrt{n_0}$. With the scaling:
\begin{equation}
\adim{\vec{x}} = \frac{\vec{x}}{L_\scal}, \quad \adim{t} = \frac{v_\scal}{L_\scal} t, \quad \adim{\psi} = \frac{\psi}{\sqrt{n_0}},
\label{eq-scaling}
\end{equation}
the dimensionless GP equation \eqref{eq-gpe-modif} (with $V_{\trap} = 0$) becomes:
\begin{equation}
i \frac{\partial}{\partial \adim{t}} \adim{\psi}(\vec{x},t) = \left( - \alpha \adim{\nabla}^2 + \beta \left(|\adim{\psi}(\adim{\vec{x}},t)|^2 -1\right)\right) \adim{\psi}(\vec{x},t),
\label{eq-gpe-num}
\end{equation}
with non-dimensional coefficients:
\begin{align}
\label{eq-scaling-ab1}
\alpha &= \frac{\hbar}{2 m} \frac{1}{L_\scal v_\scal} =  \frac{\sqrt{2}  \xi c}{2 } \frac{1}{L_\scal v_\scal} &=  \frac{1}{\sqrt{2}} \left(\frac{\xi}{L_\scal}\right)  \left(\frac{c}{v_\scal}\right), \\
\beta   &= \frac{g n_0}{\hbar} \frac{L_\scal}{v_\scal} = \frac{m c^2}{\hbar} \frac{L_\scal}{v_\scal} = \frac{c^2}{\sqrt{2}  \xi c} \frac{L_\scal}{v_\scal}
&=  \frac{1}{\sqrt{2}} \left(\frac{L_\scal}{\xi}\right)\left(\frac{c}{v_\scal}\right). 
\label{eq-scaling-ab2}
\end{align}
From \eqref{eq-scaling-ab1}-\eqref{eq-scaling-ab2} we infer that non-dimensional coefficients $\alpha$ and $\beta$ are related to physically relevant scales through:
\begin{equation}
\label{eq-scaling-ab}
\adim{\xi}=\frac{\xi}{L_\scal} =   \sqrt{\frac{\alpha}{\beta}}, \quad \adim{c}=\frac{c}{v_\scal} ={\sqrt{2 \alpha \beta}} = \frac{1}{M_\scal},
\end{equation}	
where $({\xi}/{L_\scal})$   represents the non-dimensional healing length and $({c}/{v_\scal})$ the non-dimensional sound velocity. $M_\scal$ is  the reference Mach number, defined as the ratio between the reference velocity and the sound velocity.

The last  important parameters to define when working with non-dimensional equations are the size of the computational box and the grid resolution. If the physical GP equation  \eqref{eq-gpe} is defined in a cubic computational domain of physical size $L$,  the non-dimensional size $\cal L$ of the computational box used to discretize the non-dimensional equation 
\eqref{eq-gpe-num} is then:
\begin{equation}
\label{eq-scaling-calL}
{\cal L} = \frac{L}{L_\scal}= \left(\frac{L}{\xi}\right) \left(\frac{\xi}{L_\scal}\right) =  \left(\frac{L}{\xi}\right)  \sqrt{\frac{\alpha}{\beta}}.
\end{equation}
We recall that $\xi$ is a good approximation of the radius of a quantum vortex (see \S\ref{sec:qvortices}). It follows that  the ratio $\left({L}/{\xi}\right)$  in \eqref{eq-scaling-calL}  is physically important  since it indicates how many vortices the computational domain can accommodate in one direction:
\begin{equation}
\label{eq-scaling-Nv1d}
N_v^{1d}= \frac{L}{2 \xi} =  \frac{\cal L}{2}  \sqrt{\frac{\beta}{\alpha}}.
\end{equation}
Thus, increasing the value of $\cal L$ (for fixed $\alpha$ and $\beta$) will result in a higher number of vortices present in the computational box.

 When defining the grid resolution, it is important to control the number of grid points inside the vortex core. If the numerical simulation uses $N_x$ grid points in each direction, the physical grid spacing is $\delta x=L/N_x$, or in non-dimensional units $\delta \adim{x}=(\delta x/L_\scal) = {\cal L}/N_x$. It is important to quantify the grid spacing with respect to the healing length by defining:
 \begin{equation}
 \label{eq-scaling-chi}
 \chi = \frac{\delta x}{\xi}  =  \left(\frac{\cal L}{N_x}\right) \left(\frac{L_\scal}{\xi}\right)=\left(\frac{\cal L}{N_x}\right) \sqrt{\frac{\beta}{\alpha}}.
 \end{equation}
The parameter $\chi$ defined in \eqref{eq-scaling-chi} is also important when analyzing the dispersion relation \eqref{eq-BdG-dispersion-2} presented in \S\ref{sec:dispersion} to assess on the validity of the GP model. Indeed, the maximum wave-number represented on a grid of size $\delta x$ is $k_{max} =(2\pi)/(2 \delta x)$ and, consequently, the non-dimensional quantity $(k_{max} \xi)$ is expressed as:
\begin{equation}
\label{eq-scaling-kmaxi}
k_{max} \xi =\pi \frac{\xi}{\delta x} = \frac{\pi}{\chi}.
\end{equation}
The numerical resolution will be then fixed in order to keep $(k_{max} \xi) \approx 1$, ensuring that the simulation captures the regime of phonons excitations.

Using \eqref{eq-scaling-ab}-\eqref{eq-scaling-chi} we can now recover the two different approaches  existing in the literature to set the computational parameters:
\begin{itemize}
	\item In  the first approach the value of the coefficient in front of the non-linear term is fixed to $\beta=1$. The value of  $\alpha$ is fixed to $1$ or $1/2$, as commonly used in the classical GP equation. The reference scales result then from \eqref{eq-scaling-ab} and the size of the computational domain from  \eqref{eq-scaling-calL}. For the first choice we obtain:
\begin{equation}
\label{eq-scaling-choice1}
\beta=\alpha=1  \Longrightarrow \adim{\xi}=\frac{\xi}{L_\scal} = 1, \quad v_\scal = \frac{c}{\sqrt{2}}, \quad M_\scal = \frac{1}{\sqrt{2}}, \quad \frac{L}{\xi} = {\cal L} .
\end{equation}
In this case,  a quantum vortex is of size $1$, the velocity of sound is $\sqrt{2}$ (both in non-dimensional units) and $\cal L$ is a large integer (to resolve a large number of vortices). 
A second common choice is:
\begin{equation}
\label{eq-scaling-choice2}
 \beta=1, \alpha=\frac{1}{2}   \Longrightarrow \adim{\xi}=\frac{\xi}{L_\scal} = \frac{1}{\sqrt{2}}, \quad v_\scal = c, \quad M_\scal = 1, \quad \frac{L}{\xi} = \sqrt{2} {\cal L}.
\end{equation}	
In this case, the velocity of sound is $1$ and the size of the vortex of order of $(1/ {\sqrt{2}})$. We note from \eqref{eq-scaling-choice1} and \eqref{eq-scaling-choice2} that the advantage of this approach is to keep constant $\adim{\xi}$, the size of the vortex in non-dimensional units.  In exchange, the size of the domain $\cal L$ has to be adapted in function of the resolution $N_x$, in order to keep constant the parameter  $\chi$  in \eqref{eq-scaling-chi} and, implicitly, ($k_{max} \xi$) in \eqref{eq-scaling-kmaxi}. 

\item A second approach \citep{Nore97a,Nore97b} is used in the present paper. The size of the non-dimensional computational box is first  set to ${\cal L}=2\pi$, which is convenient for spectral methods. Moreover, instead of setting independently the constants $\alpha$ and $\beta$, only the value of the reference Mach number $M_\scal$ is fixed to a relatively low value. This is equivalent to impose the value of the product $\alpha \beta$. From previous relations we infer that:
\begin{equation}
\label{eq-scaling-choice3}
{\cal L}=2\pi, \quad {\alpha}{\beta} = \frac{1}{2 M_\scal^2} \Longrightarrow
\left\{\begin{array}{lcl}
\ds \frac{\xi}{L_\scal} &=& \ds {\sqrt{2}} \alpha M_\scal,\\ \vspace{0.2cm}
 v_\scal &=& \ds M_\scal c, \\ \vspace{0.2cm}
\ds  \frac{L}{\xi} &=& \ds \frac{2 \pi}{{\sqrt{2}} \alpha M_\scal},\\ \vspace{0.2cm}
\ds k_{max} \xi  &=&\ds \left(\frac{N_x}{2}\right) {\sqrt{2}} \alpha M_\scal.
\end{array}
\right.
\end{equation}	
We note from \eqref{eq-scaling-choice3}  that the parameter $\alpha$ can be used to control the non-dimensional size of the vortex, while the grid resolution $N_x$ can be set to control the parameter $(k_{max} \xi)$. We generally set for QT simulations $M_\scal = 0.5$, equivalent to $\alpha \beta =2$.

\end{itemize} 

Particular care has to be devoted when computing non-dimensional values of different quantities appearing in integral invariants or in the hydrodynamic analogy. If the non-dimensional wave function is computed from \eqref{eq-gpe-num} as $\adim{\psi} (\adim{\vec{x}},\adim{t}) = |\adim{\psi}|  \exp(i\theta(\adim{\vec{x}},\adim{t}))$, the scaled number of atoms results from \eqref{eq-N} and \eqref{eq-rho0-N0}:
\begin{equation}\label{eq-scaling-N}
\adim{N} = \frac{N}{N_0} = \frac{1}{{\cal L}^3} \int_{\cal D} |\adim{\psi}|^2 d{\adim{\vec x}},
\end{equation}
where ${\cal D}$ is the non-dimensional computation domain. The scaled total energy (per volume unit) results from \eqref{eq-energy-mod}:
\begin{equation}\label{eq-scaling-energy}
\adim{E}(\adim{\psi}) =  \frac{E_T(\psi)}{E_\scal} = \frac{2 \alpha}{{\cal L}^3} \int_{\cal D} \left(\alpha |\adim{\nabla} \adim{\psi}|^2+{\frac{\beta}{2}} \left(|\adim{\psi}|^2 - 1\right)^2 \right)\, d{\adim{\vec x}},
\end{equation}
with energy units $E_\scal =  \rho_0 v_\scal^2 L^3 = \hbar^2 n_0 L {\cal L}^2/(4 m \alpha^2 )$.


For the hydrodynamic analogy developed in \S\ref{sec:hydro}, taking as reference the density of the background uniform flow, \ie  $\rho_\scal= \rho_0=m n_0$, results in:
\begin{equation}
\adim{\rho}  =  \frac{\rho}{\rho_\scal}  = \frac{m n_0 |\adim{\psi}|^2}{{\rho_\scal}} =  |\adim{\psi}|^2.
\label{eq-scaling-rho}
\end{equation}
The momentum is derived from \eqref{eq-vel} and thus computed in the non-dimensional code as:
\begin{equation}
\adim{\rho} \adim{\vec{v}} (\adim{\vec{x}},\adim{t})  = {2\alpha}  \frac{\adim{\psi}^\ast \adim{\nabla} \adim{\psi} - \adim{\psi} \adim{\nabla} \adim{\psi}^\ast}{2 i} =({2\alpha})\, \Imag(\adim{\psi}^\ast \adim{\nabla} \adim{\psi}).
\label{eq-scaling-rhov}
\end{equation}
The non-dimensional superflow velocity also results from \eqref{eq-vel}:
\begin{equation}\label{eq-scaling-v}
\adim{\vec{v}} (\adim{\vec{x}},\adim{t})  = \frac{\vec{v}(\vec{x}, t)}{v_\scal}   = \frac{\hbar}{m v_\scal L_\scal}\, \adim{\nabla} \theta (\adim{\vec{x}},\adim{t})  =  2\alpha\,  \adim{\nabla} \theta (\adim{\vec{x}},\adim{t}),
\end{equation}
and the non-dimensional circulation of a vortex of winding number ($\kappa=1$) from \eqref{eq-vortex-gammav}:
\begin{equation}\label{eq-scaling-Gamma}
\adim{\Gamma} = \frac{\Gamma_v}{ v_\scal L_\scal}   = 2\pi \frac{\hbar}{m v_\scal L_\scal}=  4 \pi \alpha.
\end{equation}

Finally, the hydrodynamic expression \eqref{eq-energy-mod-rho} of the total energy becomes
\begin{equation}
\label{eq-scaling-energy-mod-rho}
\adim{E}(\adim{\psi}) =\frac{E_T(\psi)}{E_\scal} = \frac{1}{{\cal L}^3} \int_{\cal D} \left(\frac{1}{2} \adim{\rho} \adim{\vec{v}}^2 + (2\alpha^2) |\nabla(\sqrt{\adim{\rho}})|^2 
+ (\alpha \beta) \left(\adim{\rho} - 1\right)^2  \right)\, d\adim{\vec x},
\end{equation}
with the same reference energy as in \eqref{eq-scaling-energy} $E_\scal = \rho_0 v_\scal^2 L^3$.
In \eqref{eq-scaling-energy-mod-rho}
the first term represents the kinetic energy $\adim{E}_{kin}(\adim{\psi})$, the second the quantum energy $\adim{E}_{q}(\adim{\psi})$ and the third the interaction energy $\adim{E}_{int}(\adim{\psi})$. Note that $2\alpha^2 =\adim{\xi}^2 \adim{c}^2$ and $\alpha \beta = 1/(2 M_\scal^2)$.

To simplify the presentation, we drop in the following the tilde notation. All the developments and results in the remaining of the paper concern non-dimensional quantities.

\subsection{Numerical method to compute stationary solutions}\label{sec:imaginary-time}

To find stationary solutions to \eqref{eq-gpe-num}, a very popular numerical method is the {\em normalized gradient flow} \citep{BEC-baow-2004-Du}. The idea is to propagate the wave function following the gradient flow corresponding to the minimization of the energy \eqref{eq-scaling-energy}. In the original method,  the solution is subsequently normalized to satisfy the constraint of the conservation of the number of atoms (equivalent to imposing the $L^2$-norm of the solution). In our case, we want to find a stationary state that mimics a classical flow with prescribed velocity $\vec{v}_\ext$. Assuming that $\nabla \cdot \vec{v}_\ext = 0$, after applying a local Galilean transformation, the non-dimensional energy of the driven field becomes  \citep[see][for details]{Nore97a}:
\begin{equation}
\label{eq-energy-num-ext-vel}
{E_{\bf v}} = \frac{2 \alpha}{{\cal L}^3} \int_{\cal D} \left(\alpha \left|\nabla \psi - i\frac{\vec{v}_\ext}{2\alpha}\psi \right|^2  +{\frac{\beta}{2} } (|\psi|^2-1)^2 \right)\, d{\vec x},
\end{equation}
or, using the hydrodynamic analogy:
\begin{equation}\label{eq-energy-num-ext-vel-hydro}
{E_{\bf v}}(\psi) = \frac{1}{{\cal L}^3} \int_{\cal D} \left( \frac{1}{2} \rho\left|{\bf v}-{\vec{v}_\ext}\right|^2 + (2 \alpha^2) \left|\nabla(\sqrt{\rho})\right|^2 +  (\alpha \beta) \left(\rho - 1\right)^2 \right)\: d{\vec{x}}.
\end{equation}
In this setting, we are searching a unconstrained minimizer of  ${E_{\bf v}}$. Owing to the previous decomposition, there is a competition between the background uniform distribution $|\psi|^2 = 1$ and a phase accommodating to $ {\bf v_{\rm ext}}$. Numerically, we solve the gradient descent equation (or Advective Real Ginzburg-Landau Equation, ARGLE):
\begin{equation}
  \dfrac{\partial}{\partial \tau}\phi(\bfx,t) =\left(\alpha \nabla^2 - i{\vec{v}_\ext}\cdot \nabla - \frac{|{\vec{v}_\ext}|^2}{4\alpha} + \beta -\beta|\phi(\bfx,t)|^{2}\right)\phi(\bfx,t), \quad
\bfx \in \calD,
\label{eqn:gradient-flow}
\end{equation}
with initial condition $\phi(\bfx,0^{+})=\phi_{0}(\bfx)$.  Note that $\tau$ is here a pseudo-time used to propagate the solution until a stationary state is reached. Hence, 
this method belongs to the class of so-called {\em imaginary time propagation} methods. 
We use a semi-implicit Backward Euler scheme to advance the solution in the pseudo-time interval $(\tau_{n}, \tau_{n+1})$:
\begin{equation}
 \dfrac{{\tilde\phi}^{n+1}(\bfx)-\phi^{n}(\bfx)}{\delta \tau}=\dfrac\alpha2\nabla^2{\tilde\phi}^{n+1} + 
 \left(\dfrac\alpha2\nabla^2- i{\bf v_{\rm ext}}\cdot \nabla - \frac{|{\bf v_{\rm ext}}|^2}{4\alpha} + \beta - \beta|\phi^{n}(\bfx)|^{2}\right){\tilde\phi}^{n}(\bfx).
\label{eqn:semi-backward-Euler}
\end{equation}
where $\delta \tau:=\tau_{n+1}-\tau_n$. The resulting system is solved with spectral accuracy using  FFTs.

The ARGLE procedure stops either after a pre-definite number of pseudo-time steps or when the convergence criterion is reached:
\begin{equation}
\dfrac{||\phi^{n+1}-\phi^{n} ||_{\infty}}{\delta \tau} \le \epsilon ,
\label{eqn:crit1}
\end{equation}
where $\epsilon$ is a user defined parameter. If this criterion is not satisfied at the end of the computation, it is still possible to check the energy convergence condition:
\begin{equation}
\dfrac{|E_{\bv}(\phi^{n+1})-E_{\bv}(\phi^{n})|}{\delta \tau E_{\bv}(\phi^{n})}   \le \epsilon,
\label{eqn:crit2}
\end{equation}
which is generally less constraining than \eqref{eqn:crit1}. Note that, even when the convergence is achieved, we can only guarantee that the \emph{Backward Euler method} provides a local minimum of  ${E_{\bf v}}$.

\subsection{Numerical method for the time evolution}\label{sec:real-time}

The simulation of QT consists of solving the GP equation \eqref{eq-gpe-num} using a pseudo-spectral scheme in space and a second order splitting for the time discretization (ADI, Alternating Direction Implicit or Strang splitting). Let us rewrite \eqref{eq-gpe-num} as:
\begin{align}\nonumber
\ds \frac{\pl}{\pl t}\psi &= i\alpha\Delta \psi - i\beta|\psi|^2\psi \\
&= \mathcal L_x\psi + \mathcal L_y\psi + \mathcal L_z\psi + N(\psi),
\end{align}
with the following definitions:
\begin{align}\nonumber
\mathcal L_x\psi &= i\alpha\partial_{xx}^2\psi, &&& \mathcal L_y\psi &=  i\alpha\partial_{yy}^2\psi, \\
\mathcal L_z\psi &= i\alpha\partial_{zz}^2\psi, &&& N(\psi) &= -i\beta|\psi|^2\psi.
\end{align}
If $H$ denotes one of the previous operators ($\mathcal L_x$, $\mathcal L_y$, $\mathcal L_z$ or $N$), and $\phi$ is a given field, we denote by $S(s,H)\phi  := \psi(s)$ the solution at time $t=s$ of the following Cauchy problem:
\begin{equation}
\begin{cases} \ds \frac{\pl}{\pl t}\psi &= H(\psi), \\ \psi(t=0) &= \phi. \end{cases}
\end{equation}
Then the second order ADI time scheme could be presented as:
\begin{equation}\label{eq:adi_looptime}
\psi_{n+1} = %
S\left(\dtsd,\mathcal L_x\right)%
S\left(\dtsd,\mathcal L_y\right)%
S\left(\dtsd,\mathcal L_z\right)%
S(\dt,N)%
S\left(\dtsd,\mathcal L_z\right)%
S\left(\dtsd,\mathcal L_y\right)%
S\left(\dtsd,\mathcal L_x\right)%
\psi_n.
\end{equation}
This scheme is second order accurate, provided that we solve the partial problems exactly, \ie each term $S(s,H)$ is computed exactly. This is achieved using the spectral representation. For $j\in\{x,y,z\}$, using $\four[j]$ the Fourier transform in the $j$ direction, we obtain
\begin{align}
S\left(s,\mathcal L_j\right)\phi &= \four[j]^{-1}\left( e^{-i\alpha k_j^2s} \four[j]\phi\right).
\end{align}
For the non linear operator $S(\delta t,N)$, we notice that, if $\psi^N$ is such that $\partial_t \psi^N = -i\beta |\psi^N|^2\psi^N$, then $\partial (|\psi^N|^2)/\pl t=0$.
Consequently, this step can be solved analytically and:
\begin{equation}
S\left(s,N\right)\phi = e^{-i\beta|\phi|^2 s}\phi.
\end{equation}
In conclusion, using the spectral discretization we obtain a second order accurate scheme for the time integration.

\section{Initial data preparation and benchmarks}\label{sec:prepar}

As in numerical studies of classical turbulence, the preparation of the initial state is crucial in investigating statistical properties  of QT. We describe in this section four different approaches to generate the initial field for the simulation of decaying GP-QT.  Each method is associated to a benchmark for the GP-QT simulation. The first two methods are classical \citep{Nore97a,Nore97b} and 
	 inspired from CT. They start from defining a velocity field containing vortices. The Taylor-Green or the Arnold-Beltrami-Childress (ABC) model flows are used for this step. A wave function field is then constructed such that its nodal lines correspond to vortex lines of the velocity field.  This initial wave function is then used in the ARGLE procedure described in \S\ref{sec:imaginary-time} to generate an initial field for the real-time GP simulations.  The role of the ARGLE step is to reduce the acoustic emission of the initial field.  The last two methods are new and based on the direct manipulation of the wave function. We prescribe either a random phase field or we manufacture an initial field containing many quantum vortex rings. The four methods are described in detail below.

 
\subsection{Taylor-Green (TG) flow} \label{sec:init-tg}
 
The velocity ${\bf v}_{\rm TG}$ of the Taylor-Green (TG) three-dimensional vortex flow is defined as:
\begin{align}
\begin{split}
v_{{\rm TG},x}(x,y,z) &= \sin(x) \cos(y) \cos(z), \\
v_{{\rm TG},y}(x,y,z) &=-\cos(x) \sin(y) \cos(z), \\
v_{{\rm TG},z}(x,y,z)  &= 0 .
\end{split}
\label{tg3d}
\end{align}
To create a wave function field  $\psi_{\rm TG}$ with zeros
along vortex lines of ${\bf v}_{\rm TG}$, we make use of the Clebsch
representation of the velocity field \citep{Nore97a,Nore97b}:
\begin{equation}
\nabla \times {\bf v}_{\rm TG} = \nabla \lambda \times \nabla \mu,
\label{clebsch}
\end{equation}
with Clebsch potentials
\begin{align}
\begin{split}
\lambda(x,y,z) &= \cos(x) \sqrt{2 \;|\cos(z)|} , \\
\mu(x,y,z) &= \cos(y)  \sqrt{2 \; |\cos(z)|} \;
{\mbox{sgn}}(\cos(z) ),
\end{split}
\label{clebsch3d}
\end{align}
where sgn is the sign function. Note that a zero in the $(\lambda,\mu)$ plane corresponds to a
vortex line of ${\bf v}_{\rm TG}$ (see \cite{Nore97a,Nore97b} for
details).

\begin{figure}[!h]
	\begin{center}
	\begin{minipage}{0.72\textwidth}
		a)\\
		\includegraphics[width=\textwidth]{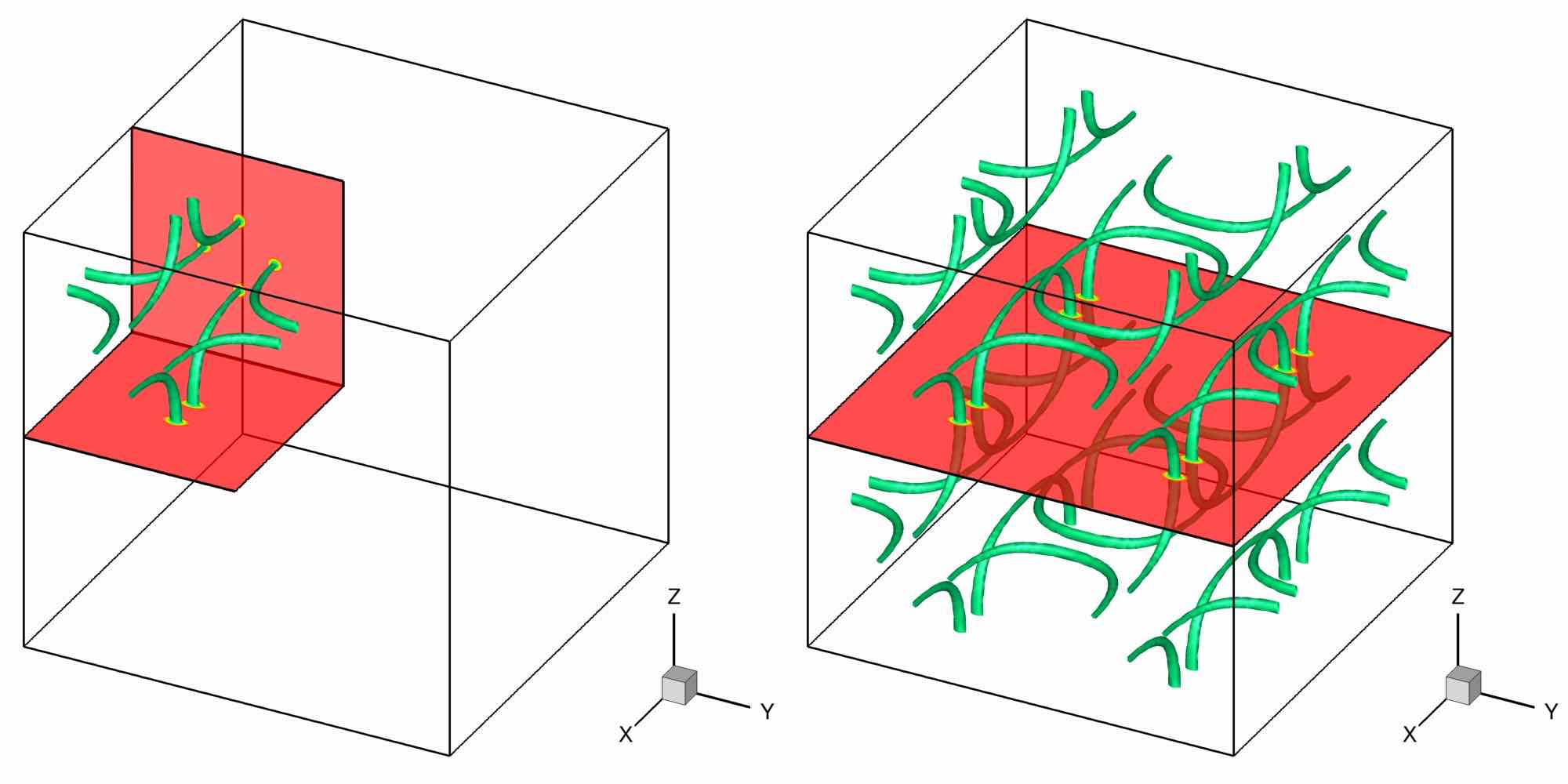}
	\end{minipage}\\
	\begin{minipage}{0.72\textwidth}
	b)\\
	\includegraphics[width=\textwidth]{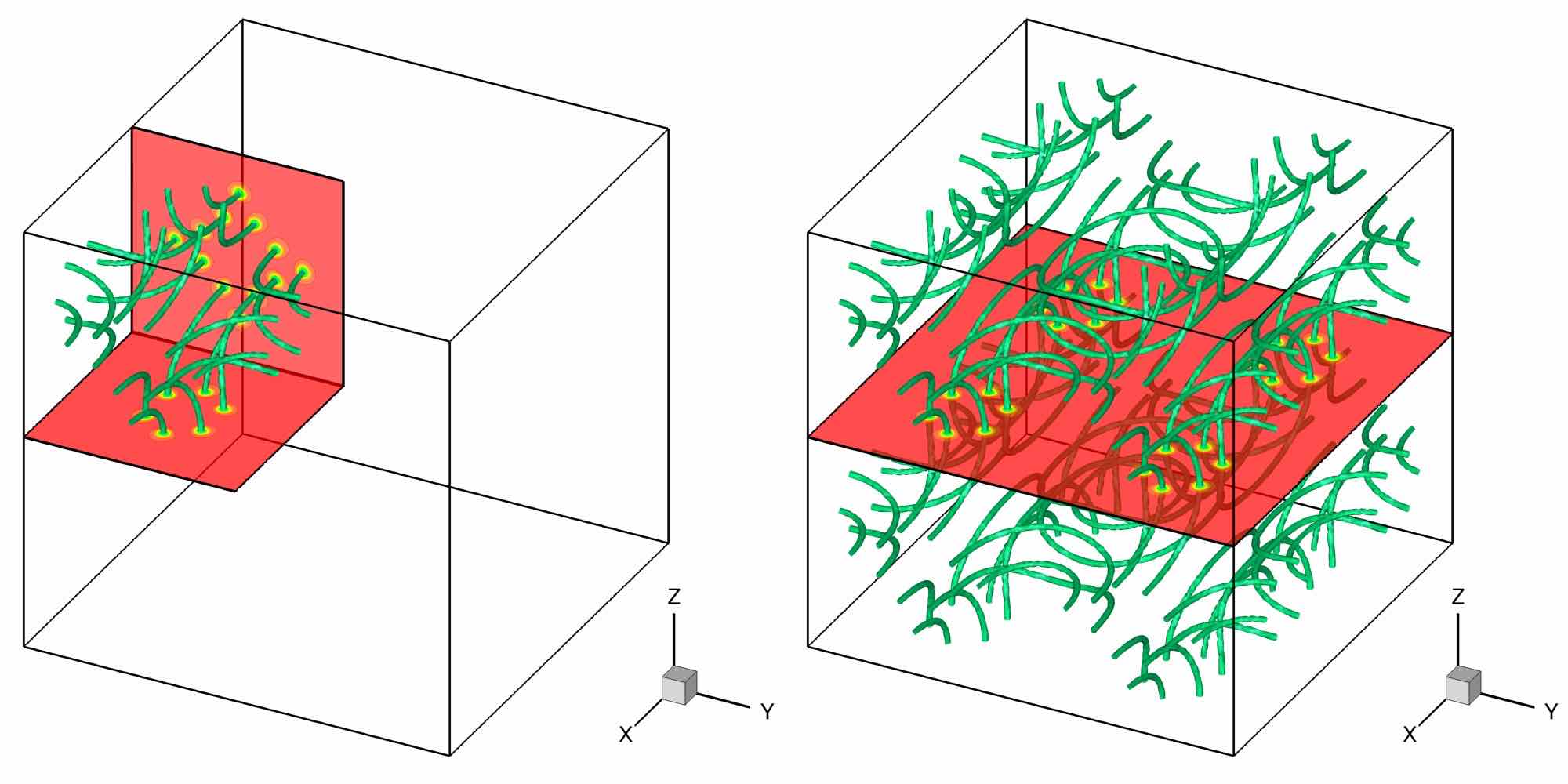}
\end{minipage}\\
	\begin{minipage}{0.72\textwidth}
	c)\\
	\includegraphics[width=\textwidth]{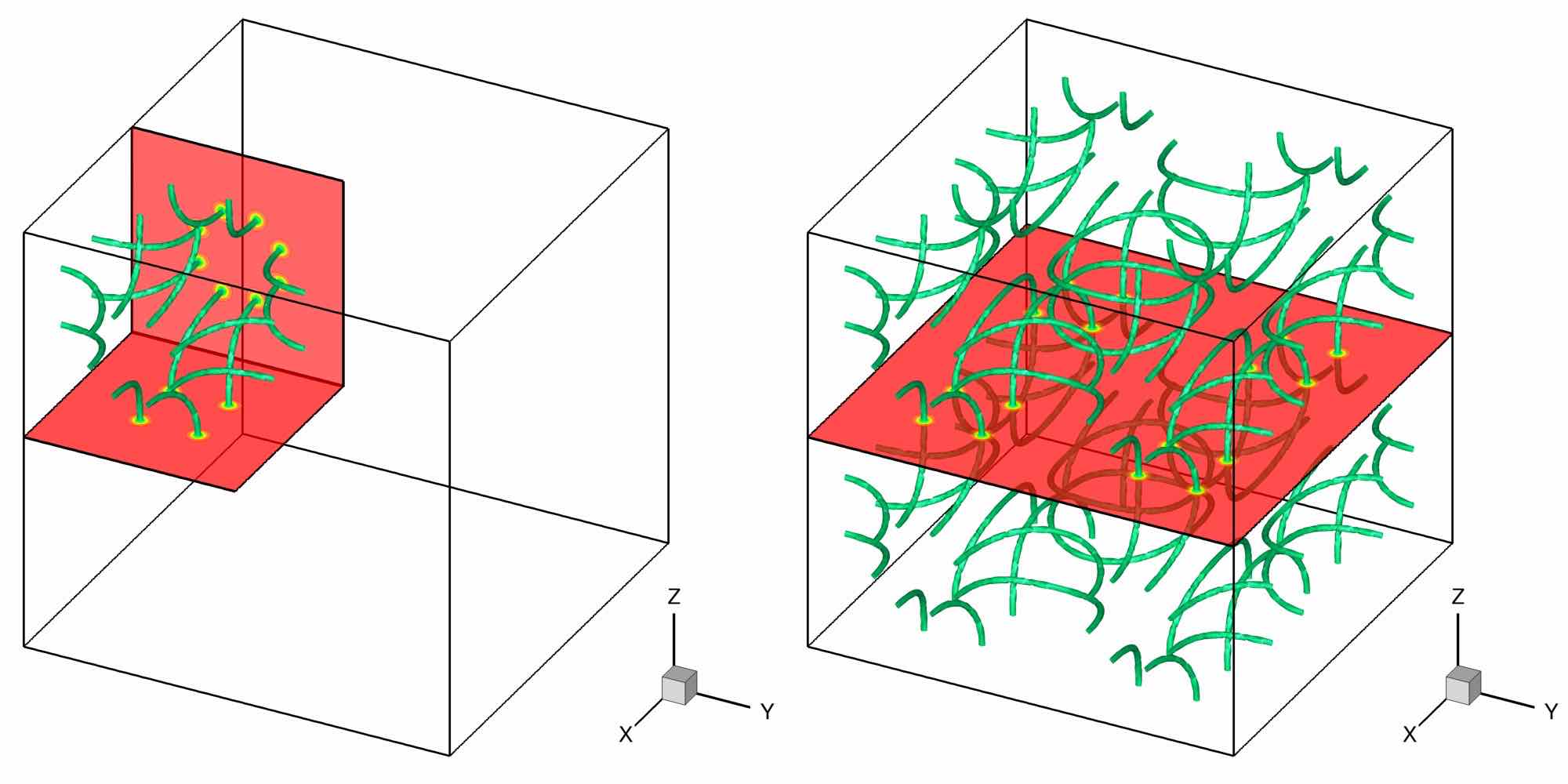}
\end{minipage}\\
		\end{center}
	\caption{Illustration of the initial field preparation using the Taylor-Green vortex flow. Imaginary time evolution of quantized vortices (iso-surfaces of low $\phi_{\rm TG}$) during the ARGLE calculation. Case $N_x=128$ with $[\gamma_d/4]=3$ (see Table \ref{tab:TG-params}). Two different views, on the subdomain $[0,\pi]^3$ (left) and the entire domain $[0,2\pi]^3$ (right), illustrating the symmetry of the flow. Panels from top to bottom: (a) $\tau=0$, the initial condition $\phi(t=0^+)=\psi_{\rm ARGLE}$, Eq. (\ref{psitot3d}) with multiply quantized (thick) vortices, (b) $\tau=1$ when each initial vortex line splits in 3 singly quantized vortices and (c)  $\tau=60$ for the final converged ARGLE field, with closed loops inside the domain.}
	\label{fig:TG_psiTG}
\end{figure}

In practice, we start by defining in the ($\lambda,\mu$) plane a complex field $\psi_e$ with a simple zero at the origin:
\begin{equation}
\psi_e(\lambda,\mu)=(\lambda + i \mu) 
\frac{\tanh(\sqrt{\lambda^2 + \mu^2}/\sqrt{2} \xi)}
{\sqrt{\lambda^2 + \mu^2}}.
\label{psie}
\end{equation}
When replacing \eqref{clebsch3d} into \eqref{psie}, a three-dimensional complex field is obtained, with one nodal line. We can further define 
on $[0,\pi]^3$:
\begin{align}
\begin{split}
\psi_4 (x,y,z)  = \psi_4 (\lambda(x,y,z),\mu(x,y,z)) &= \psi_e(\lambda -\frac{1}{\sqrt{2}}, \mu)
\psi_e(\lambda, \mu-\frac{1}{\sqrt{2}}) \\
&\times \psi_e(\lambda +\frac{1}{\sqrt{2}}, \mu)
\psi_e(\lambda, \mu + \frac{1}{\sqrt{2}}),
\end{split}
\label{psi4}
\end{align}
which now contains four nodal lines (see Fig. \ref{fig:TG_psiTG} a, left). When $\psi_e$ is extended by mirror reflection to the entire domain $[0,2\pi]^3$, the obtained wave function field contains closed rings inside the domain (see Fig. \ref{fig:TG_psiTG} a, right). 

The last manipulation  of the wave function is intended to match the circulation of the velocity field
${\bf v}_{\rm TG}$. From \eqref{clebsch} and  \eqref{clebsch3d} we compute the circulation on the face $z=0, (x,y) \in[0,\pi]\times[0,\pi]$ using the Stokes' theorem:
\begin{equation}
	\Gamma_{z=0} = \int_{0}^{\pi} \int_{0}^{\pi} (\nabla \times {\bf v}_{\rm TG})\cdot \vec{e}_z dx\, dy = \int_{0}^{\pi} \int_{0}^{\pi} 2 \sin(x) \sin(y)  dx\, dy = 8.
\end{equation}
Defining the ratio of the total circulation to the circulation \eqref{eq-scaling-Gamma} of a single vortex as $\gamma_d=\Gamma/\Gamma_v={2}/({\pi\alpha})$, the wave-function field matching the circulation of the TG velocity field is \citep{Nore97a} is
\begin{equation}
\psi_{\rm ARGLE}(x,y,z)=\psi_4
(\lambda(x,y,z),\mu(x,y,z))^{[\gamma_d/4]},
\label{psitot3d}
\end{equation}
where $[.]$ denotes the integer part. 
In this setting,  each vortex line corresponds to a multiple zero line (see Fig. \ref{fig:TG_psiTG} a).
The next step in the preparation of the initial field is to  use the ARGLE imaginary time procedure \eqref{eqn:gradient-flow} with ${\bf v_{\rm ext}}={\bf v}_{\rm TG}$ and initial condition $\phi(t=0^+)=\psi_{\rm ARGLE}$. During the
ARGLE dynamics the multiple zero lines in $\psi_{\rm ARGLE}$ will
spontaneously split into $[\gamma_d/4] = \left[1/(2\pi \alpha)\right]$ single zero lines (see Fig. \ref{fig:TG_psiTG} b).  The system will
finally converge to initial conditions for the GPE, compatible with
the TG flow, and with minimal sound emission. We denote the resulting
converged state as $\phi_{\rm TG}$ (see Fig. \ref{fig:TG_psiTG} c).

\clearpage

\clearpage

\subsection{Arnold-Beltrami-Childress (ABC) flow} \label{sec:init-abc}

The TG vortex flow \eqref{tg3d} has zero helicity. To obtain a a helical flow at large scales, we use the method suggested by \cite{Clark17} in their study of helical quantum turbulence at zero
temperature. The external velocity field is defined as the superposition of two Arnold-Beltrami-Childress (ABC) flows:
 \begin{equation}
 	{\bf v}_{\rm ABC}={\bf v}_{\rm ABC}^{(1)}+{\bf v}_{\rm ABC}^{(2)},
 \end{equation}
with
\begin{align}
\begin{split}
v^{(k)}_{{\rm ABC},x}(x,y,z) &= B \cos(k y) + C \sin(k z) , \\
v^{(k)}_{{\rm ABC},y}(x,y,z) &= C \cos(k z) + A \sin(k x) , \\
v^{(k)}_{{\rm ABC},z}(x,y,z)  &= A \cos(k x) + B \sin(k y) .
\end{split}
\label{ABC}
\end{align}
Unless stated otherwise, we use $(A,B,C)=(0.9,1,1.1)/\sqrt{3}$. As for the Taylor-Green flow, we use the ARGLE procedure \eqref{eqn:gradient-flow} with ${\bf v_{\rm ext}} = {\bf v}_{\rm ABC}$ and initial condition:
\begin{equation}
\phi(t=0^+) =\psi_{\rm ABC}^{(1)} \times \psi_{\rm ABC}^{(2)}.
\label{ABCphi0}
\end{equation}
The wave functions $\psi_{\rm ABC}$ are defined as:
\begin{equation}
\psi_{\rm ABC}^{(k)}= \psi_{A,k}^{x,y,z}  \times \psi_{B,k}^{y,z,x}  
\times \psi_{C,k}^{z,x,y},\quad \psi_{A,k}^{x,y,z} =\exp\left( i \left[\frac{A \sin(k x)}{2\alpha}\right] y
+i \left[\frac{A \cos(k x)}{2\alpha}\right] z\right),
\label{ABCphi}
\end{equation}
where $[a]$ stands for the nearest
integer to $a$. The ARGLE procedure has the role to minimize the amount of energy of acoustic modes in the initial condition. 
Details of the quantum ABC flow are discussed by
\cite{Clark16,Clark17}. We denote the resulting
converged state as $\phi_{\rm ABC}$. 


\subsection{Smoothed random phase (SRP) initial wave function field}\label{sec:init-srp}

Previous initial fields for the simulation of the QT were built based on the analogy with classical flows (TG and ABC) with vortices. We present in this section the first method to set an initial field by direct manipulation of the wave function. A smoothed random phase (SRP) is assigned to the initial wave function $\psi_{SRP}$. Initially, there are no vortices present in the field. Vortices nucleate during the time evolution and their interaction generate a QT field.   
In practice, to obtain the nucleation of enough vortices for QT, we initialize the field as follows:
\begin{equation}
\label{SRP_Init}
\psi_{\rm SRP} = e^{i\theta({\bf x})},
\end{equation}
where $\theta$ is a smooth random periodic function in the computational box.
To create this initial phase, we first generate the random phase $\theta_{i,j,k} \in [- K, K]$ at $N_r^3$ points ${\bf x}_{i,j,k} = N_s \times (i,j,k)$ where $N_s = N / N_r$ and $i,j,k \in \{0,1,2,\cdots,N_r-1\}$.
Then, $\theta$ is obtained by cubic spline interpolation (with periodicity) using the points $({\bf x}_{i,j,k}, \theta_{i,j,k})$.
The one-dimensional cubic (and uniform) spline interpolation is expressed as:
\begin{align}
\begin{split}
& \theta_{N_s i_r + i_s, N_s j_r, N_s k_r} = A_{i_s} \theta_{N_s i_r, N_s j_r, N_s k_r} + B_{i_s} \theta_{N_s (i_r + 1), N_s j_r, N_s k_r} \\
& \phantom{\theta_{N_s i_r + i_s, N_s j_r, N_s k_r} =} + C_{i_s} \theta^{\prime\prime}_{N_s i_r, N_s j_r, N_s k_r} + D_{i_s} \theta^{\prime\prime}_{N_s (i_r + 1), N_s j_r, N_s k_r}, \\
& A_{i_s} = \frac{N_s - i_s}{N_s}, \quad B_{i_s} = \frac{i_s}{N_s}, \quad C_{i_s} = \frac{(A_{i_s}^3 - A_{i_s}) N_s^2}{6}, \quad D_{i_s} = \frac{(B_{i_s}^3 - B_{i_s}) N_s^2}{6},
\end{split}
\end{align}
for $i_r, j_r, k_r \in \{0, 1, \cdots, N_r - 1\}$ and $i_s \in \{1, 2, \cdots, N_s-1\}$.
The second derivative $\theta^{\prime\prime}$ is obtained by solving the following linear system with tridiagonal matrix:
\begin{align}
\begin{split}
&\quad N_s^2 (\theta^{\prime\prime}_{N_s i_r - 1, N_s j_r, N_s k_r} + 4 \theta^{\prime\prime}_{N_s i_r, N_s j_r, N_s k_r} + \theta^{\prime\prime}_{N_s i_r + 1, N_s j_r, N_s k_r}) \\
&= 6 \left(\theta_{N_s i_r - 1, N_s j_r, N_s k_r} - 2 \theta_{N_s i_r, N_s j_r, N_s k_r} + \theta_{N_s i_r + 1, N_s j_r, N_s k_r}\right).
\end{split}
\end{align}
After the interpolation along the $i$-direction, we compute the spline interpolation along the $j$-direction
\begin{align}
\begin{split}
& \theta_{i, N_s j_r + j_s, N_s k_r} = A_{j_s} \theta_{i, N_s j_r, N_s k_r} + B_{j_s} \theta_{i, N_s (j_r + 1), N_s k_r} \\
& \phantom{\theta_{i, N_s j_r + j_s, N_s k_r} =} + C_{j_s} \theta^{\prime\prime}_{i, N_s j_r, N_s k_r} + D_{j_s} \theta^{\prime\prime}_{i, N_s (j_r + 1), N_s k_r},
\end{split}
\end{align}
for $i \in \{0, 1, \cdots, N - 1\}$, $j_r, k_r \in \{0, 1, \cdots, N_r - 1\}$, and $j_s \in \{1, 2, \cdots, N_s-1\}$, and, finally, that along the $k$-direction
\begin{align}
\begin{split}
& \theta_{i, j, N_s k_r + k_s} = A_{k_s} \theta_{i, j, N_s k_r} + B_{k_s} \theta_{i, j, N_s (k_r + 1)} + C_{k_s} \theta^{\prime\prime}_{i, j, N_s k_r} + D_{k_s} \theta^{\prime\prime}_{i, j, N_s (k_r + 1)},
\end{split}
\end{align}
for $i, j \in \{0, 1, \cdots, N - 1\}$, $k_r \in \{0, 1, \cdots, N_r - 1\}$, and $k_s \in \{1, 2, \cdots, N_s-1\}$.

With this method, the characteristic variation of the phase $\theta$ is $K N_r / \pi$. The characteristic velocity results from \eqref{eq-scaling-v}: ${v} = 2\alpha  (K N_r / \pi)$. The Mach number of the system is computed using \eqref{eq-scaling-ab} as $M=v/c=\sqrt{2 \alpha} K N_r / \pi \sqrt{\beta}$.

\begin{figure}[!h]
		\begin{center}
	\begin{minipage}{0.4\textwidth}
	\includegraphics[width=\textwidth]{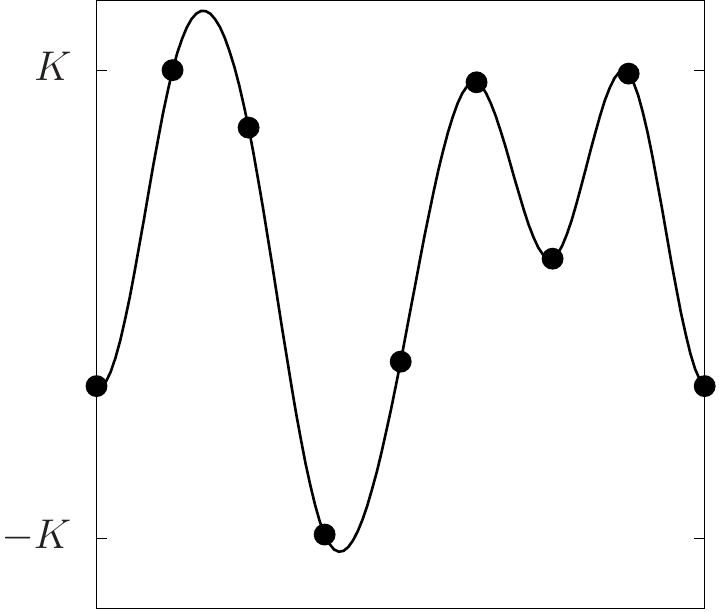}
	\end{minipage}	\hfill
	\begin{minipage}{0.5\textwidth}
	\includegraphics[width=\textwidth]{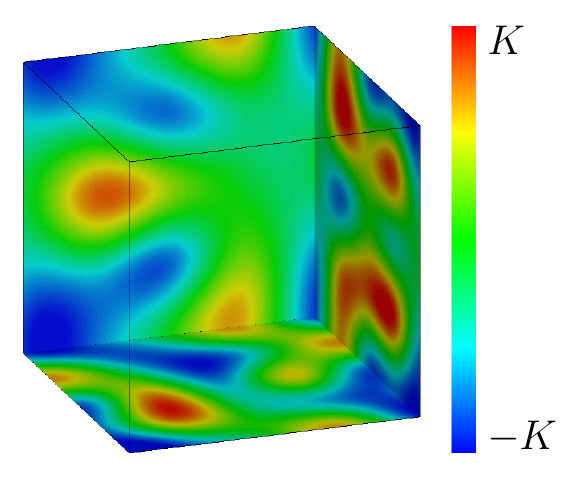}
\end{minipage}	
	\end{center}
\caption{Illustration of the initial field preparation using the SRP (smoothed random phase) method. 
		Spline interpolation in one dimension using random values for the phase (left) and density contours (right) of the final 3D wave function $\psi_{\rm SRP}$. }
\label{fig:SRP_psi}
\end{figure}

We denote the resulting
	converged state as $\psi_{\rm SRP}$. An example of the resulting flow is shown in Fig. \ref{fig:SRP_psi}.


\subsection{Random vortex rings (RVR) initial wave function field}\label{sec:init-rvr}

The main idea for this last initial condition is to prepare an initial state containing enough vortices to lead to QT. 
We derive in the following a method to fill the computational bow with vortex rings. The challenge is to obtain a physically acceptable ansatz.
We start from the single vortex ring solution to the GP equation \citep{BEC-book-2003-pita}. A vortex ring of radius $R$ and constant translational speed
can be approximated as:
\begin{align}
\psi_{\rm VR}(x,y,z,R) &= f\left(\sqrt{(r-R)^2+\tilde{z}^2}\right) e^{\pm i \tan^{-1} \left(\frac{\tilde{z}}{r-R}\right)}, \label{eq-one-vortex-ring} \\
f(r) &= \sqrt{\frac{a_1 (r/\xi)^2 + a_2 (r/\xi)^4}{1 + b_1 (r/\xi)^2 + a_2 (r/\xi)^4}}, \label{eq-Pade} \\
a_1 &= \frac{73 + 3 \sqrt{201}}{352} , \quad a_2 = \frac{6 + \sqrt{201}}{528}, \quad b_1 = \frac{21 + \sqrt{201}}{96}, \label{eq-Pade-coefficient} \\
\tilde{{\bf x}} &= {\bf x} - \pi, \\
r &= \sqrt{\tilde{x}^2 + \tilde{y}^2}, \\
\xi &= \sqrt{\alpha / \beta}
\end{align}
where $f(r)$ is the solution to the GP equation \eqref{eq-gpe-num} written in cylindrical coordinates for   $\psi = f(r) e^{i \kappa \tan^{-1} (y / x)}$ with $\kappa = 1$:
\begin{equation}
- \frac{\alpha}{r} \frac{d}{d r} \left( r \frac{d f}{d r} \right) + \frac{\alpha \kappa^2 f}{r^2} + \beta (f^2 - 1) f = 0.
\label{eq-single-vortex-GPE}
\end{equation}
The form in \eqref{eq-Pade} is obtained as the Pad\'e approximation of this solution.
Coefficients $a_1$, $a_2$, and $b_1$ in \eqref{eq-Pade-coefficient} are fixed by satisfying \eqref{eq-single-vortex-GPE} to the order of $(r / \xi)^3$ for both $r / \xi \ll 1$ and $r / \xi \gg 1$.
This expression stands for a vortex ring centered in the origin.
Note that this definition is consistent with a vortex core size of the order of $\xi$.

The vortex ring ansatz $\psi_{\rm R}$ has the finite net momentum ${\bf j}$ (see Eq. \ref{eq-j}).
To eliminate this momentum, we add an opposite-symmetrical ring by setting the wave function for a vortex-ring pair (VRP) as:
\begin{equation}
\psi_{\rm VRP}(x,y,z,R,d) = \psi_{\rm VR}(x,y,z-d/2,R)\psi^\ast_{\rm VR}(x,y,z+d/2,R),
\end{equation}
where $d$ is the inter-vortex distance.
Because the ansatz $\psi_{\rm VRP}$ for a vortex-ring pair does not satisfy the periodic boundary condition, we rewrite it as
\begin{align}
\begin{split}
\psi_{\rm VRP}(x,y,z,R,d) \to &\: \psi_{\rm VRP}(x,y,z,R,d) \\
\times &\: \psi^\ast_{\rm VRP}(2 {\cal L} - x, y, z, R, d)\, \psi^\ast_{\rm VRP}(- x, y, z, R, d) \\
\times &\: \psi^\ast_{\rm VRP}(x, 2 {\cal L} - y, z, R, d)\, \psi^\ast_{\rm VRP}(x, - y, z, R, d) \\
\times &\: \psi^\ast_{\rm VRP}(x, y, 2 {\cal L} - z, R, d)\, \psi^\ast_{\rm VRP}(x, y, - z, R, d).
\end{split}
\label{psiRP}
\end{align}

\begin{figure}[!h]
	\begin{center}
	\begin{minipage}{\textwidth}
		\includegraphics[width=0.3\textwidth]{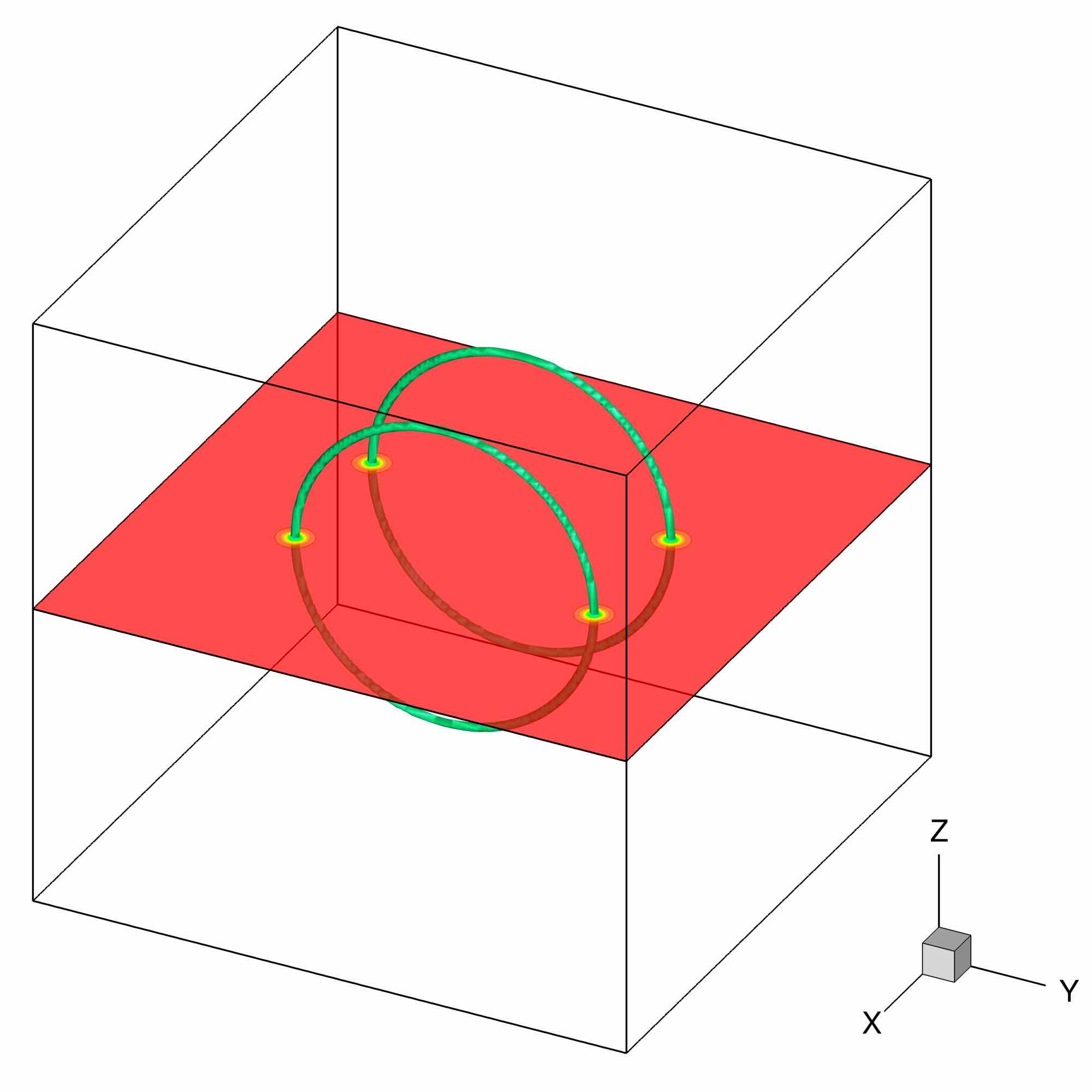}
		\includegraphics[width=0.3\textwidth]{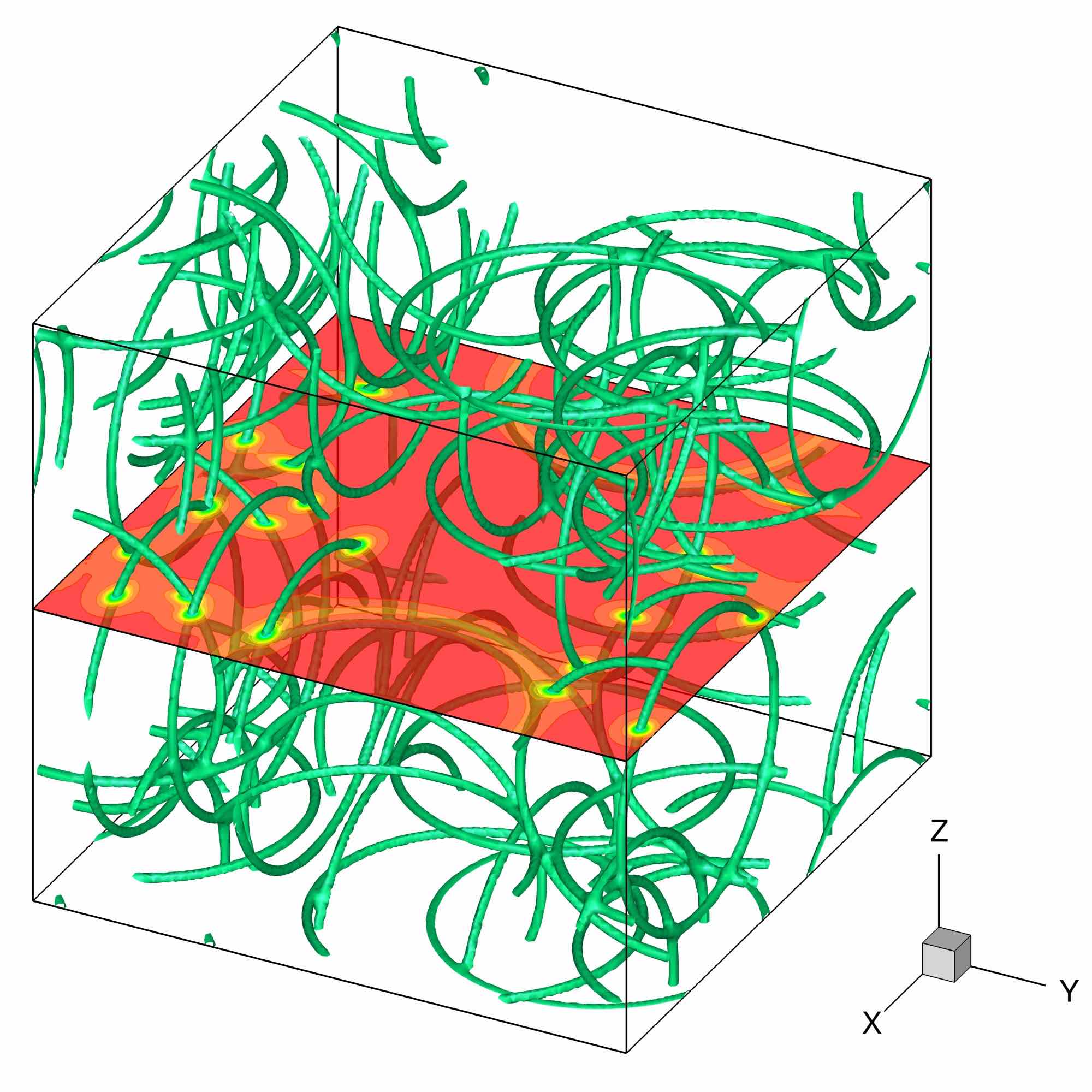}
		\includegraphics[width=0.3\textwidth]{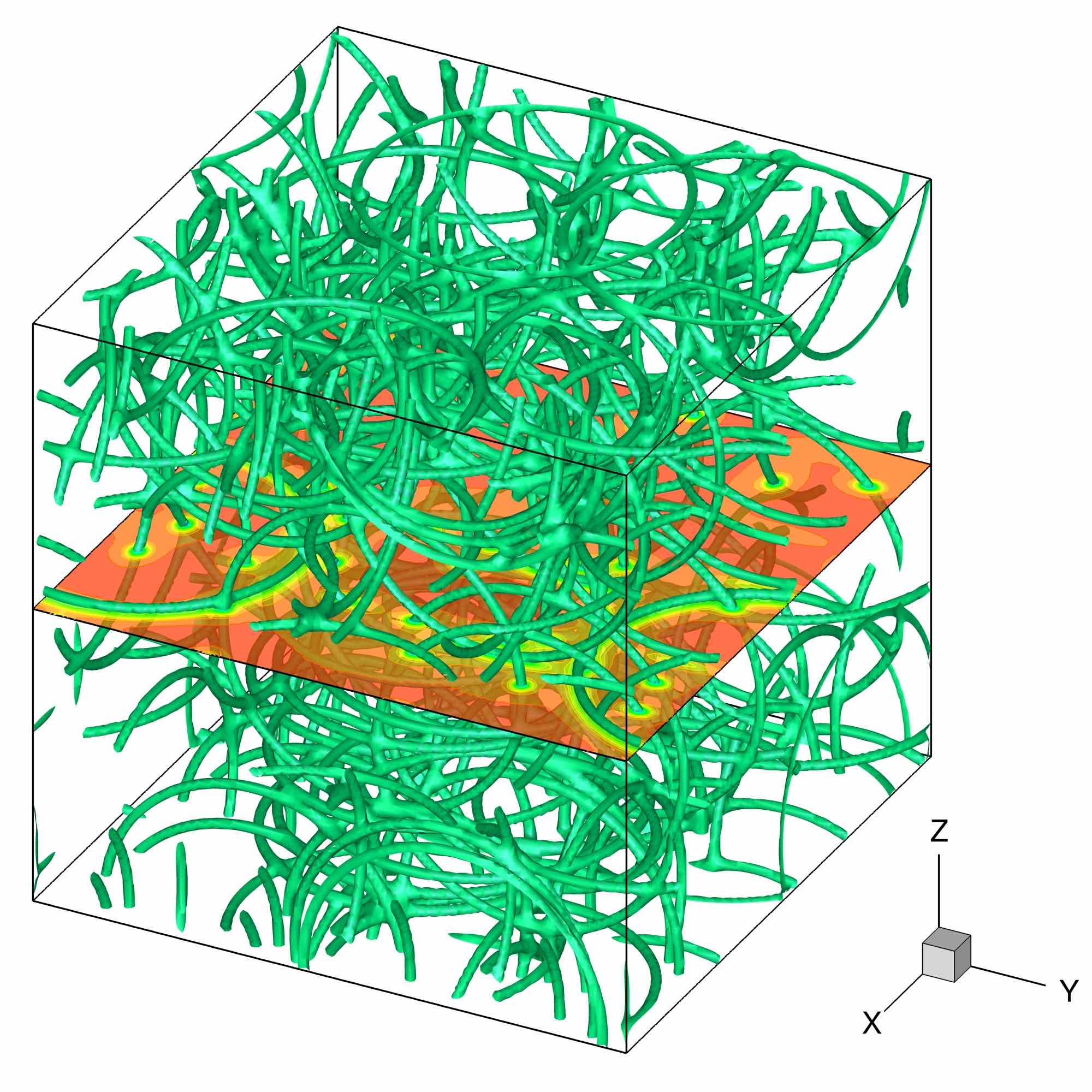}
	\end{minipage}
\end{center}
	\caption{Illustration of the initial field preparation using random vortex ring pairs. Vortex lines (iso-surfaces of low $\rho$) for the wave function $\psi_{\rm RVR}$ (Eq. \ref{psiRVR})   with, from left to right, $N_{\rm V}$ =1, 20 and 50 vortex ring pairs.}
	\label{fig:RPR}
\end{figure}

The last step to prepare the initial state $\psi_{\rm RVR}$ (random vortex rings) is obtained by randomly putting vortex-ring pairs in the domain.
First, we randomly translate the ansatz $\psi_{\rm VRP}$ \eqref{psiRP} as
\begin{align}
\psi_{\rm RVR}(x,y,z,R,d) \equiv \mathcal{F}_{\bf x}^{-1}\left( e^{i {\bf k} \cdot {\bf X}} \mathcal{F}_{\bf k}(\psi_{\rm VRP}(x,y,z,R,d))\right),
\end{align}
where ${\bf X} = (X, Y, Z) \in [0, 2 \pi]^3$ are uniform random numbers.
After that, we randomly rotate the ansatz by:
\begin{align}
\psi_{\rm RVR}(x,y,z,R,d) \to \left\{ \begin{array}{c}
\psi_{\rm RVR}(x,y,z,R,d) \\
\psi_{\rm RVR}(x,z,y,R,d) \\
\psi_{\rm RVR}(y,x,z,R,d) \\
\psi_{\rm RVR}(y,z,x,R,d) \\
\psi_{\rm RVR}(z,x,y,R,d) \\
\psi_{\rm RVR}(z,y,x,R,d)
\end{array} \right\}.
\label{psiRVR}
\end{align}
Finally, the initial state $\psi_{\rm RPR}$ is obtained by preparing $N_{\rm V}$ different ansatze $\psi_{\rm RVR}$ and multiplying them.
Changing the radius of the ring $R$, the inter-vortex distance $d$ or the number of vortex rings pairs $N_{\rm V}$ will impact the behaviour of QT.
An example of the resulting flow is shown in Fig. \ref{fig:RPR}.

\clearpage

\section{Numerical results}\label{sec:Results}

We describe in this section quantum turbulence flows simulated using the four initial conditions described in the previous section: Taylor-Green (TG), Arnold-Beltrami-Childress (ABC), smoothed random phase (SRP) and  random vortex rings (RVR). We  present values, spectra and structure functions of main quantities of interest (energy, helicity, etc.)  that could be useful to benchmark numerical codes simulating QT. 

The main physical and numerical parameters of the runs were fixed following the scaling analysis provided in \S\ref{sec:scaling} and are summarized in Table \ref{tab:ALL-params}. Runs are identified using the abbreviation of the initial condition, followed by the identifier of the space resolution, \eg {TG\_a} is the run using the Taylor-Green initial condition and a $128^3$ grid. Resolutions up to $1024^3$ grid points (runs "\_d") were considered for some cases.  For all simulations, the grid is equidistant in each space direction, covering a domain of the same size $[0, {\cal L}]^3$, with ${\cal L}=2\pi$. We recall that a Fourier spectral spatial discretization with periodic boundary conditions is used in the GP solvers.
\begin{table}[h!]\centering
	{\small	\begin{tabular}{|c|c|c|c|c|c|c|c|c|c|c|c|c|}\hline
			Run  & ${\cal L}$ & $N_x$ & $M_\scal$ & $c$ & $\alpha$ & $\beta$  & $k_{max} \xi$ &$\delta x/\xi$ &$\xi$ & $N_v^{1d}$  \\ \hline
			\_a & $2\pi$&	128  &0.5&2.0&0.05000   & 40  &   2.26& 1.388&   0.035355 &88 \\
			\_b & $2\pi$&	256  &0.5&2.0&0.02500   &80   &   2.26& 1.388&   0.017678 &177\\
			\_c & $2\pi$&	512  &0.5&2.0&0.01250   &160 &   2.26& 1.388& 0.008839 &355\\
			\_d & $2\pi$&	1024&0.5&2.0& 0.00625  &320 &   2.26& 1.388& 0.004419 &710\\ \hline
		\end{tabular}
	}
	\caption{Numerical and physical parameters used in the QT simulations.}\label{tab:ALL-params}
\end{table}

Using the contribution by \cite{Nore97a} as guideline, the reference Mach number was fixed to $M_\scal=0.5$, equivalent to a non-dimensional speed of sound $c=2$.  Consequently, $\alpha \beta=2$ for all cases.  Following \eqref{eq-scaling-choice3}, when increasing the grid resolution $N_x$ by a factor of 2, the value of the parameter $\alpha$ is diminished by the same factor in order to keep constant the value of $k_{max} \xi =8\sqrt{2}/5=2.26$. There are two main consequences of this setting:  the non-dimensional value of the healing length $\xi=\sqrt{2}\alpha M_\scal$ diminishes when $N_x$ is increased, while the 
 grid resolution of a vortex is kept constant $\delta x/\xi=\pi/(k_{max} \xi )=1.388$. We check  for the TG case that this grid resolution is enough to accurately resolve vortices in our QT simulations. Since the size $\cal L$ of the computational box is kept constant, the higher the grid resolution $N_x$, the larger is the number of vortices present in the domain (see values of $N_v^{1d}$ in Table \ref{tab:ALL-params}).

\subsection{Benchmark \#1: Taylor-Green quantum turbulence (TG-QT)}

The Taylor-Green initial field was prepared as described in \S \ref{sec:init-tg}.  We display in  Table \ref{tab:TG-params}  the values of the time step $\delta t$ used in the GP solver (see \S\ref{sec:real-time}) and the final time $T_f$ of each simulation. The parameters of the corresponding imaginary-time (IT) run cases preparing the initial condition using the ARGLE solver are also presented, with $\delta \tau$ and $\tau_f$ the imaginary-time step and final value at convergence, respectively, and $[\gamma_d/4]$ the winding number of initial TG vortices seed at $\tau=0$ (see Eq. \eqref{psitot3d} and Figure \ref{fig:TG_psiTG}). 
\begin{table}[h!]\centering
	{\small	\begin{tabular}{|c|c|c|c|}\hline
		Run  &  $N_x$  & $\delta t$&$T_f$\\ \hline
		TG\_a &	128    &1.250e{-3} & 12\\
		TG\_b &	256    &6.250e{-4} & 12\\
        TG\_c &	512    &3.125e{-4} & 12\\
		TG\_d &	1024  &3.125e-4& 10\\ \hline
	\end{tabular}
}	
	{\small	\begin{tabular}{|c|c|c|c|c|}\hline
		Run  &  $N_x$ & $[\gamma_d/4]$ & $\delta \tau$&$\tau_f$\\ \hline
		TG\_aIT &	128   & 3 &1.2500e{-2} & 60\\
		TG\_bIT &	256  & 6   &6.2500e{-3} & 60\\
		TG\_cIT &	512  &12   &3.1250e{-3} & 60\\
		TG\_dIT &	1024& 25 &1.5625e{-3}& 60\\ \hline
	\end{tabular}
}	
	\caption{Runs for the TG-QT case. Parameters used in the GP solver (cases TG\_a to TG\_d)  and the imaginary-time (IT) ARGLE solver (cases TG\_aIT to TG\_dIT). For each space resolution $N_x$, the corresponding physical and numerical parameters are displayed in Table \ref{tab:ALL-params}. }\label{tab:TG-params}
\end{table}

\subsubsection{Results for the imaginary time (ARGLE) procedure}

\begin{figure}[!h]
	\begin{center}
		\begin{minipage}{\textwidth}
	                	\includegraphics[width=0.33\textwidth]{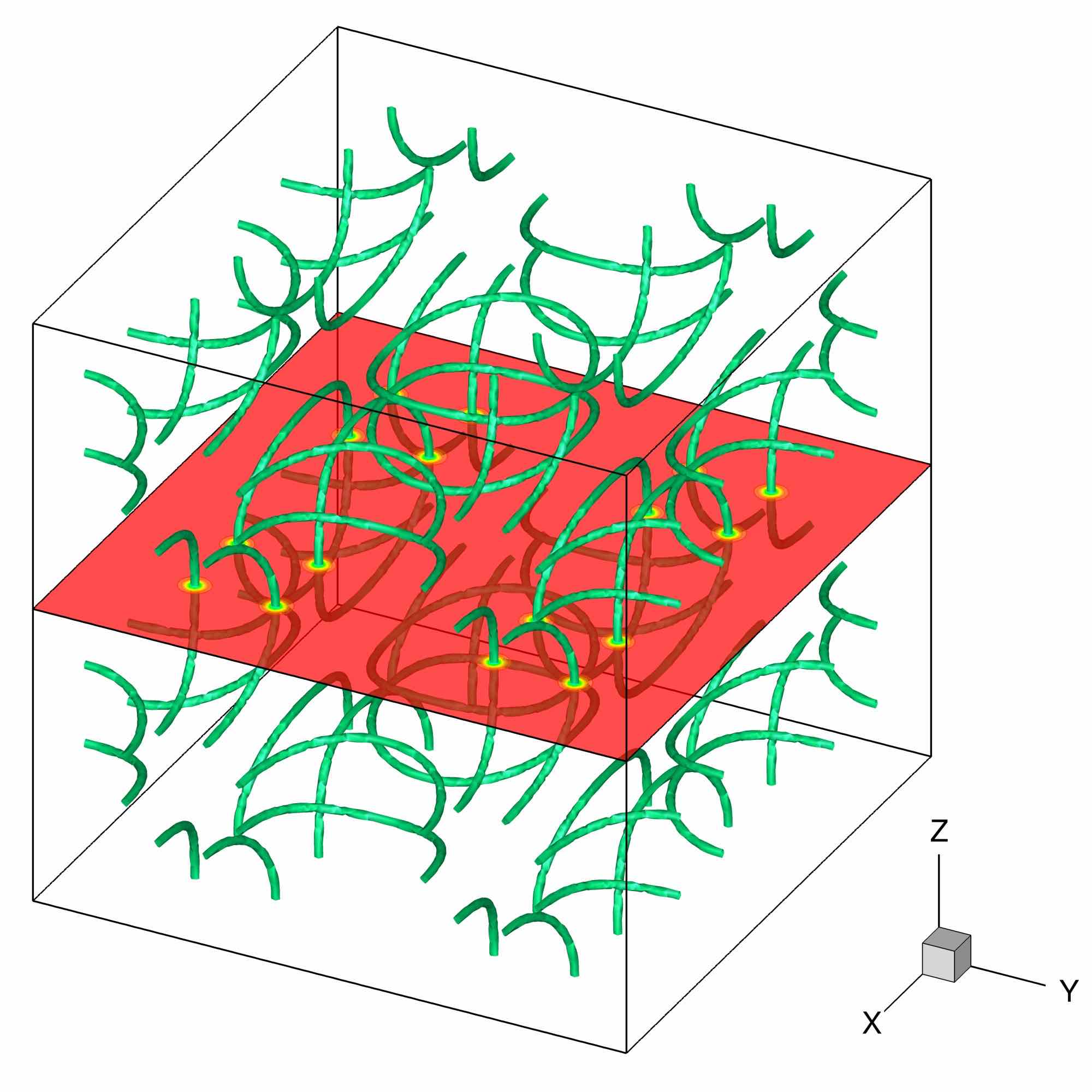}
						\includegraphics[width=0.33\textwidth]{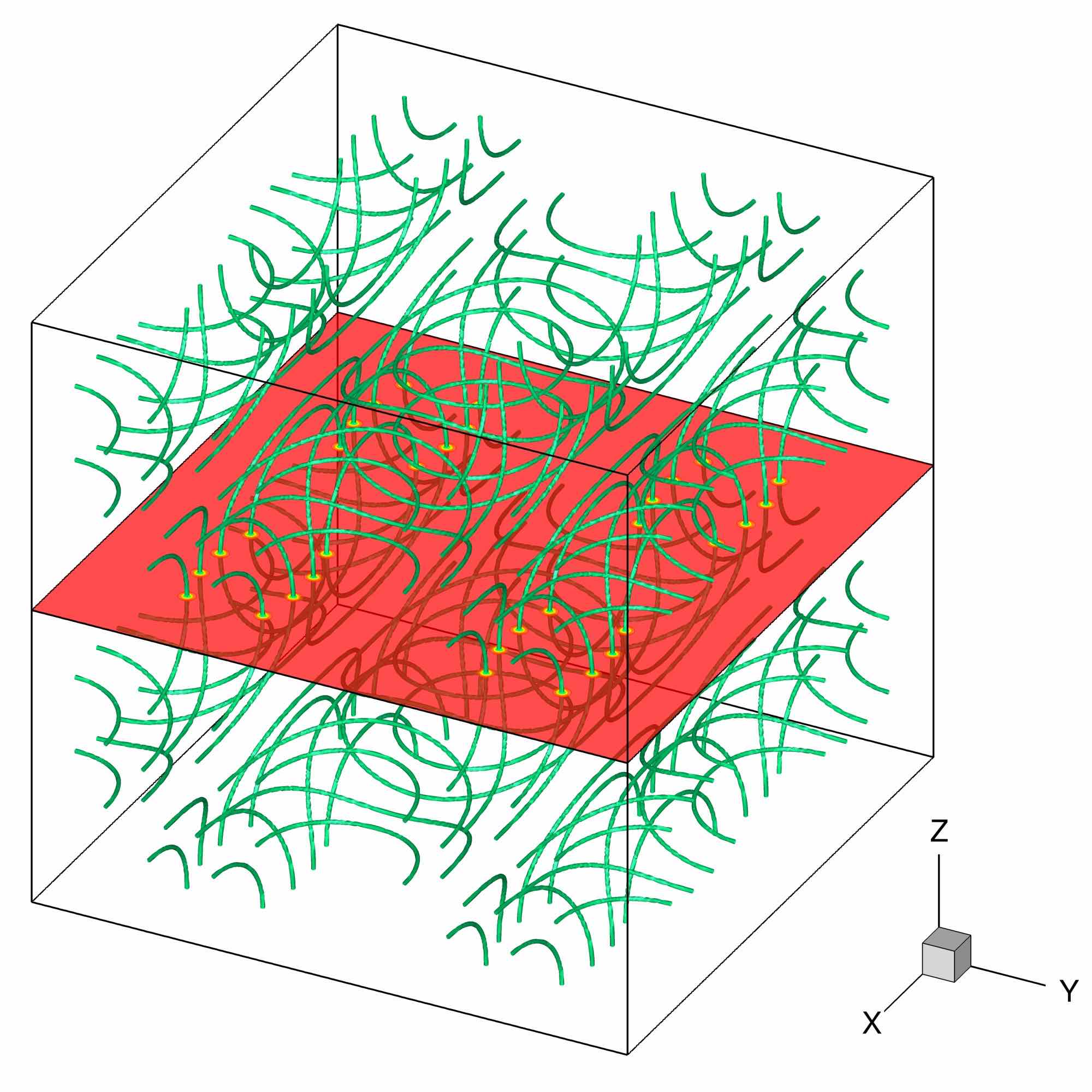}
						\includegraphics[width=0.33\textwidth]{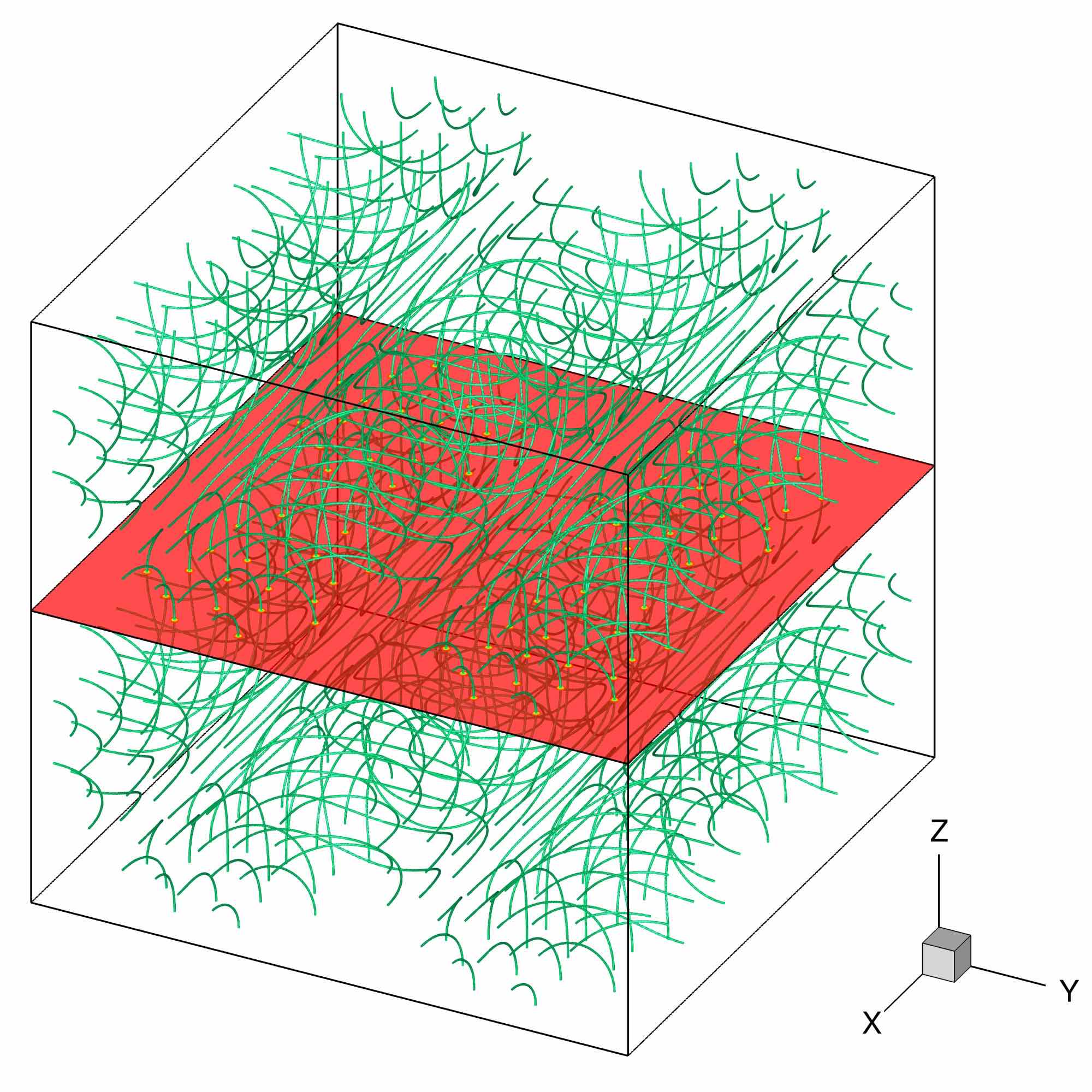}
		\end{minipage}
	\end{center}	
	\caption{TG-QT. Initial condition computed with the imaginary-time ARGLE solver. Vortex lines (iso-surfaces of low $\rho$) of the  converged wave function $\phi_{\rm TG}$ at final imaginary-time $\tau_f$.	From left to right: grid resolutions $N_x=$ 128, 256, 512 (corresponding to runs TG\_aIT, TG\_bIT and TG\_cIT  in Table \ref{tab:TG-params}).}
	\label{fig:TG-at-tauf}
\end{figure}

In defining this benchmark, it is important to describe in detail the initial field obtained after the imaginary time (ARGLE) procedure. We recall that this procedure starts from the ansatz $\psi_{\rm ARGLE}$ (\ref{psitot3d}) containing multiple  zero TG vortices that split into $[\gamma_d/4]=\left[1/(2\pi \alpha)\right]$  singly quantized vortices during the imaginary time propagation (see Fig. \ref{fig:TG_psiTG} illustrating the case TG\_aIT).  Note from Table \ref{tab:TG-params} that when increasing the grid resolution $N_x$, the ansatz TG vortices split in a larger number of individual quantized vortices (up to 25 for $N_x=1024$). This is illustrated in Fig. \ref{fig:TG-at-tauf} showing vortex configurations obtained at the end of the ARGLE procedure for runs TG\_aIT, TG\_bIT and TG\_cIT.

To validate the ARGLE runs, we report in Table \ref{tab:TG_ARGLE_Energy} the values of different energies (see \S\ref{sec:energ-decomp}) computed  for the final field (at $\tau_f$). The results are in good agreement with those reported by \cite{Nore97a,Nore97b}.  The values of the helicity (see \S\ref{sec:helicity}) are also reported in Table \ref{tab:TG_ARGLE_Energy}. Note that in this particular case, we expect the helicity to be zero, which is satisfied for regularized helicity. As already stated by \cite{Clark17}, the regularized helicity is smoother and less noisy, which explains the discrepancies for helicity in runs TG\_cIT and TG\_dIT.
\begin{table}[!h]\centering
{\small	\begin{tabular}{|c|c|c|c|c|c|c|}\hline
		Run & $E_{kin}^i$ & $E_{kin}^c$ & $E_q$ & $E_{int}$ & $H$ & $H_{\rm reg}$ \\ \hline
TG\_aIT	 & 0.12901707	 & 4.8667051e-04	 & 7.9239425e-03	 & 1.2995235e-02	 & 1.37e-13	 & -1.87e-11 \\ 
TG\_bIT	 & 0.11334487	 & 2.2334712e-04	 & 4.0373670e-03	 & 6.8223665e-03	 & 2.96e-07	 & -5.52e-07 \\ 
TG\_cIT	 & 0.12884207	 & 1.5065059e-04	 & 2.4895687e-03	 & 4.2864757e-03	 & 3.83e-03	 & -5.63e-07 \\ 
TG\_dIT	 & 0.12968555	 & 9.5590716e-05	 & 1.3476259e-03	 & 2.3466209e-03	 & -3.94e-04	 & -9.71e-08 \\ 
\hline
	\end{tabular}
}
	\caption{TG-QT. Values of different energies and helicity at $\tau_f$ for the runs preparing  the Taylor-Green initial condition, using the imaginary-time ARGLE solver. }\label{tab:TG_ARGLE_Energy}
\end{table}

\subsubsection{Results for the TG-QT}

\begin{figure}[!h]
	\begin{center}
		\begin{minipage}{\textwidth}
						\includegraphics[width=0.33\textwidth]{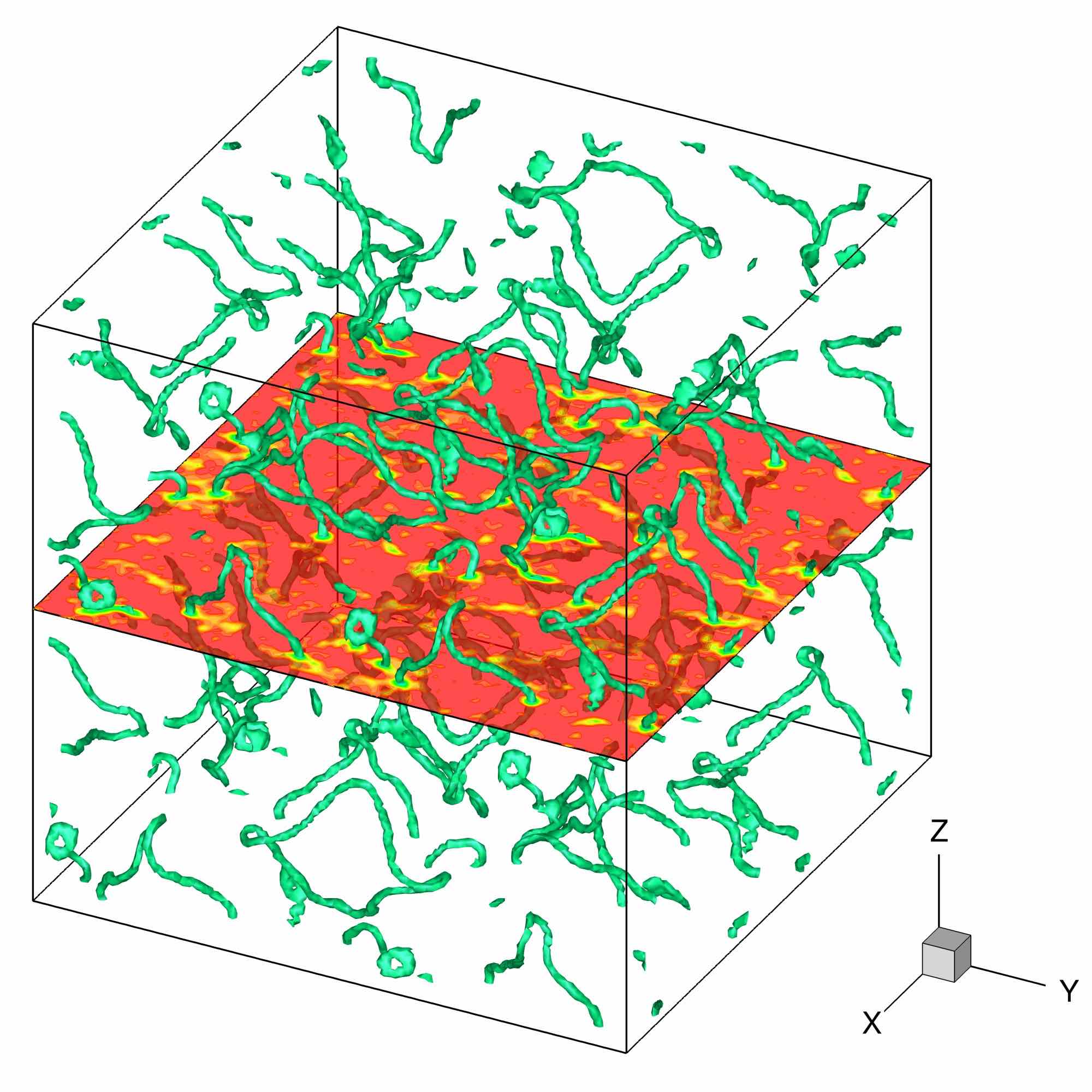}
						\includegraphics[width=0.33\textwidth]{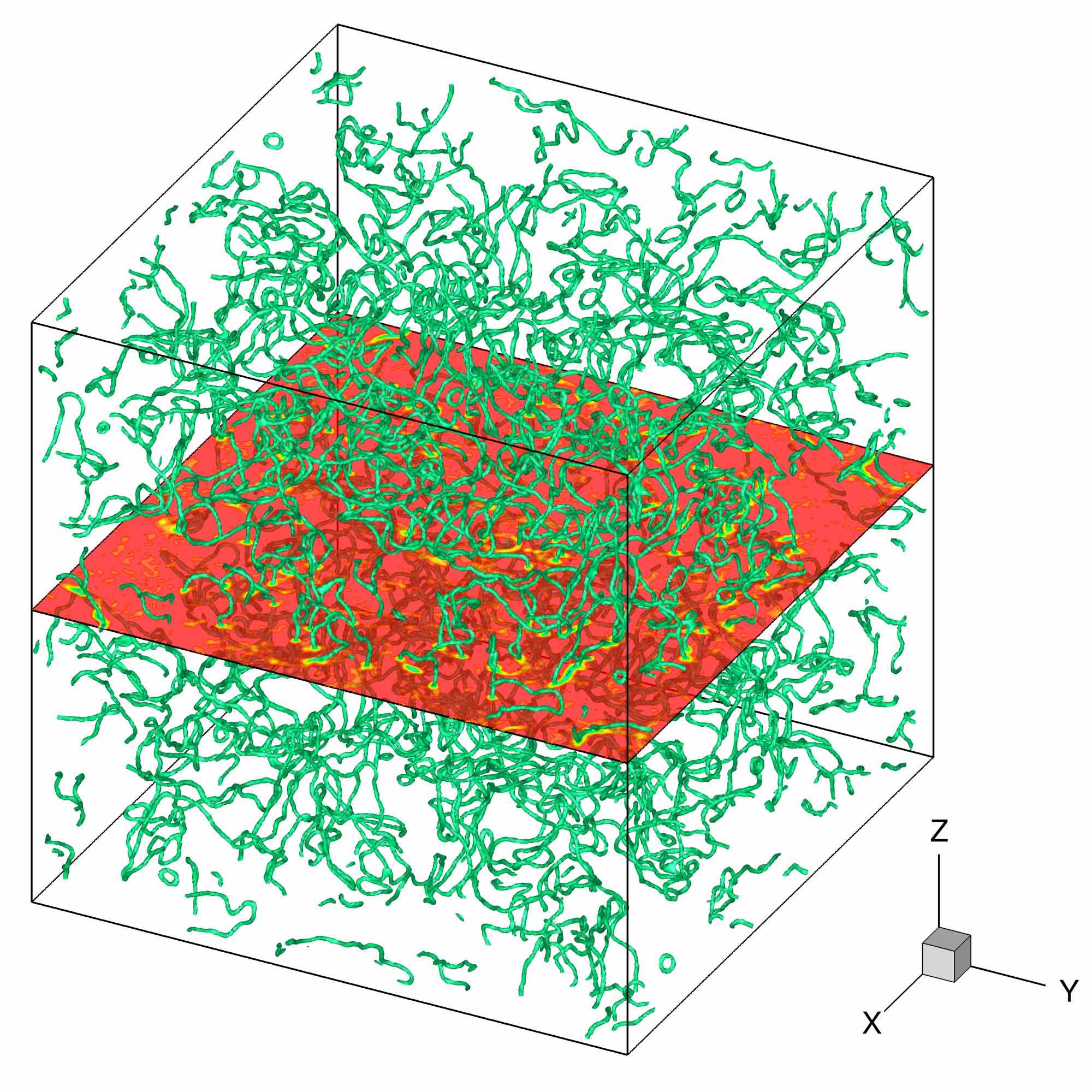}
			            \includegraphics[width=0.33\textwidth]{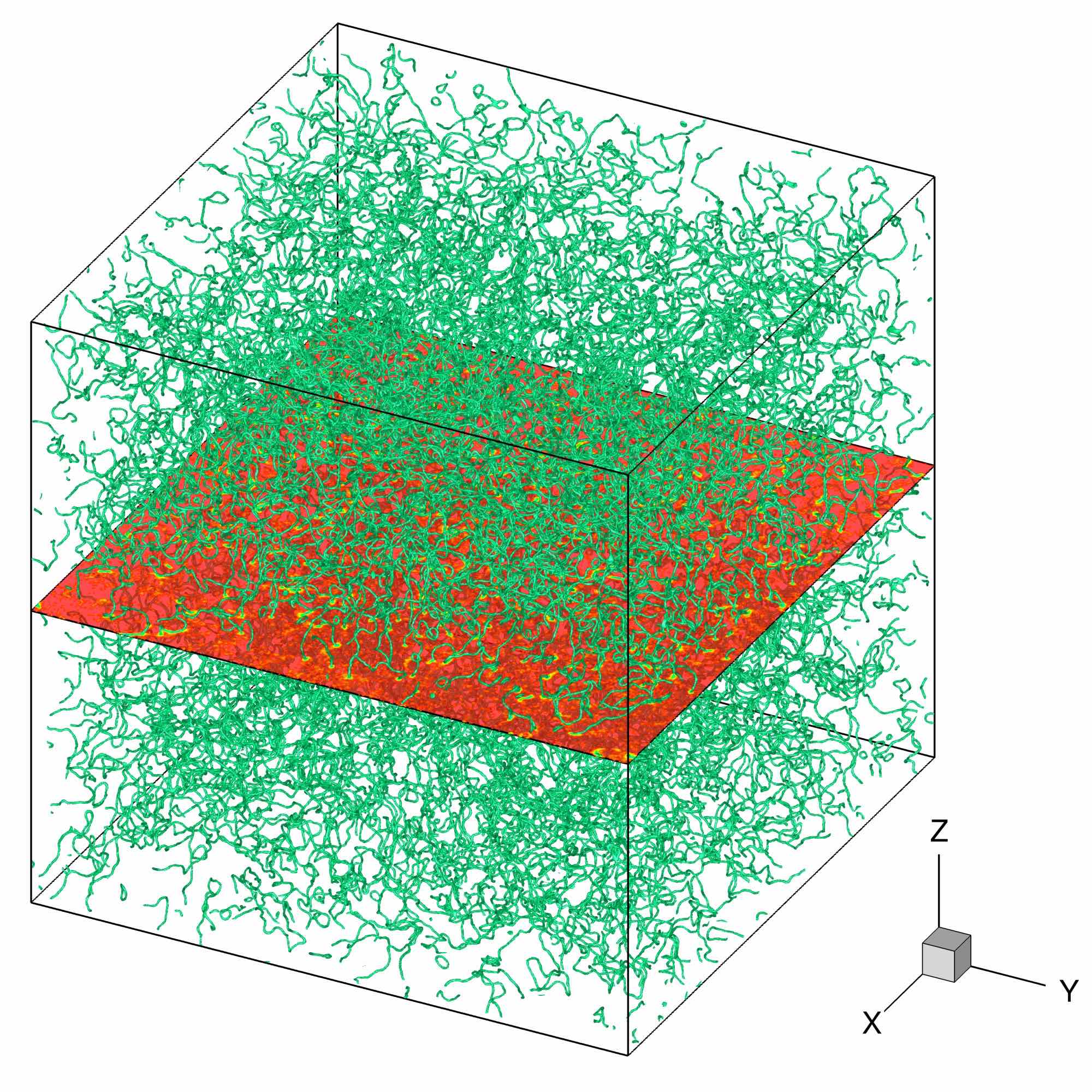}
		\end{minipage}
	\end{center}	
	\caption{TG-QT. Instantaneous fields computed with the real-time GP solver, starting from the initial condition presented in Fig. \ref{fig:TG-at-tauf}. Vortex lines (iso-surfaces of low $\rho$) of the  wave function at final time $T_f$.
                From left to right: grid resolutions $N_x=$ 128, 256, 512 (corresponding to runs TG\_a, TG\_b and TG\_c in Table \ref{tab:TG-params}).}
	\label{fig:TG-at-Tf}
\end{figure}

Starting from the initial condition presented in Fig. \ref{fig:TG-at-tauf}, we used the Strang--splitting GP solver (see \S\ref{sec:real-time})  to advance the wave function in real time. The final (at $t=T_f$) QT field is displayed in Fig. \ref{fig:TG-at-Tf} for runs TG\_a, TG\_b and TG\_c. As explained before, when the grid resolution $N_x$ is increased, the size of a vortex core $\xi$ diminishes and, consequently, the density of the tangled vortex lines is increased in the computational box. Meanwhile, we recall that the grid resolution of the vortex core ($\delta x/\xi$) is the same for all simulations.

\begin{figure}[!h]
	\begin{center}	
		\begin{minipage}{0.5\textwidth} 
			a)\\
			\includegraphics[width=\textwidth]{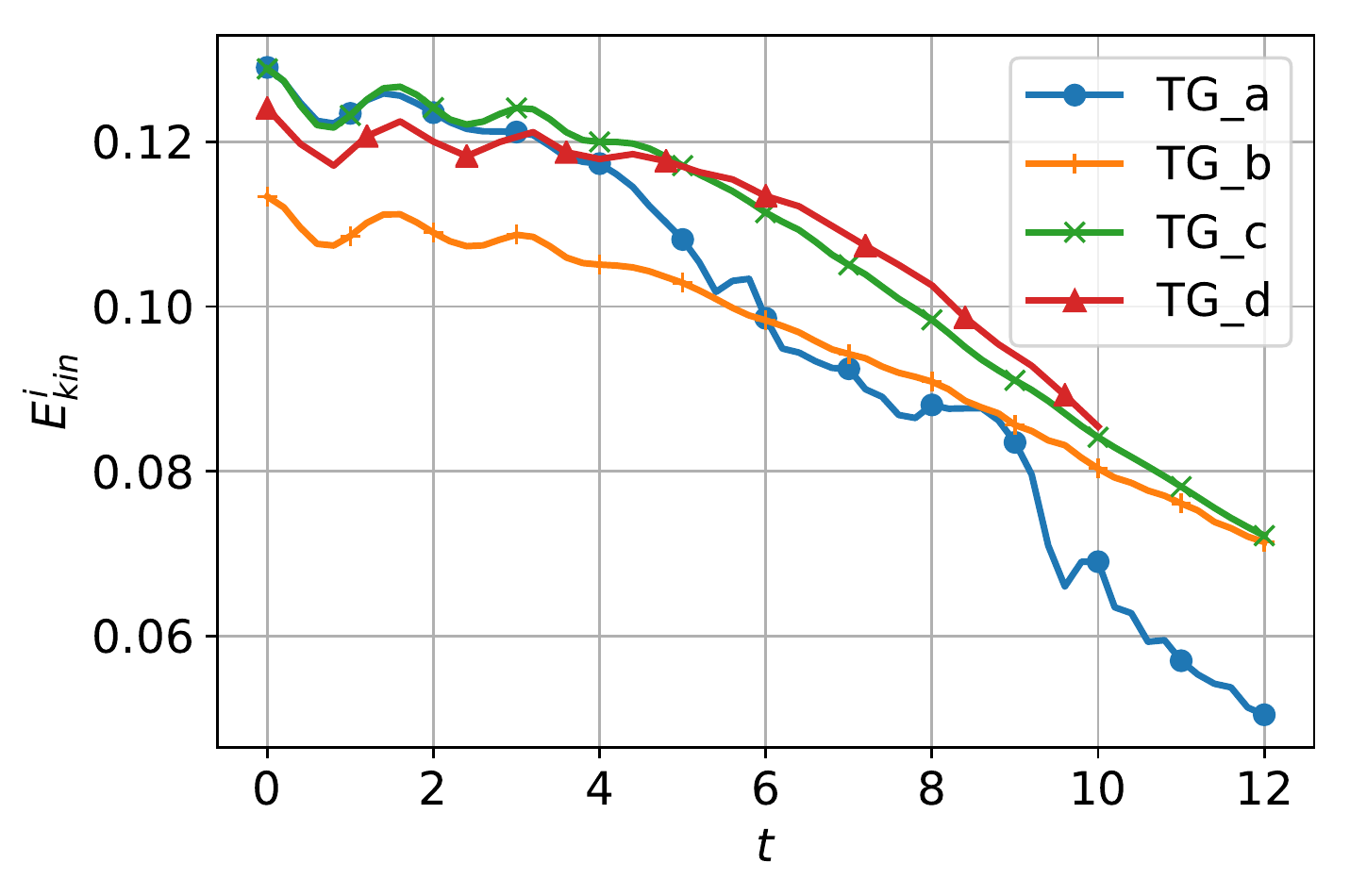}
		\end{minipage}\hfill
		\begin{minipage}{0.5\textwidth} 
			b)\\
			\includegraphics[width=\textwidth]{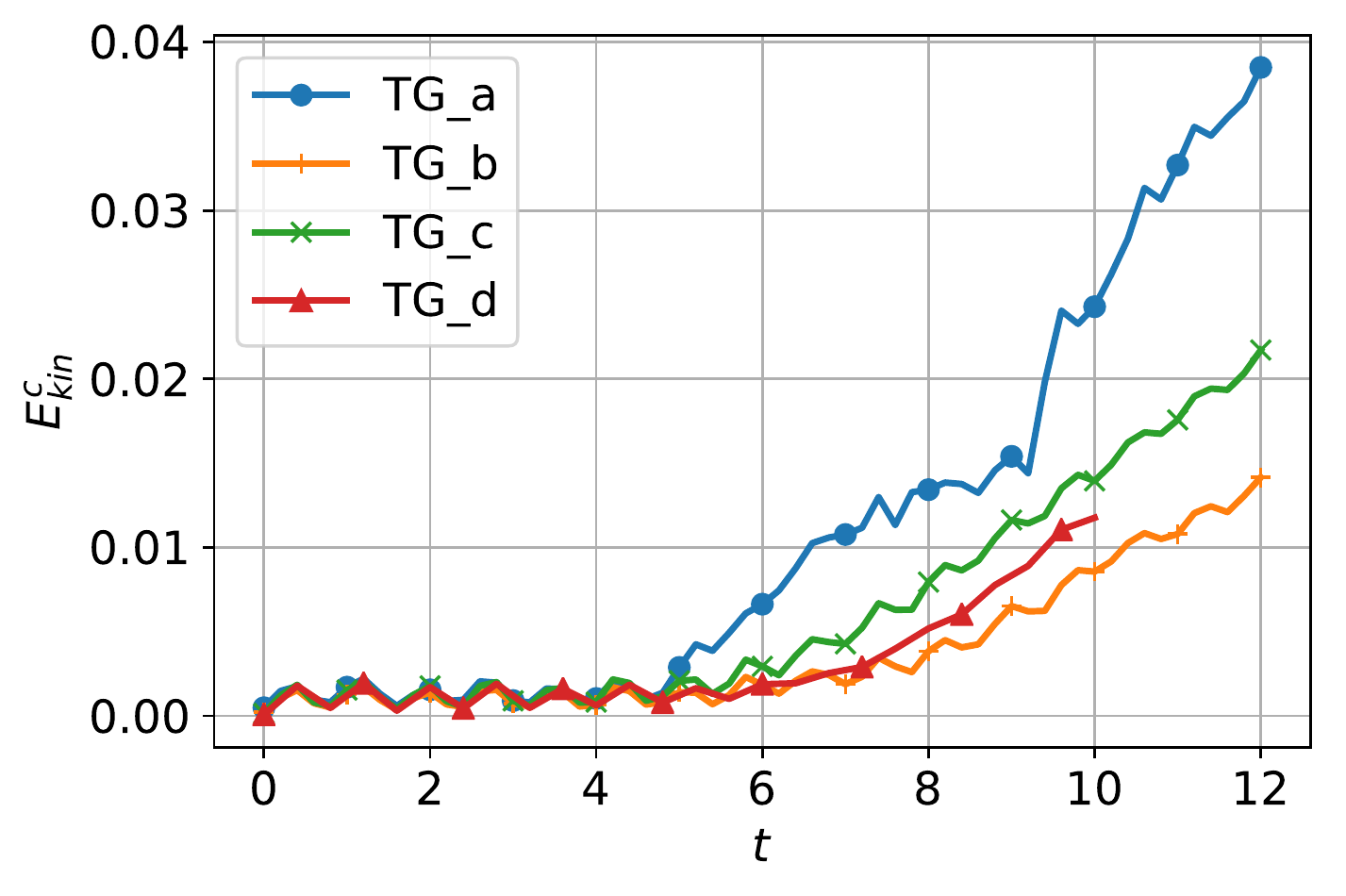}
		\end{minipage}
	\end{center}	
	\caption{TG-QT. Time evolution of incompressible kinetic energy  $\Ekini$ (a) and compressible kinetic energy $\Ekinc$ (b) for runs TG\_a to TG\_d (see Table \ref{tab:TG-params}).}\label{fig:evol_energy_TG}
\end{figure}

\begin{figure}[h!]
	\begin{center}	
		\begin{minipage}{0.5\textwidth} 
			a)\\
			\includegraphics[width=\textwidth]{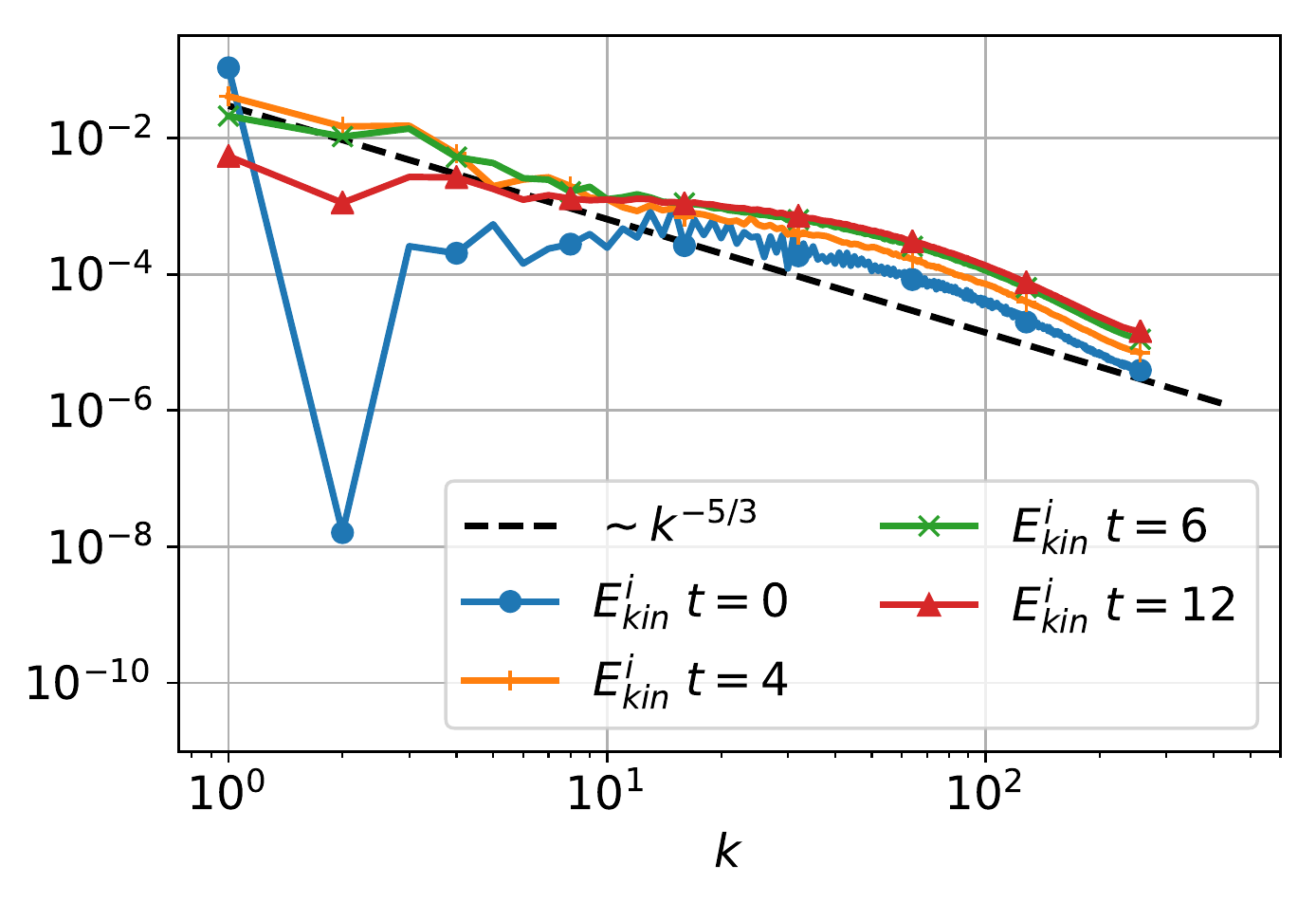}
		\end{minipage}\hfill
		\begin{minipage}{0.5\textwidth} 
			b)\\
			\includegraphics[width=\textwidth]{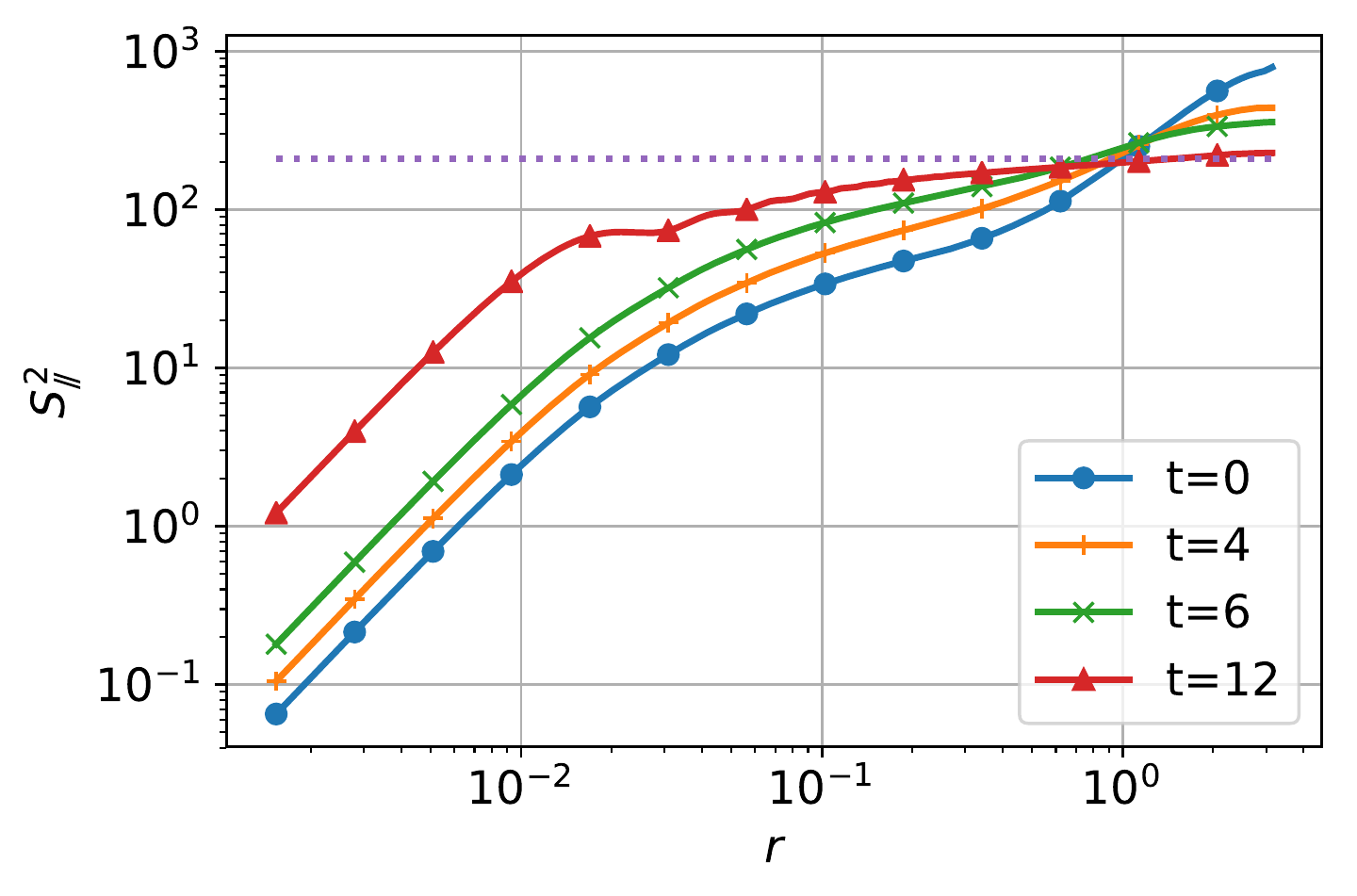}
		\end{minipage}
	\end{center}		
	\caption{TG QT. Spectrum of $\Ekini$ (a) and second--order structure function (b) for the case TG\_c (see Table \ref{tab:TG-params}).} \label{fig:TG_spectrum_and_sf}
\end{figure}

To  compare our results with those reported by \cite{Nore97a,Nore97b}, the TG-QT fields are analyzed  by providing in Fig. \ref{fig:evol_energy_TG} the time evolution of the incompressible  ($\Ekini$) and compressible  ($\Ekinc$) parts of the kinetic energy \eqref{eq-scaling-energy-mod-rho}
for cases TG\_a to TG\_d. For each case, the incompressible kinetic energy is dominant at the beginning of the simulation, and slowly decreases in time, while the compressible part increases.  We report in  Fig. \ref{fig:TG_spectrum_and_sf} (a) the spectrum of $\Ekini$ for the case TG\_c  at different time instants of the computation. For small $k$, the spectrum follows a (Kolmogorov-like) power law $\Ekini(k)\sim k^{-5/3}$ (dashed line in Fig. \ref{fig:TG_spectrum_and_sf} a), especially for early times of the simulation. These  results concerning the incompressible energy evolution and its spectrum are in very good agreement with the numerical results reported by \cite{Nore97a,Nore97b} for the grid resolution $N_x=512$.  As a novel diagnostic tool of the turbulent field (not presented in \cite{Nore97a,Nore97b}), we show in Fig. \ref{fig:TG_spectrum_and_sf} (b) the time-evolution of the second-order structure function $S^2_{\para}(r)$  (see Eq. \eqref{strfunc-def}).  For a developed QT field at 
$t=12$, the slope of the structure function curve at the origin is close to 2, while for length scales larger than 0.1, the slope evolves  to 1/3. Using Eq. \eqref{strfunc-verif} to check the structure function calculation, we also plot in Fig. \ref{fig:TG_spectrum_and_sf} (b) as a dotted line the value  $2\int v_x^2$ which is reached for large length scales (see Eq. \eqref{strfunc-verif}).

\subsubsection{Accuracy of numerical results}

There are  two important check--points in validating a QT-GP simulation: the accuracy in verifying conservation laws (see \S \ref{sec:GPEModel}) and the grid convergence.   In our numerical simulations, we monitor the time variation of the number of particles $N$ (see Eq. \eqref{eq-scaling-N}) and  total energy per volume unit  E (see Eq. \eqref{eq-scaling-energy}). These two quantities should be conserved by the GP solver. For the TG-QT simulations, we report in Table \ref{tab:TG-conservation} initial and final values for the norm and normalized energy, as well as their relative maximum variation during the time evolution. Note from Table \ref{tab:TG-conservation} that $N$ is perfectly conserved, and energy relative fluctuations $\delta(E)$ are less than 0.02\%, which is sufficiently small value to guarantee the validity of the computation.
\begin{table}[h!]\centering
	{\small
	\begin{tabular}{|c|c|c|c||c|c|c|}\hline
		Run & ${N}_{|t=0}$ & ${N}_{|t=T_f}$ & $\delta({N})$ & $E_{|t=0}$ & $E_{|t=T_f}$ & $\delta(E)$ \\ \hline
TG\_a	 & 0.9789997 	 & 0.9789997 	 & 0.0 	 & 0.1504230 	 & 0.1504111 	 & 7.97e-05  \\ 
TG\_b	 & 0.9873160 	 & 0.9873160 	 & 0.0 	 & 0.1244280 	 & 0.1244235 	 & 3.60e-05  \\ 
TG\_c	 & 0.9920083 	 & 0.9920083 	 & 0.0 	 & 0.1357688 	 & 0.1357597 	 & 6.67e-05  \\ 
TG\_d	 & 0.9955732 	 & 0.9955732 	 & 0.0 	 & 0.1275738 	 & 0.1275550 	 & 1.47e-04  \\  \hline
	\end{tabular}
}
	\caption{TG-QT. Conservation of the number of particles $N$ and  energy per volume unit. 
		Initial (at $t=0$) and final values (and $t=T_f$) and relative maximum variation, defined following \eg $\delta(E)=\max_{t\in[0;T_f]} \left| E(t) - E_{t=0}\right|/E_{t=0}$.}\label{tab:TG-conservation}
\end{table}

The second important check-point is the grid convergence. To correctly capture vortices of radius $\xi$, we need enough discretization points in each vortex core. We recall that the grid step size was fixed to $\delta x/\xi=1.338$ for all runs, corresponding to $\xi k_{\max}=\frac{8\sqrt{2}}{5}\simeq 2.26$ (see Table \ref{tab:ALL-params}). To check the influence of this parameter on the accuracy of the QT simulation, we performed two other runs reported in Table \ref{tab:TG_cases_grid}, with double (TG\_g) or half (TG\_h) grid step size $\delta x/\xi$.
\begin{table}[h!]\centering
{\small	\begin{tabular}{|c|c|c|c|c|c|}\hline
		Run  & $N_x$ & $\alpha$ & $\beta$ & $\xi k_{\max}$ & $\delta x/\xi$ \\ \hline
		TG\_a  & 128 & $0.05$ & $40$ &  $2.26$ & 1.388\\
		TG\_g  & 64 & $0.05$ & $40$ &  $1.13$& 2.776\\
		TG\_h  & 256 & $0.05$ & $40$ &  $4.52$&0.694\\ \hline
	\end{tabular}
}
	\caption{TG-QT. Supplementary runs used to check the grid convergence of the results (to be compared to runs in Table \ref{tab:ALL-params}). }\label{tab:TG_cases_grid}
\end{table}

\begin{figure}[h!]
		\begin{center}	
		\begin{minipage}{0.5\textwidth} 
			a)\\
			\includegraphics[width=\textwidth]{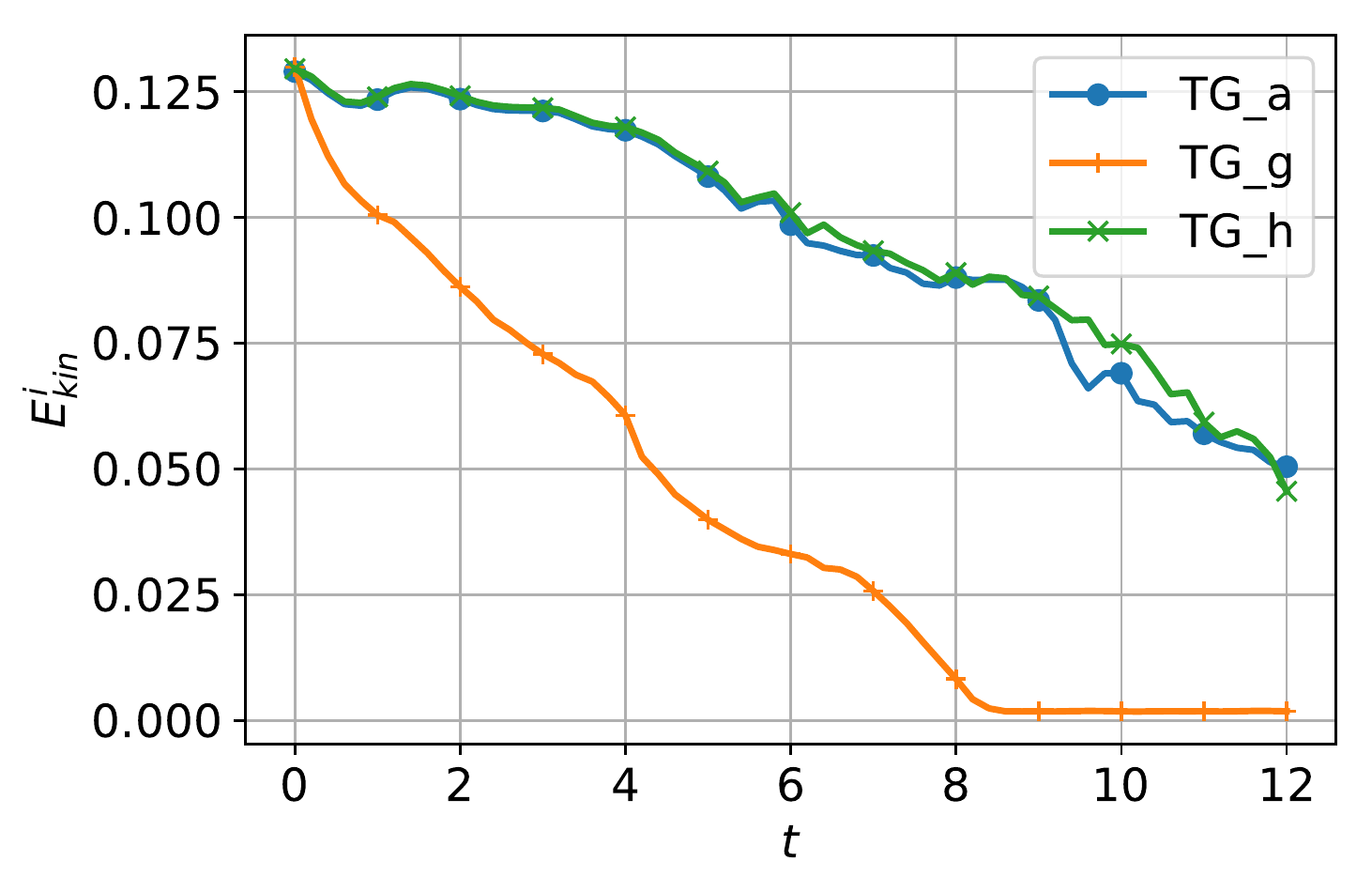}
		\end{minipage}\hfill
		\begin{minipage}{0.5\textwidth}
			b)\\ 
			\includegraphics[width=\textwidth]{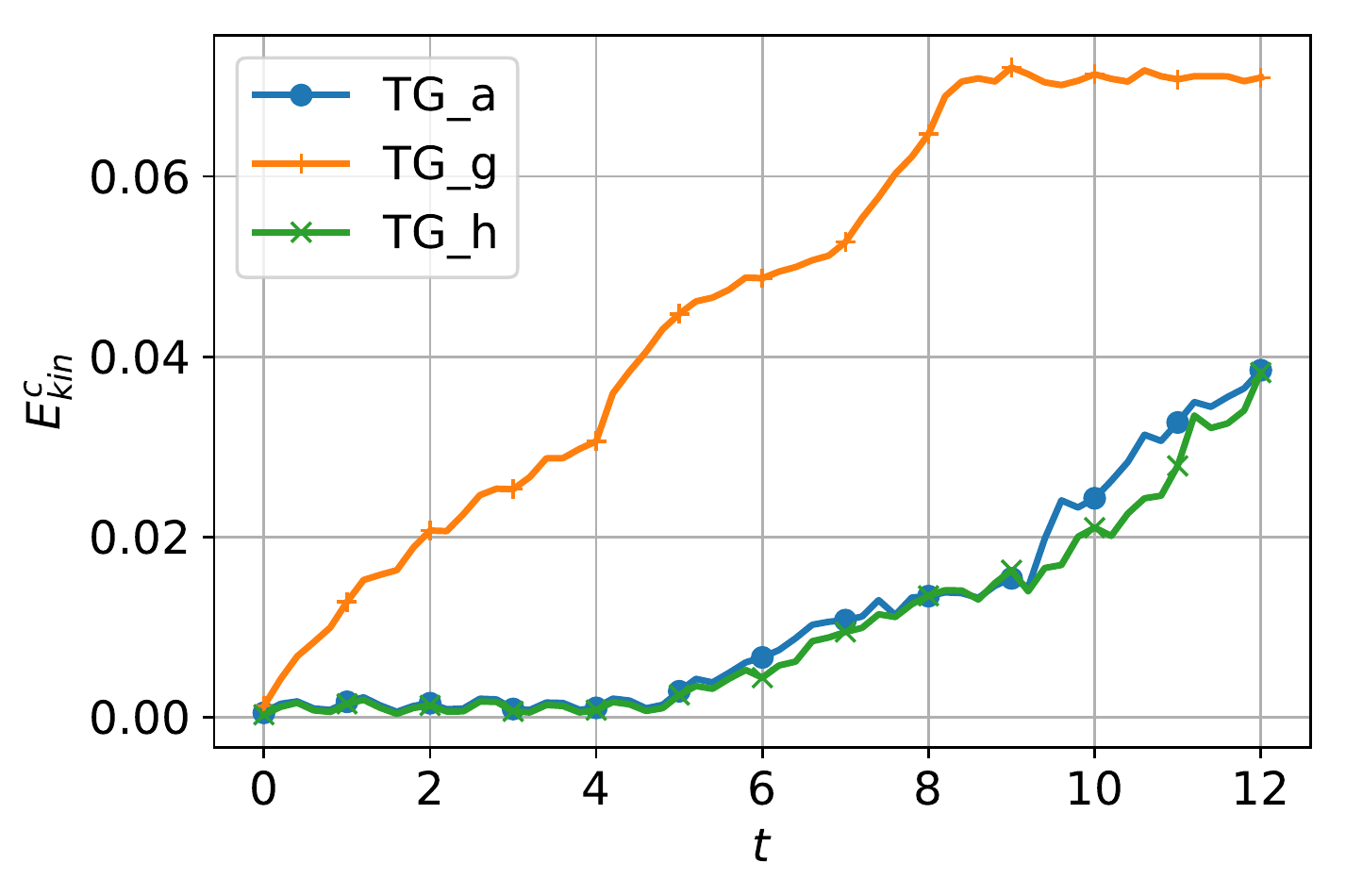}
		\end{minipage}
	\end{center}
	\caption{TG-QT. Time evolution of incompressible $\Ekini$ (a) and compressible $\Ekinc$ (b) energies for cases TG\_a, TG\_g, TG\_h, used to check grid convergence.}\label{fig:TG-grid-evol}
\end{figure}

In Table \ref{tab:TG_grid_energy}, we report the values of different energies obtained at the end of the imaginary-time ARGLE procedure for these new cases. Relative errors were computed with respect to reference values of the case TG\_a. We conclude that a value of $\xi k_{\max}\simeq 2$, \ie $\delta x/\xi =\pi/2$, is sufficient to ensure the grid convergence and good accuracy of numerical results. To further check this assessment, we also simulated the QT evolution starting from these runs. We report in Fig. \ref{fig:TG-grid-evol} the time evolution of incompressible and compressible kinetic energies. The similarities between cases TG\_a and TG\_h  suggest that the resolution used for the case TG\_a is fine enough to capture the vortices in the QT field. This validates the choice of parameters in Table \ref{tab:ALL-params}. 
\begin{table}[h!]\centering
{\small	\begin{tabular}{|c|c|cl|cl|}\hline
	 & TG\_a & TG\_g &(rel. err.) & TG\_h &(rel. err.) \\ \hline
$\Ekini$	 & 1.29017e-01 & 1.29896e-01 & (6.82e-03) & 1.29562e-01 & (4.23e-03) \\ 
$\Ekinc$	 & 4.86671e-04 & 1.24066e-03 & (1.55e+00) & 2.72478e-04 & (4.40e-01) \\ 
$\Eq$		 & 7.92394e-03 & 1.08017e-02 & (3.63e-01) & 7.80383e-03 & (1.52e-02) \\ 
$\Eint$		 & 1.29952e-02 & 9.94301e-03 & (2.35e-01) & 1.30274e-02 & (2.48e-03) \\ 
$E_{\bv}$ 	 & 7.10060e-01 & 6.98672e-01 & (1.60e-02) & 7.10095e-01 & (4.98e-05) \\  \hline
\end{tabular}
}
	\caption{TG-QT. Energies computed from $\phi_{TG}$, the wave function obtained at the end of the imaginary-time ARGLE procedure for cases TG\_a, TG\_g and TG\_h. Relative errors (rel. err.) were computed with respect to reference values of the case TG\_a. } \label{tab:TG_grid_energy}
\end{table}

\clearpage

\subsection{Benchmark \#2: Arnold-Beltrami-Childress quantum turbulence (ABC-QT)}
The ABC initial field was prepared as described in \S \ref{sec:init-abc}.  We display in  Table \ref{tab:ABC-params}  the values of the time step $\delta t$ used in the GP solver (see \S\ref{sec:real-time}) and the final time $T_f$ of each simulation. The parameters of the corresponding imaginary-time (IT) run cases preparing the initial condition using the ARGLE solver are also presented, with $\delta \tau$ and $\tau_f$ the imaginary-time step and final value at convergence, respectively. 
\begin{table}[h!]
	\centering
	{\small	\begin{tabular}{|c|c|c|c|}\hline
			Run  &  $N_x$  & $\delta t$&$T_f$\\ \hline
			ABC\_a &	128    &8.0e{-4} & 10\\
			ABC\_b &	256    &4.0e{-4} & 10\\
			ABC\_c &	512    &2.0e{-4} & 10\\ \hline
		\end{tabular}
	}	
	{\small	\begin{tabular}{|c|c|c|c|}\hline
			Run  &  $N_x$ & $\delta \tau$&$\tau_f$\\ \hline
			ABC\_aIT &	128  &4.0e{-3} & 30\\
			ABC\_bIT &	256  &2.0e{-3} & 30\\
			ABC\_cIT &	512  &1.0e{-3} & 30\\ \hline
		\end{tabular}
	}	
	
	\caption{Runs for the ABC-QT case. Parameters used in the GP solver (cases ABC\_a to ABC\_c)  and the imaginary-time (IT) ARGLE solver (cases ABC\_aIT to ABC\_cIT). For each space resolution $N_x$, the corresponding physical and numerical parameters are displayed in Table \ref{tab:ALL-params}. }\label{tab:ABC-params}
\end{table}

\subsubsection{Results for the imaginary time (ARGLE) procedure}

\begin{figure}[!h]
	\begin{center}
		\begin{minipage}{\textwidth}
			\includegraphics[width=0.33\textwidth]{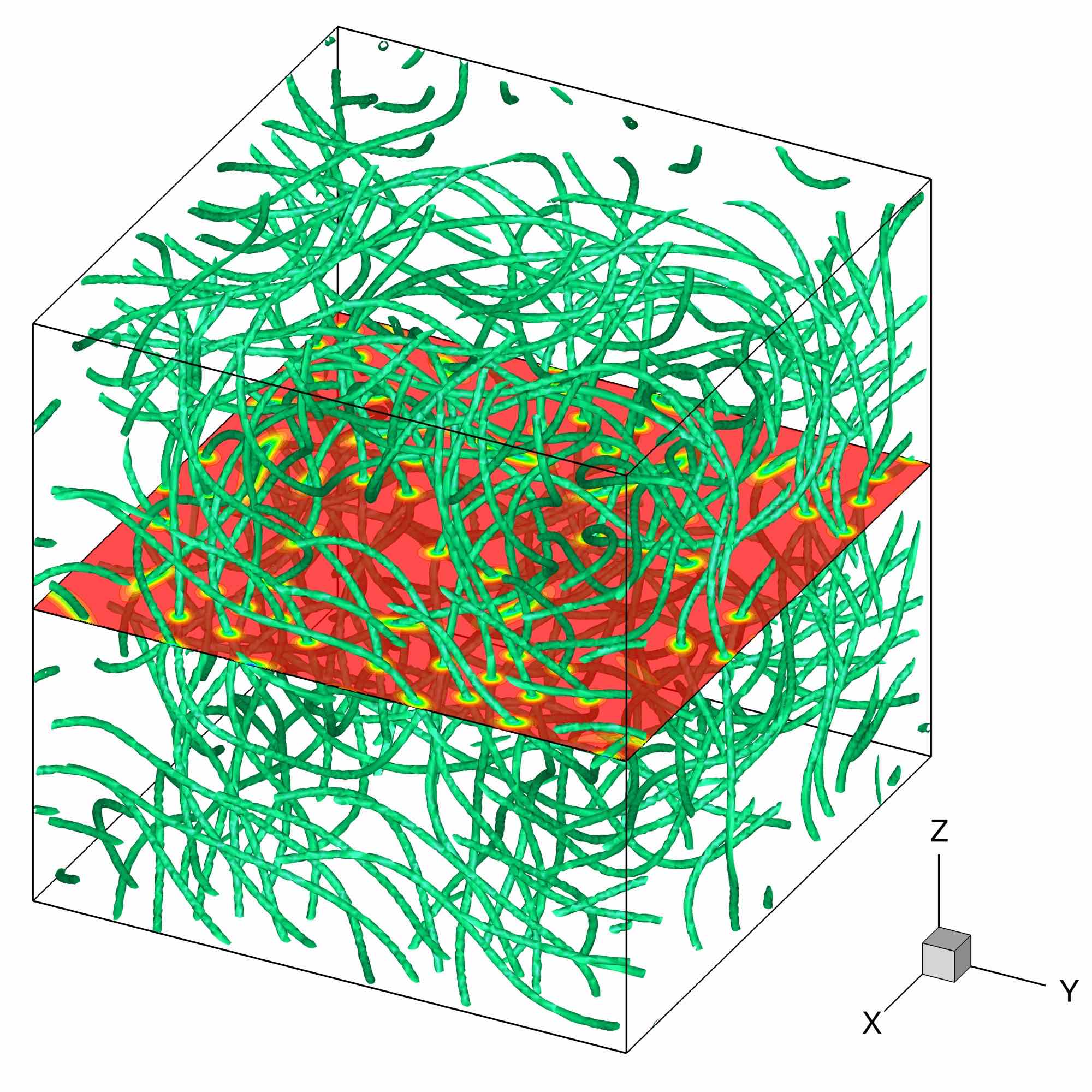}
			\includegraphics[width=0.33\textwidth]{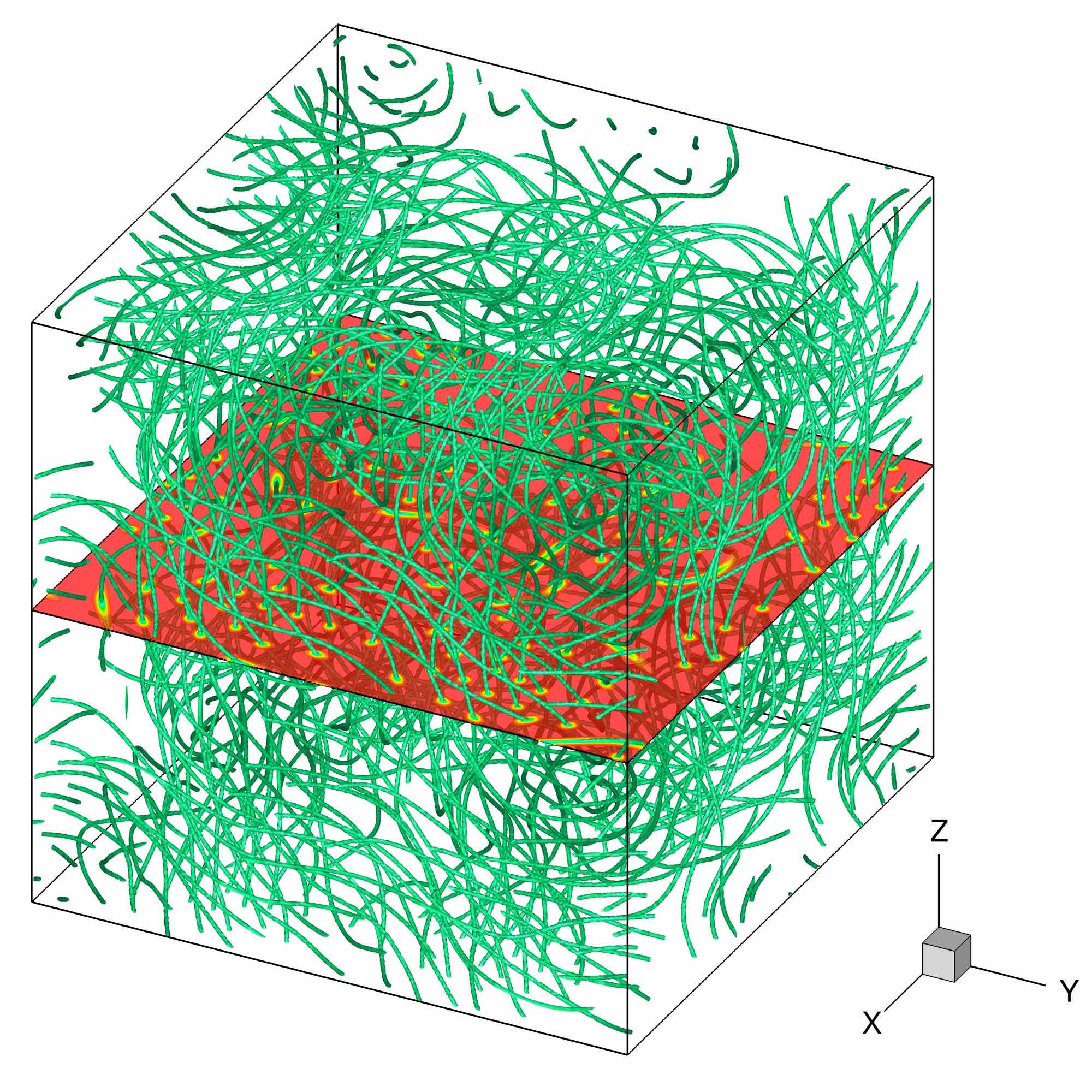}
			\includegraphics[width=0.33\textwidth]{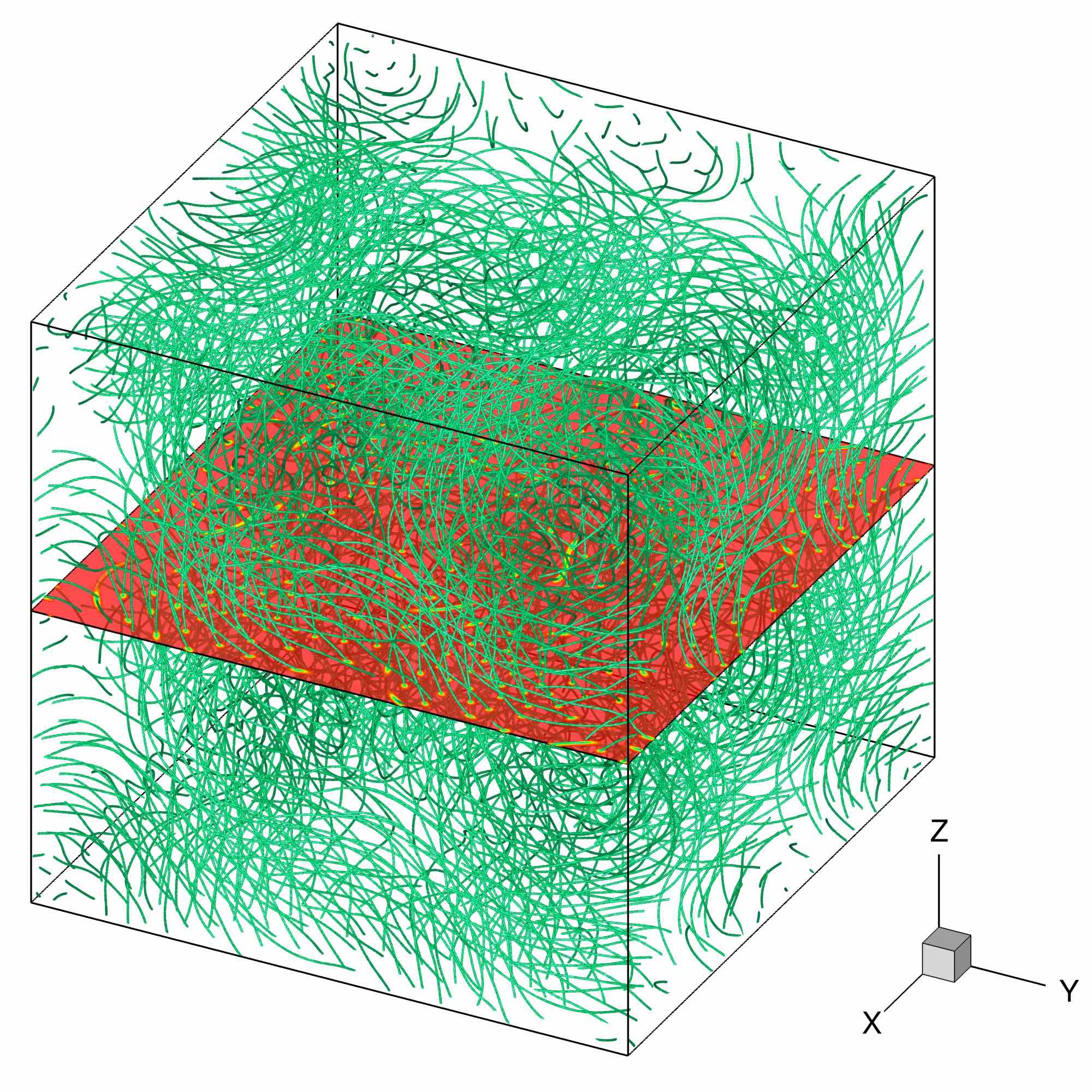}
		\end{minipage}
	\end{center}	
	\caption{
		ABC-QT. Initial condition computed with the imaginary-time ARGLE solver. Vortex lines (iso-surfaces of low $\rho$) of the  converged wave function $\phi_{\rm ABC}$ at final imaginary-time $\tau_f$.	From left to right: grid resolutions $N_x=$ 128, 256, 512 (corresponding to runs ABC\_aIT, ABC\_bIT and ABC\_cIT  in Table \ref{tab:ABC-params}).}
	\label{fig:ABC-at-tauf}
\end{figure}

Following \eqref{ABCphi0} and \eqref{ABCphi}, the initial condition for the imaginary--time ARGLE procedure is obtained only by phase manipulations of the wave function. Therefore, vortices are not present at $\tau=0$, but they nucleate during the imaginary-time evolution, which a  dissipative process.  The obtained fields with vortices at the end of the ARGLE procedure are illustrated in Fig. \ref{fig:ABC-at-tauf}. Note that, compared to the TG fields in Fig.  \ref{fig:TG-at-tauf}, 
the distribution of vortices in the computational box displays no symmetries with respect to central planes. This is the first feature that makes the ABC case different from the TG case.

The second differentiating feature is the presence of helicity in the ABC flow obtained after the ARGLE procedure. We recall that the helicity of the TG flow is strictly zero (see Table \ref{tab:TG_ARGLE_Energy}).  We report in Table \ref{tab:ABC_ARGLE_Energy} the values of different energies (see \S\ref{sec:energ-decomp}) and helicity computed  for the final ABC field (at $\tau_f$).  As expected, the value of the incompressible kinetic energy is close to 1, which corresponds to the energy of the classical ABC flow. For the helicity, the theoretical value for  the classical ABC flow is 3. A close value to 3  is obtained only for large grid resolutions ($N_x \geq 256$), \ie for sufficiently small values of the vortex size $\xi$. 
\begin{table}[h!]\centering
{\small	\begin{tabular}{|c|c|c|c|c|c|c|}\hline
		Run & $E_{kin}^i$ & $E_{kin}^c$ & $E_q$ & $E_{int}$ & $H$ & $H_{\rm reg}$ \\ \hline
ABC\_aIT	 & 0.9485114	 & 0.0014218	 & 0.0277287	 & 0.0430379	 & 2.4106982	 & 2.4726091 \\ 
ABC\_bIT	 & 0.9792042	 & 0.0008142	 & 0.0144931	 & 0.0237201	 & 2.7439625	 & 2.6532860 \\ 
ABC\_cIT	 & 0.9884992	 & 0.0006486	 & 0.0073802	 & 0.0124975	 & 2.7217161	 & 2.7365301 \\ 
 \hline	
	\end{tabular}
}
			\caption{ABC-QT. Values of different energies and helicity at $\tau_f$ for the runs preparing  the ABC initial condition, using the imaginary-time ARGLE solver. }\label{tab:ABC_ARGLE_Energy}
\end{table}

\begin{figure}[h!]
				\begin{center}	
		\begin{minipage}{0.5\textwidth} 
			a)\\
			\includegraphics[width=\textwidth]{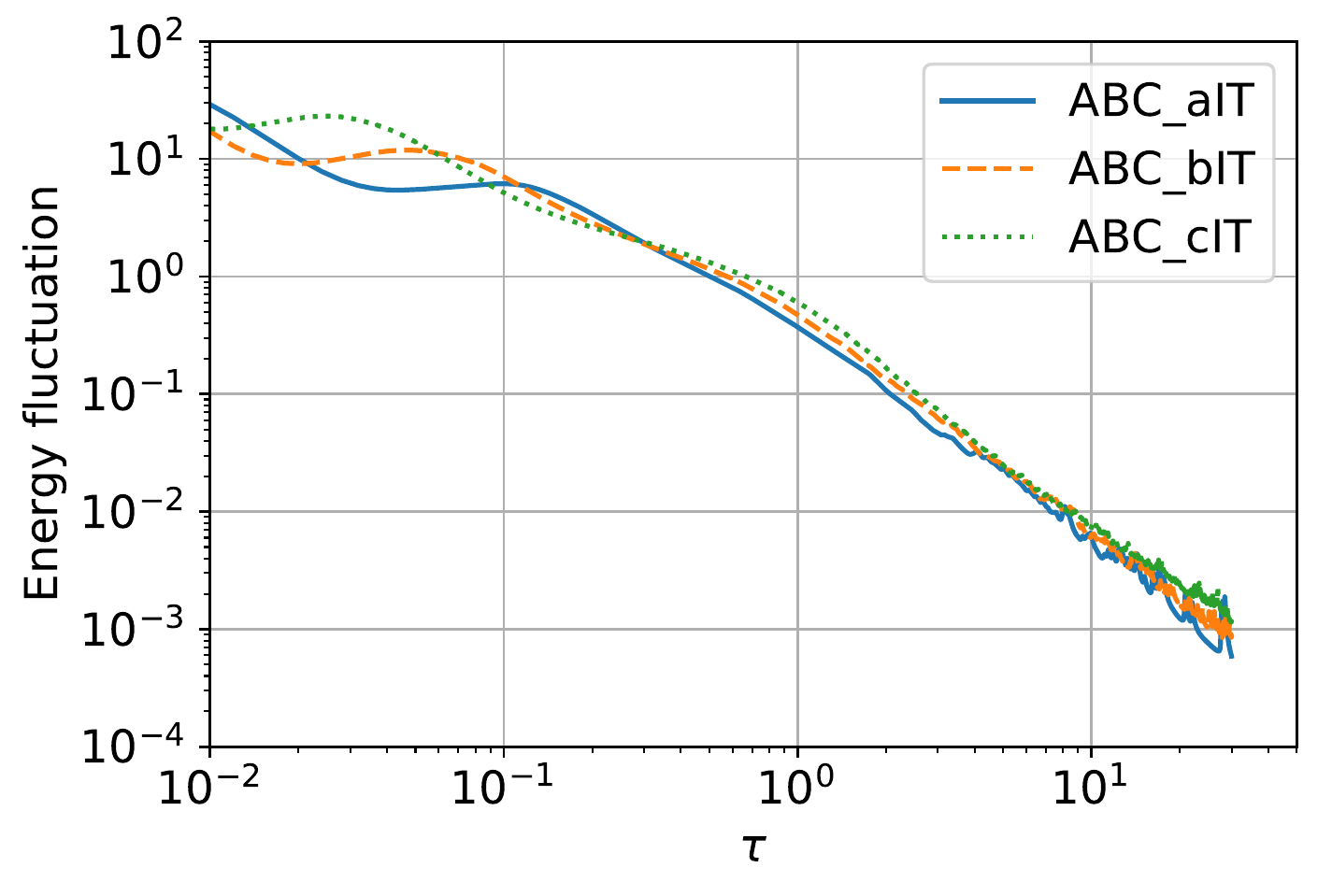}
		\end{minipage}\hfill
		\begin{minipage}{0.5\textwidth} 
			b)\\
			\includegraphics[width=\textwidth]{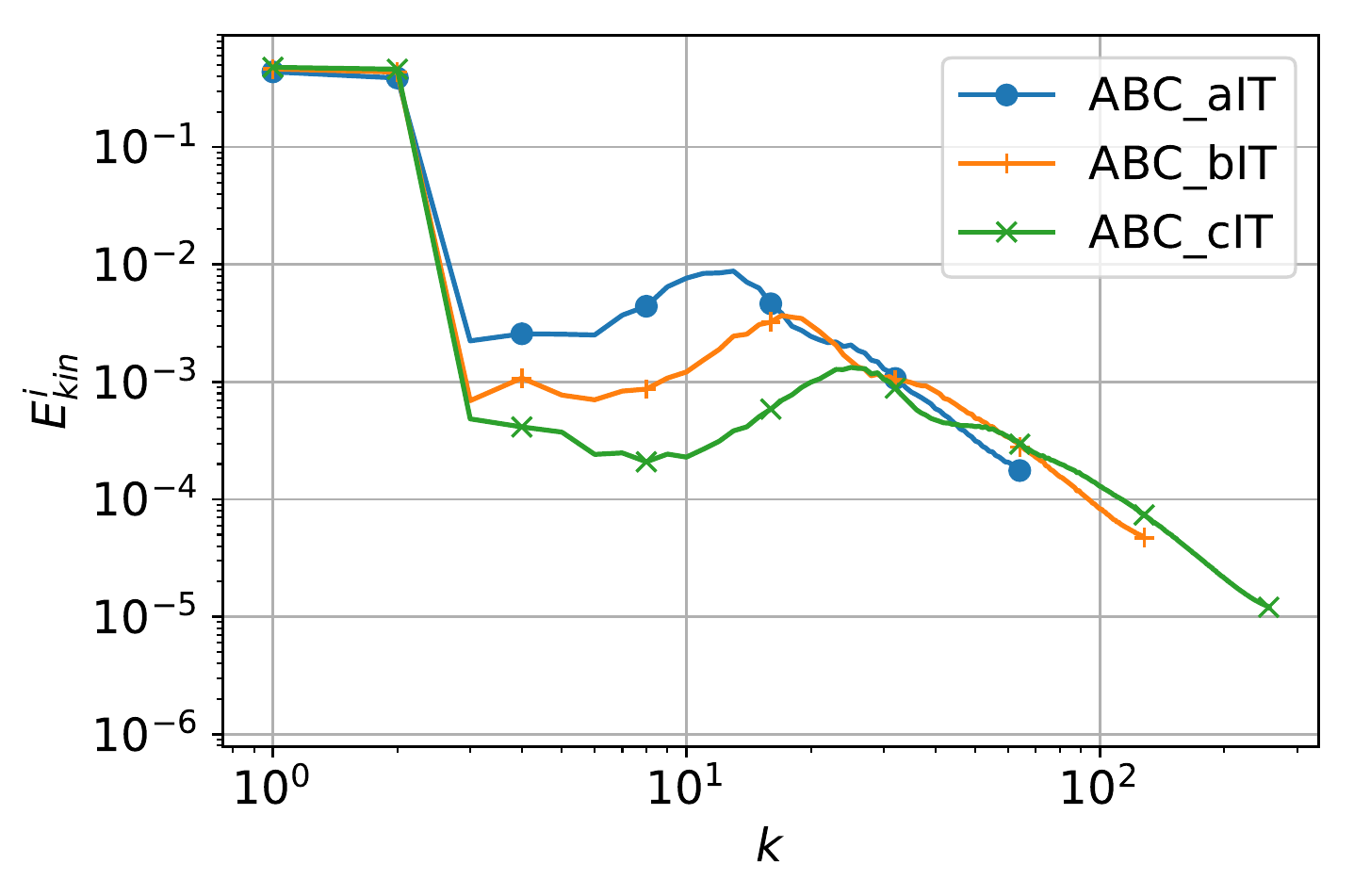}
		\end{minipage}
	\end{center}
	\caption{ABC-QT. (a) Relative fluctuation of the total energy (${|E_{\bv}(\phi^{n+1})-E_{\bv}(\phi^{n})|}/\left({\delta \tau E_{\bv}(\phi^{n})}\right)$) during the ARGLE computation. (b) Spectrum of $\Ekini$,  the  incompressible kinetic energy  of ARGLE solutions. Results for cases ABC\_aIT, ABC\_bIT and ABC\_cIT  described in Table \ref{tab:ABC-params}.}
	\label{fig:ABC-conv-energy}
\end{figure}

For these computations, the ARGLE procedure required a significant computational time and was therefore stopped before the convergence criterion \eqref{eqn:crit1} was satisfied. To ensure the validity of the ARGLE solution, we estimated the criterion  \eqref{eqn:crit2} by monitoring in Fig. \ref{fig:ABC-conv-energy}a the imaginary-time evolution of energy fluctuations defined as ${|E_{\bv}(\phi^{n+1})-E_{\bv}(\phi^{n})|}/\left({\delta \tau E_{\bv}(\phi^{n})}\right)$, with $E_{\bv}$ expressed by \eqref{eq-energy-num-ext-vel}. The convergence criterion  \eqref{eqn:crit2} is satisfied to a fairly good degree of precision ($10^{-3}$). Figure \ref{fig:ABC-conv-energy}b shows the spectra of $\Ekini$, the incompressible kinetic energy of ARGLE solutions. This is an important benchmark verification, since $\Ekini$ represents the most important part in the total energy of the ABC super-flow (see Table \ref{tab:ABC_ARGLE_Energy}). The similar slopes of the spectra for large wave numbers $k$ indicate that the energy distribution of the three ABC flows are similar at small scales. The low-$k$ part of the spectrum ($k \ll k_\xi$) reproduces the classical spectrum which has only $2$ nonzero modes, $k=1, 2$. The slope for $ k \gg k_\xi$ is $-3$ (\ie $\Ekini (k)\sim k^{-3})$. This feature of the high-$k$ spectrum is detailed in \cite{Krstulovic2010}.

\subsubsection{Results for the ABC-QT}
\begin{figure}[!h]
	\begin{center}
		\begin{minipage}{\textwidth}
			\includegraphics[width=0.33\textwidth]{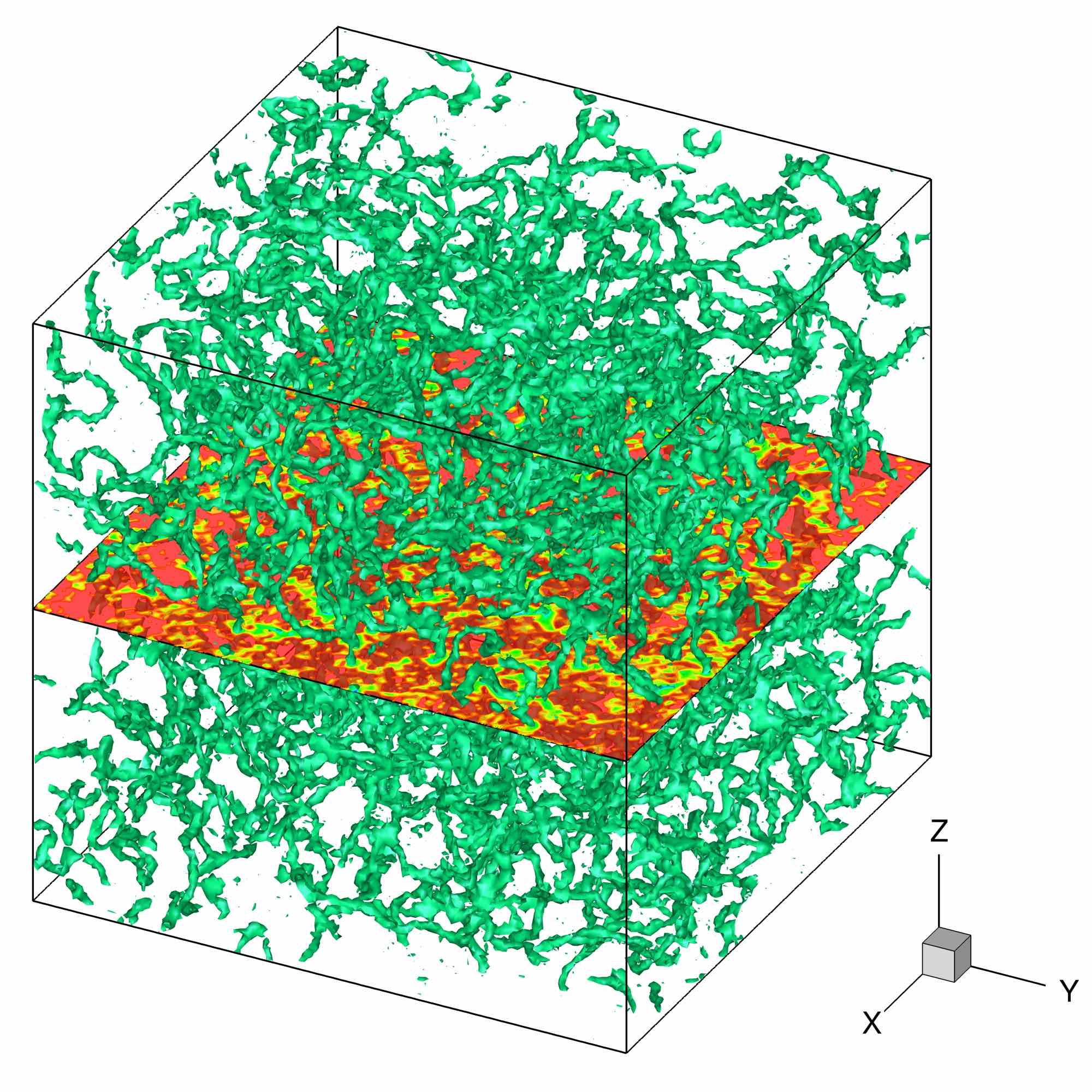}
			\includegraphics[width=0.33\textwidth]{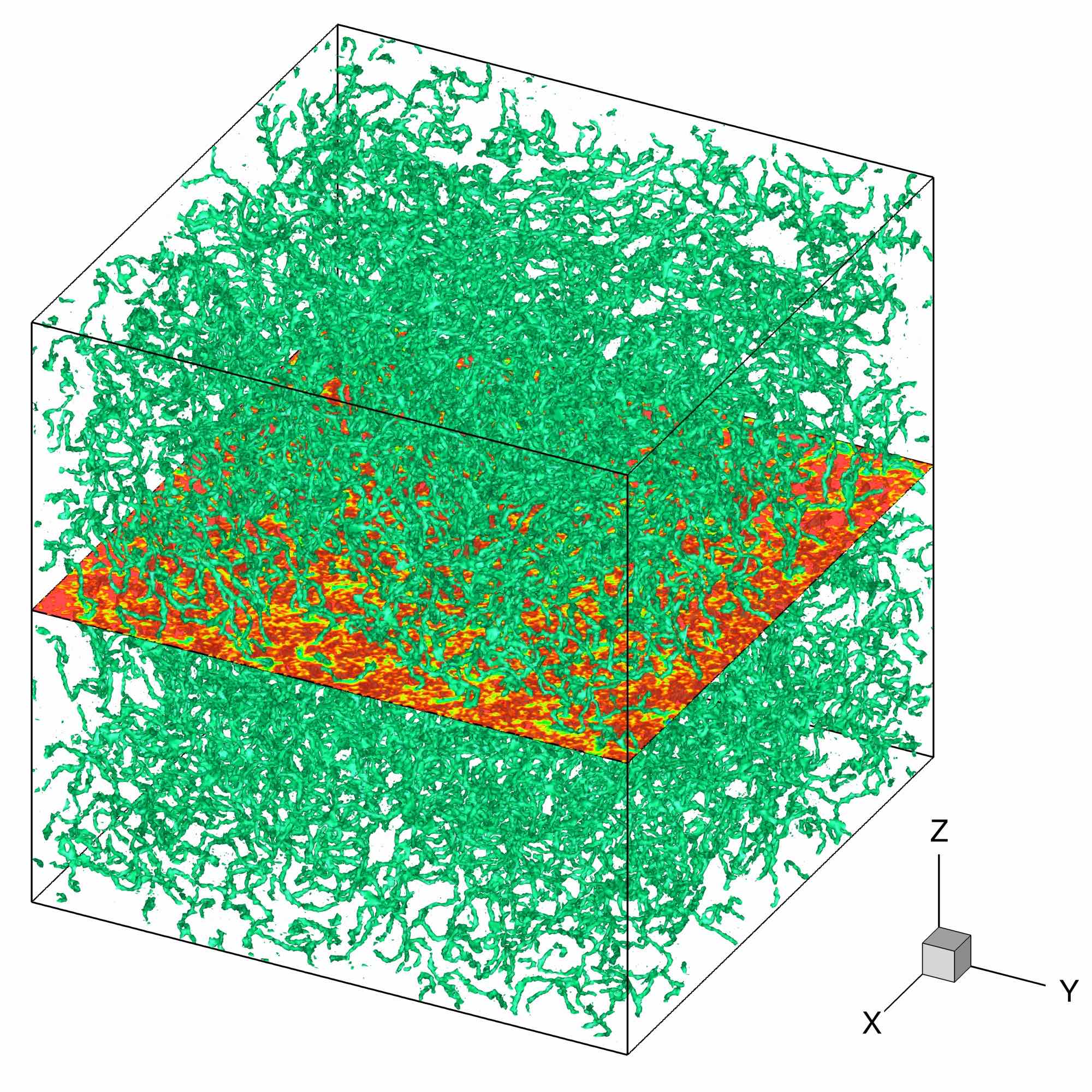}
			\includegraphics[width=0.33\textwidth]{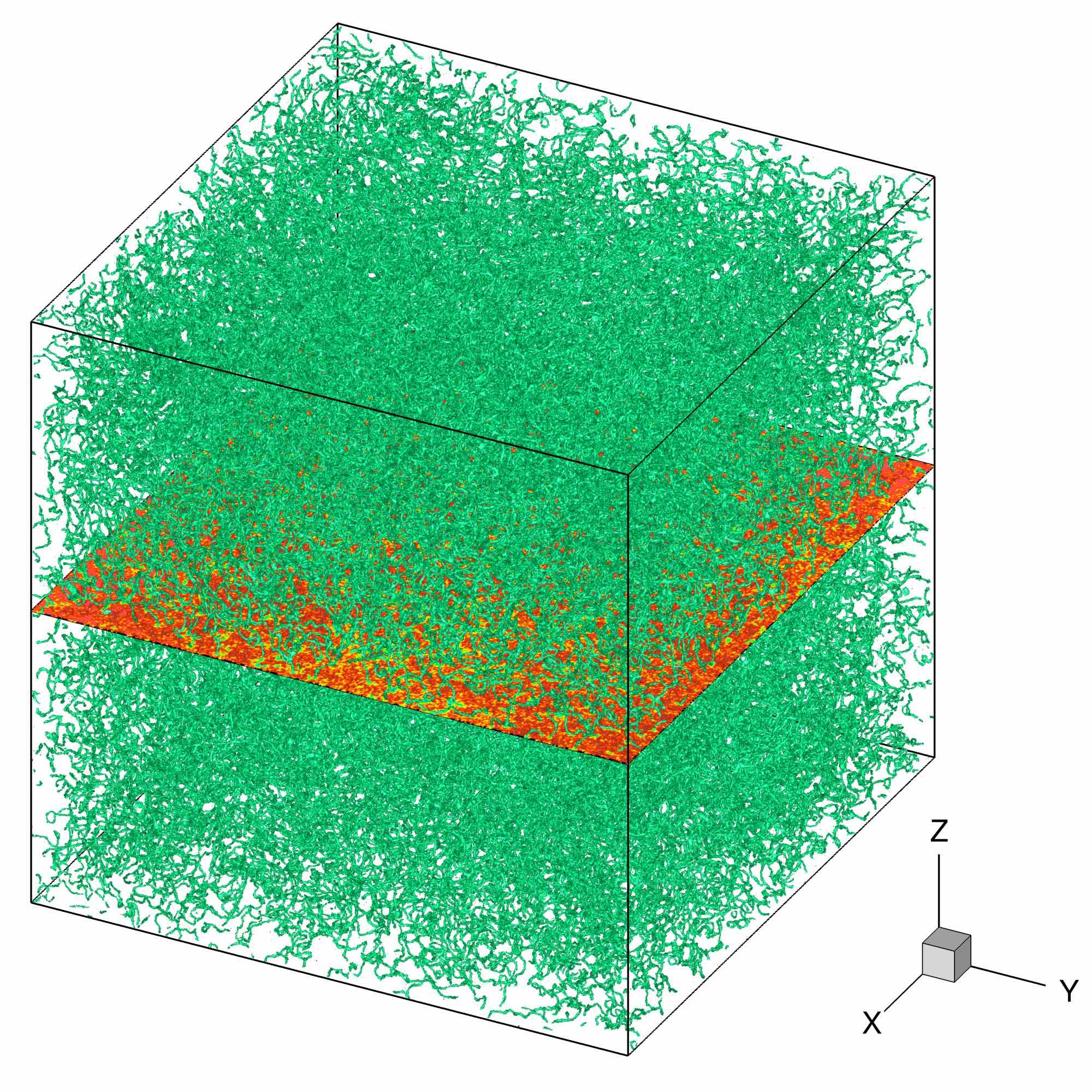}
		\end{minipage}
	\end{center}
	\caption{ABC-QT. Instantaneous fields computed with the real-time GP solver, starting from the initial condition presented in Fig. \ref{fig:ABC-at-tauf}. Vortex lines (iso-surfaces of low $\rho$) of the  wave function at final time $T_f$.
		From left to right: grid resolutions $N_x=$ 128, 256, 512 (corresponding to runs ABC\_a, ABC\_b and ABC\_c in Table \ref{tab:ABC-params}).}
	\label{fig:ABC-at-Tf}
\end{figure}
Starting from the initial condition presented in Fig. \ref{fig:ABC-at-tauf}, we used the Strang--splitting GP solver (see \S\ref{sec:real-time})  to advance the wave function in real time. The final (at $t=T_f$) QT field is displayed in Fig. \ref{fig:ABC-at-Tf} for runs ABC\_a, ABC\_b and ABC\_c. As for the TG case, when the grid resolution $N_x$ is increased, the size of a vortex core $\xi$ diminishes and, consequently, the density of the tangled vortex lines is increased.

\begin{figure}[!h]
				\begin{center}	
		\begin{minipage}{0.5\textwidth} 
			a)\\
			\includegraphics[width=\textwidth]{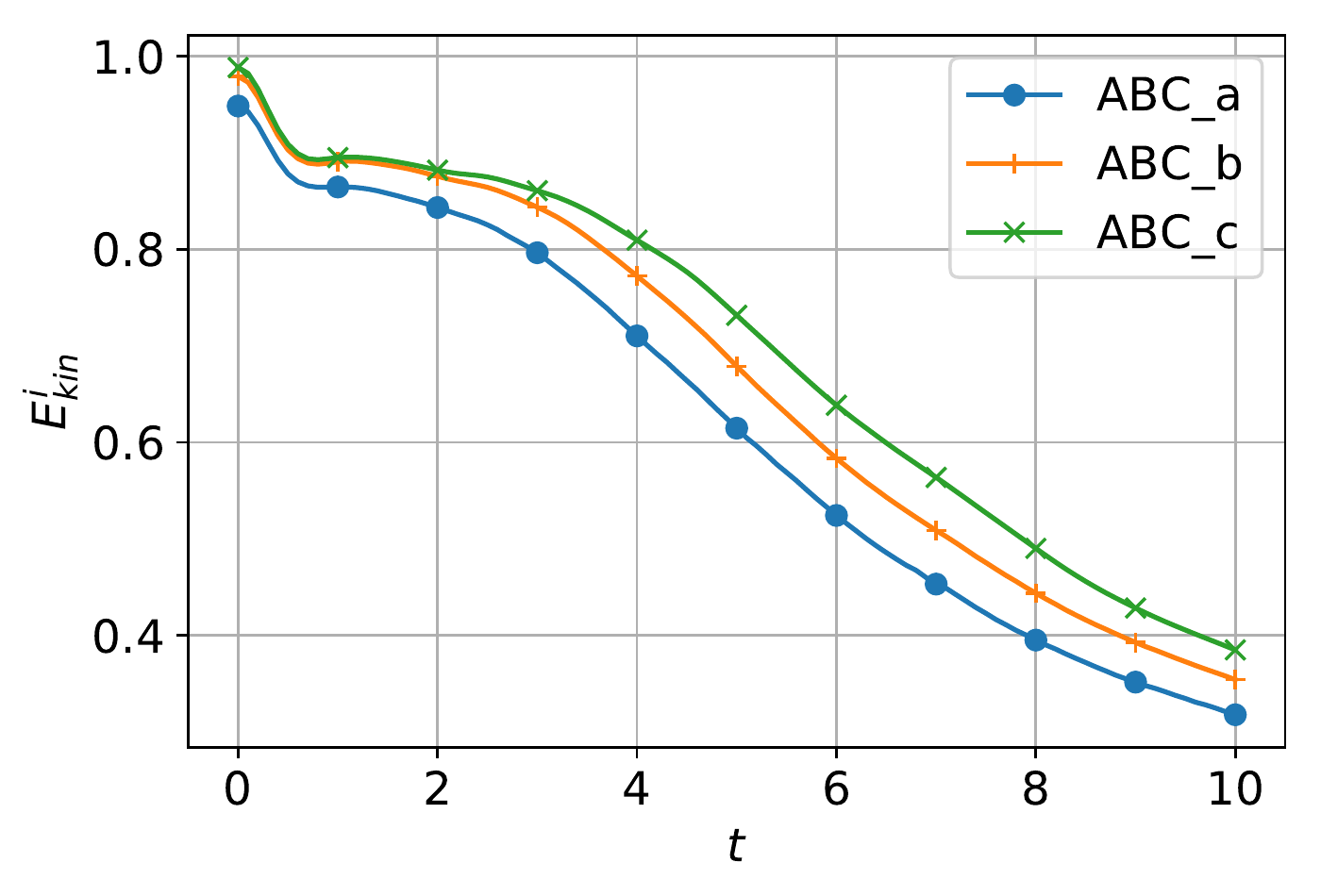}
		\end{minipage}\hfill
		\begin{minipage}{0.5\textwidth} 
			b)\\
			\includegraphics[width=\textwidth]{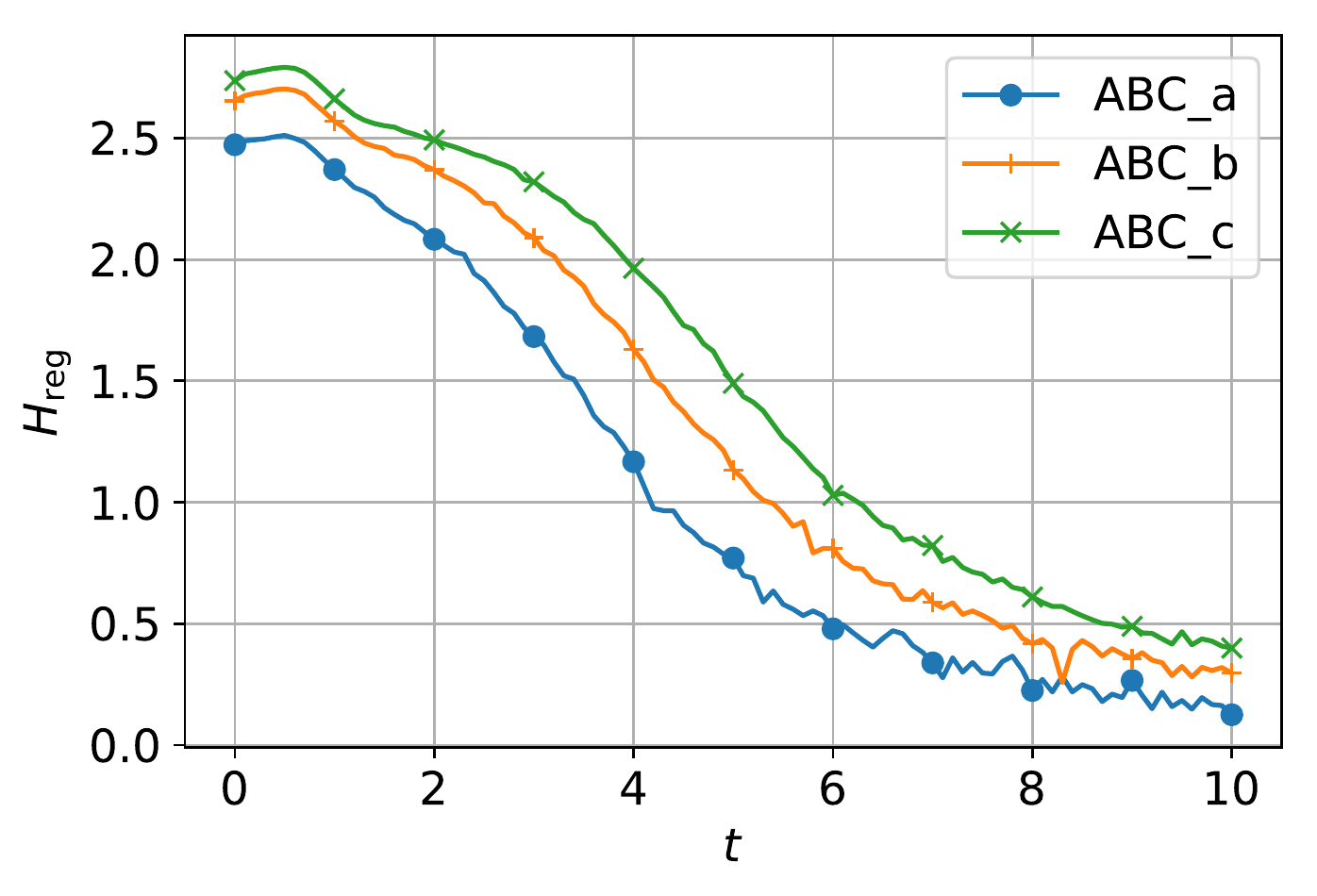}
		\end{minipage}
	\end{center}
	\caption{ABC-QT. Time evolution of incompressible kinetic energy $\Ekini$ (a) and  regularized helicity $H_{\rm reg}$  (b) (see Eq. \eqref{Hreg}).}\label{fig:evol_energy_ABC}
\end{figure}

The time evolution of the incompressible kinetic energy $\Ekini$ and the regularized helicity $H_{\rm reg}$ (see Eq. \eqref{Hreg}) are shown in Fig. \ref{fig:evol_energy_ABC}. These results are in good agreement with those reported by  \cite{Clark17}. 

\begin{figure}[!h]
	\begin{center}	
		\begin{minipage}{0.5\textwidth} 
			a)\\
			\includegraphics[width=\textwidth]{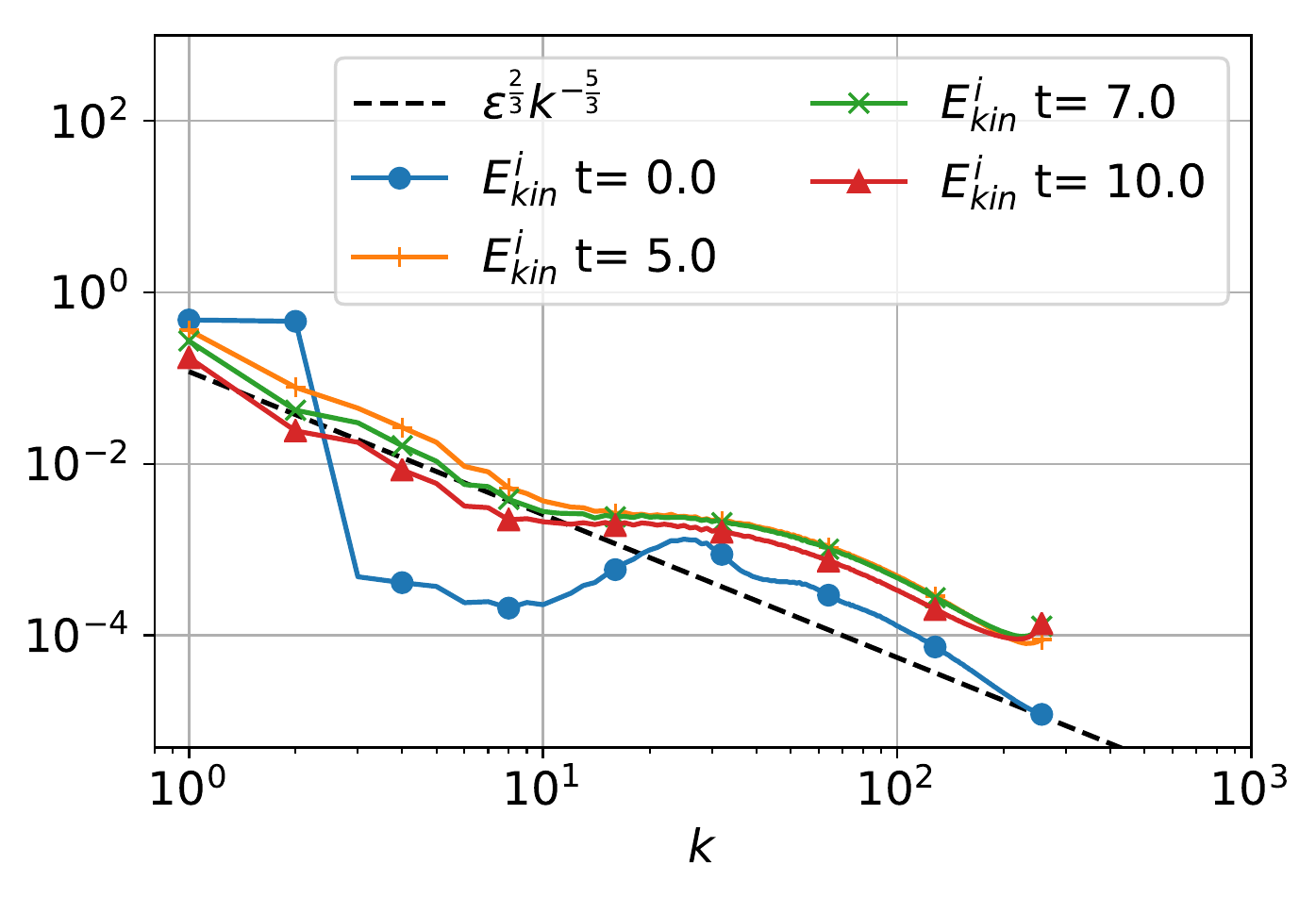}
		\end{minipage}\hfill
		\begin{minipage}{0.5\textwidth} 
			b)\\
			\includegraphics[width=\textwidth]{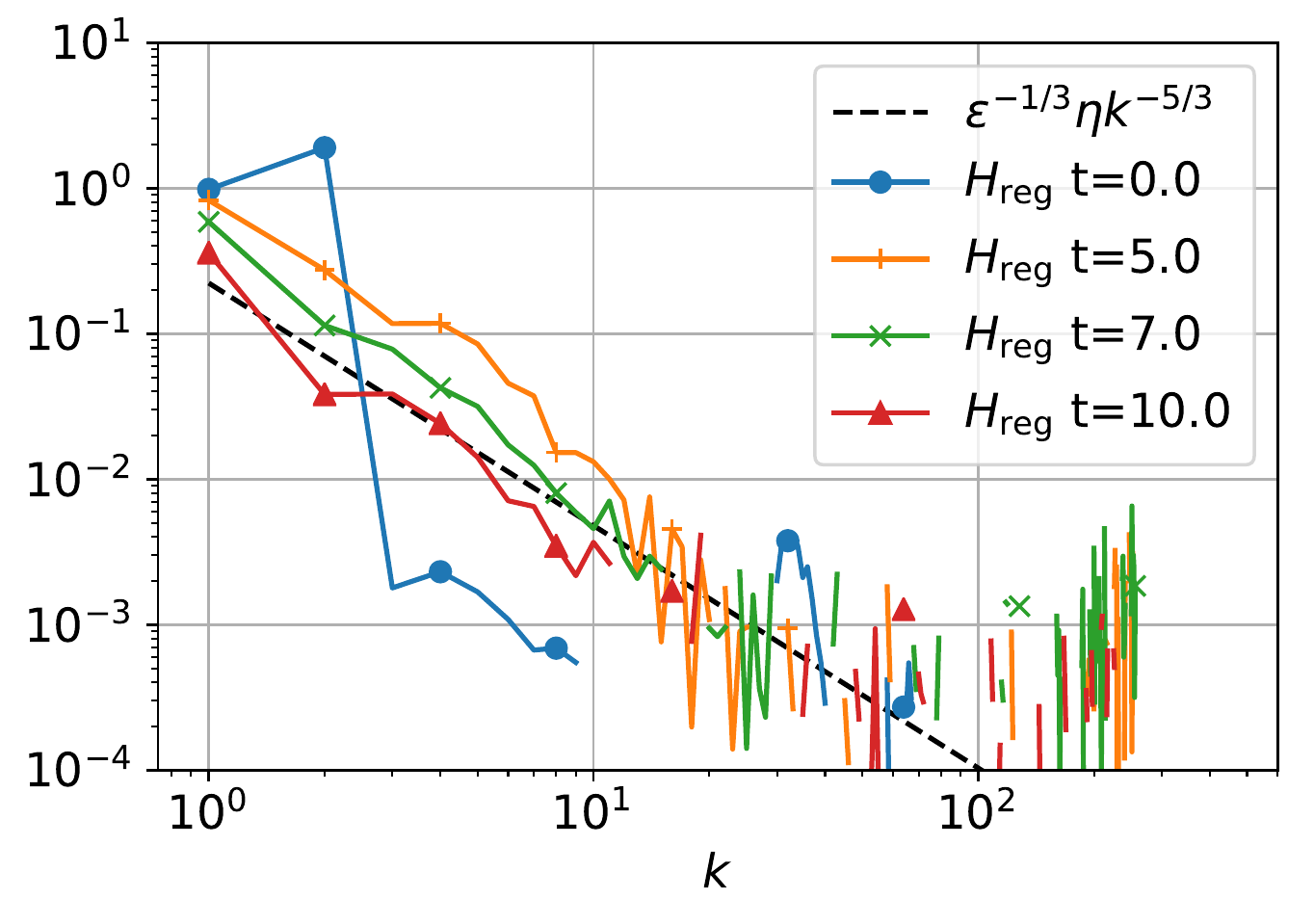}
		\end{minipage}
		\begin{minipage}{0.5\textwidth} 
			c)\\
			\includegraphics[width=\textwidth]{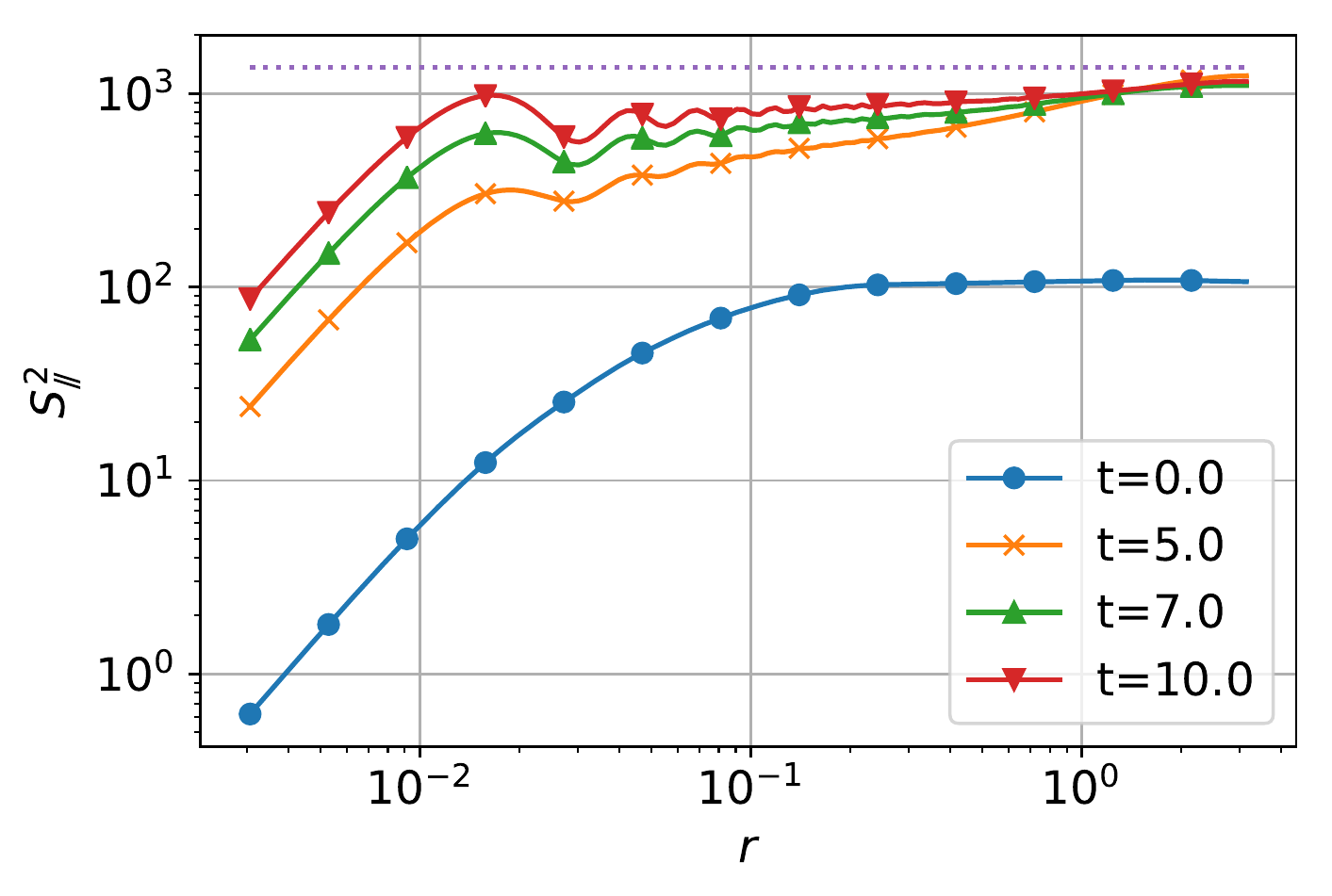}
		\end{minipage}
	\end{center}
	\caption{ABC-QT. Analysis of the turbulent super-flow. Spectrum of incompressible kinetic energy $\Ekini$  (a) and regularized  helicity $H_{\rm reg}$ (b) at different time instants for case ABC\_c. Dashed lines represent reference power laws with  slope $-5/3$. Panel (c) displays the second order structure function for the same case ABC\_c and same time instants.}\label{fig:ABC_spectra}
\end{figure}

To analyze the turbulent super-flow, we plot in  Fig. \ref{fig:ABC_spectra} spectra for the incompressible kinetic energy $\Ekini$ (panel a) and the regularized helicity $H_{\rm reg}$ (panel b) at different time instants and the second-order structure function $S^2_{\para}(r)$ (panel c). We plot with dashed lines in Figs. \ref{fig:ABC_spectra}a and  \ref{fig:ABC_spectra}b the reference (Kolmogorov-like)  power laws $\varepsilon^{2/3} k^{-5/3}$ for $\Ekini$  and $\eta\varepsilon^{-1/3}k^{-5/3}$ for helicity, respectively. The constants $\varepsilon$ and $\eta$ were computed  as:
\begin{equation}
\varepsilon = -\left.\frac{d\Ekini}{dt}\right|_{t=10},\qquad \eta = -\left.\frac{dH}{dt}\right|_{t=10}.
\end{equation}

We note from Figs. \ref{fig:ABC_spectra}a and  \ref{fig:ABC_spectra}b that for $t>5$, both energy and helicity $H_{\rm reg}$ spectra exhibit at large scales a power law variation with exponent $-5/3$, compatible with a dual energy and helicity cascade. Again, this result is in good agreement with the results  of \cite{Clark17}. The novel diagnostic tool introduced in the previous section for the TG flow is also performed with the ABC flow by computing the second--order structure function $S^2_{\para}(r)$  (see Eq. \eqref{strfunc-def}). Figure \ref{fig:ABC_spectra}c displays the structure function for the same case ABC\_c and same time instants considered for plotting spectra. A similar evolution as noted for the TG case (see Fig. \ref{fig:TG_spectrum_and_sf}) is observed:  the slope of the structure function curve at the origin is close to 2, and, for  large length scales, the asymptotic value $2\int v_x^2$ (dotted line).

\clearpage

\subsubsection{Accuracy of numerical results and influence of the Mach number}

As for the TG case, we monitor the time variation of the number of particles $N$ (see Eq. \eqref{eq-scaling-N}) and  total energy per volume unit  E (see Eq. \eqref{eq-scaling-energy}). The accuracy to which these two quantities are conserved by the GP solver is reported in  Table \ref{tab:ABC-conservation} initial and final values for the norm and normalized energy, as well as their relative maximum variation during the time evolution. Note from Table \ref{tab:TG-conservation} that $N$ is perfectly conserved, and energy relative fluctuations $\delta(E)$ are less than 0.01\%, which is sufficiently small value to guarantee the validity of the computation.
\begin{table}[h!]\centering
	\begin{tabular}{|c|c|c|c|c|c|c|}\hline
		Run & ${N}_{|t=0}$ & ${N}_{|t=T_f}$ & $\delta({N})$ & $E_{|t=0}$ & $E_{|t=T_f}$ & $\delta E$ \\ \hline
ABC\_a	 & 0.9410138 	 & 0.9410138 	 & 0.0 	 & 1.0206996 	 & 1.0206400 	 & 5.98e-05  \\ 
ABC\_b	 & 0.9647786 	 & 0.9647786 	 & 0.0 	 & 1.0182316 	 & 1.0181689 	 & 6.18e-05  \\ 
ABC\_c	 & 0.9796053 	 & 0.9796053 	 & 0.0 	 & 1.0090255 	 & 1.0089631 	 & 6.19e-05  \\ 
 \hline
	\end{tabular}
	\caption{ABC-QT. Conservation of the number of particles $N$ and  energy per volume unit. 
		Initial (at $t=0$) and final values (and $t=T_f$) and relative maximum variation, defined following \eg $\delta(E)=\max_{t\in[0;T_f]} \left| E(t) - E_{t=0}\right|/E_{t=0}$.}\label{tab:ABC-conservation}
\end{table}

\begin{figure}[h!]
	\begin{center}	
		\begin{minipage}{0.5\textwidth} 
			a)\\
			\includegraphics[width=\textwidth]{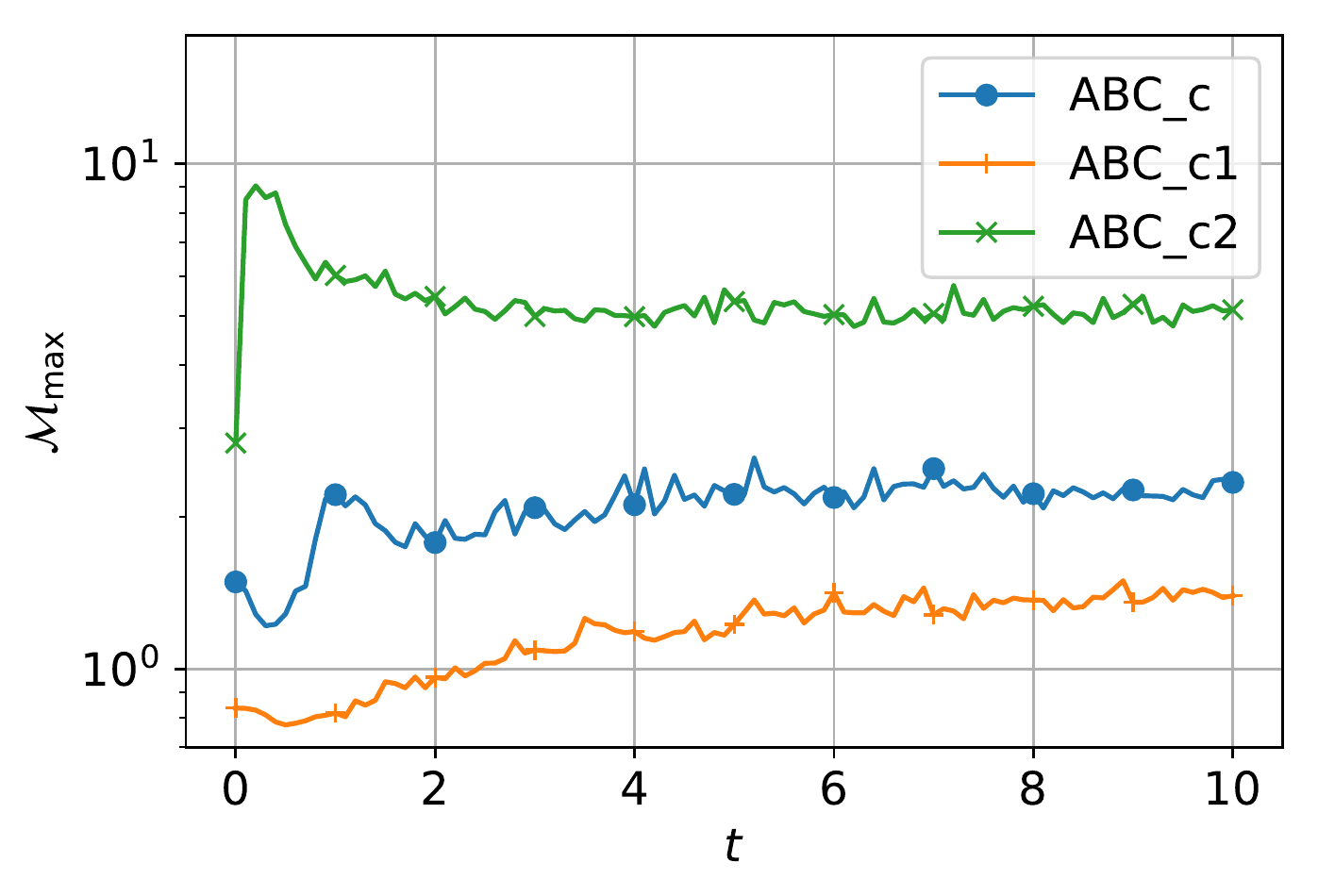}
		\end{minipage}\hfill
		\begin{minipage}{0.5\textwidth} 
			b)\\
			\includegraphics[width=\textwidth]{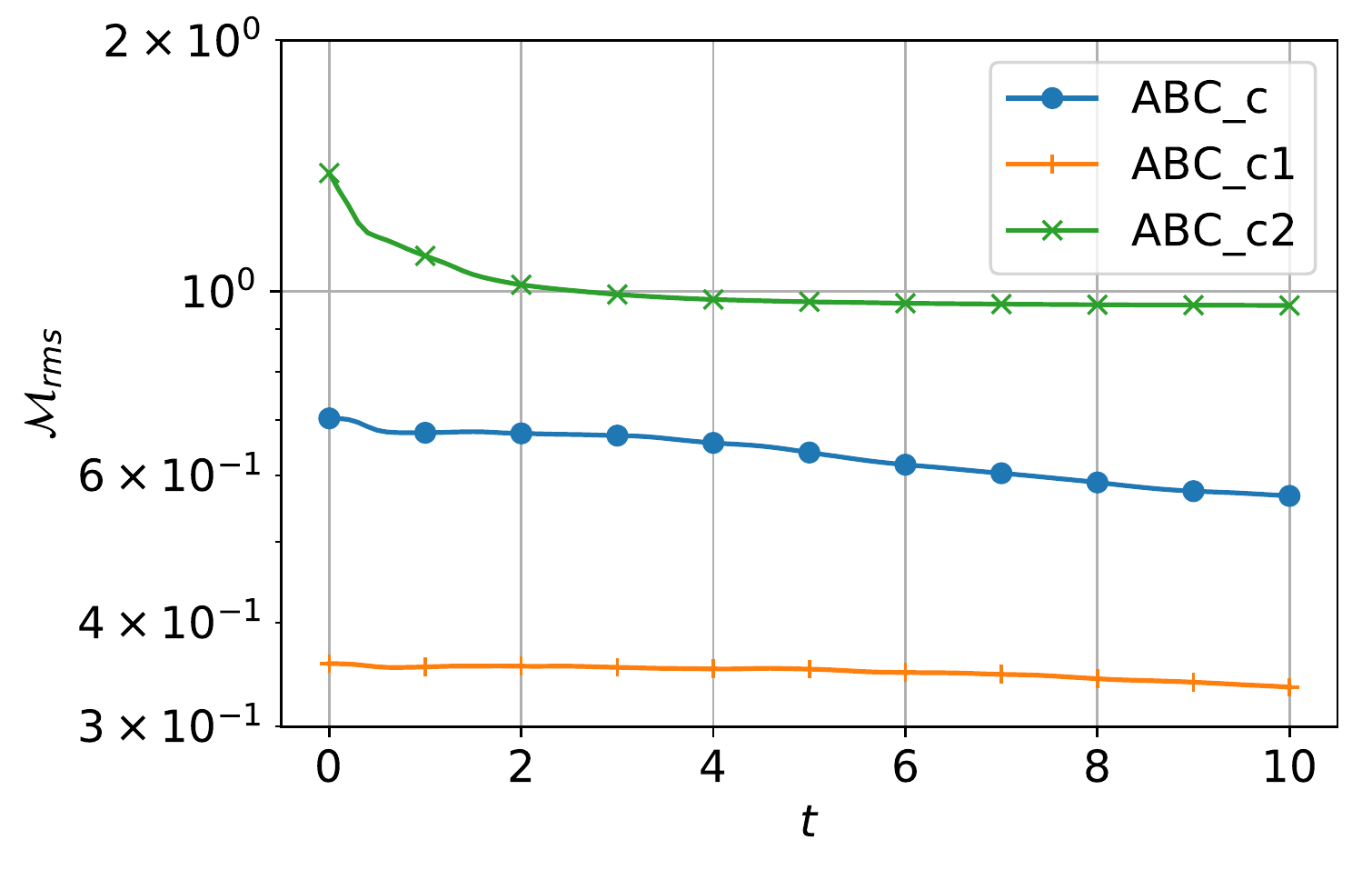}
		\end{minipage}
	\end{center}
	\caption{ABC-QT. 
		 Time evolution of the Mach number $\M_{\max}$ (a) and $\M_{rms}$ (a) for cases ABC\_c, ABC\_c1 and ABC\_c2 (see Table \ref{tab:ABC_Mach_table_case}). }\label{fig:ABC_Mach_Mach}
\end{figure}

Another interesting question that can be addressed using the ABC flow is the influence of the Mach number on the QT dynamics.  Since the velocity $\vec{v}$ is singular at the vortex center  $r=0$, we considered in defining the local Mach number the quantity $\sqrt{\rho}  \vec{v}$ which is not singular ($\vec{v} \sim 1/r$ and $\sqrt{\rho} \sim r$, see \S \ref{sec:qvortices}). We thus computed two representative values: a maximum Mach number $\M_{\max}$ based on the maximum superfluid velocity, and a Mach number $\M_{rms}$ based on averaged values:
\begin{equation}\label{eq:Mach-Number}
\M_{\max} := \frac{ \| \sqrt{\rho} \bv\|_{L^\infty(D)}}{c}, \qquad \M_{rms} :=  \frac{ \| \sqrt{\rho} \bv\|_{L^2(D)}}{c\sqrt{{\cal L}^3}} = \frac{\sqrt{2 E_{\rm{kin}}}}{c}.
\end{equation}
Keeping $c$ and $\xi$ constant, one can change the Mach number in the ABC flow by tuning the values of the parameters $A,B,C$ in \eqref{ABCphi}. Using as reference the case ABC\_c ($N_x=512$) we performed two new runs for which the parameters are displayed in  Table \ref{tab:ABC_Mach_table_case}. The values of constants $A, B, C$ were divided (ABC\_c1) or multiplied (ABC\_c2) by a factor of 2.   As a result, compared to case ABC\_c, the velocities are  divided (resp. multiplied) by 2 for case ABC\_c1 (resp. ABC\_c2). The values for the Mach number reported in Table \ref{tab:ABC_Mach_table_case} were computed at the end of the ARGLE procedure  preparing the initial condition.  
Figure \ref{fig:ABC_Mach_Mach} shows the time evolution for the two values of the Mach number, $\M_{\max}$ and $\M_{rms}$ computed by the (real-time) GP solver. The ratio of 2 is well conserved in time, though the values are varying significantly. This proves that tuning the values of constants $A, B, C$ is a simple and practical approach in modifying the Mach number of the QT super-flow.
\begin{table}[h!]\centering
	\begin{tabular}{|c|c|c|c|c|}\hline
		name & $(A,B,C)$ & $N_x$ & $\M_{\max}$ & $\M_{rms}$ \\ \hline
		ABC\_c & $(0.9,1,1.1)/\sqrt{3}$ & 512 & 1.486860 &	0.703259 \\ 
		ABC\_c1 & $(0.9,1,1.1)/(2\sqrt{3})$ & 512 & 0.836509 &	0.357021 \\ 
		ABC\_c2 & $2(0.9,1,1.1)/\sqrt{3}$& 512 & 2.800959 &	1.385344 \\  \hline
	\end{tabular}
	\caption{ABC-QT. Runs used to test of the influence of the Mach number.  Compared to run ABC\_c (see Table \ref{tab:ABC-params}), only the constants $A, B, C$ in defining the ABC flow were modified (see Eq. \eqref{ABCphi}).}\label{tab:ABC_Mach_table_case}
\end{table}

\begin{figure}[h!]
						\begin{center}	
		\begin{minipage}{0.5\textwidth} 
			a)\\
			\includegraphics[width=\textwidth]{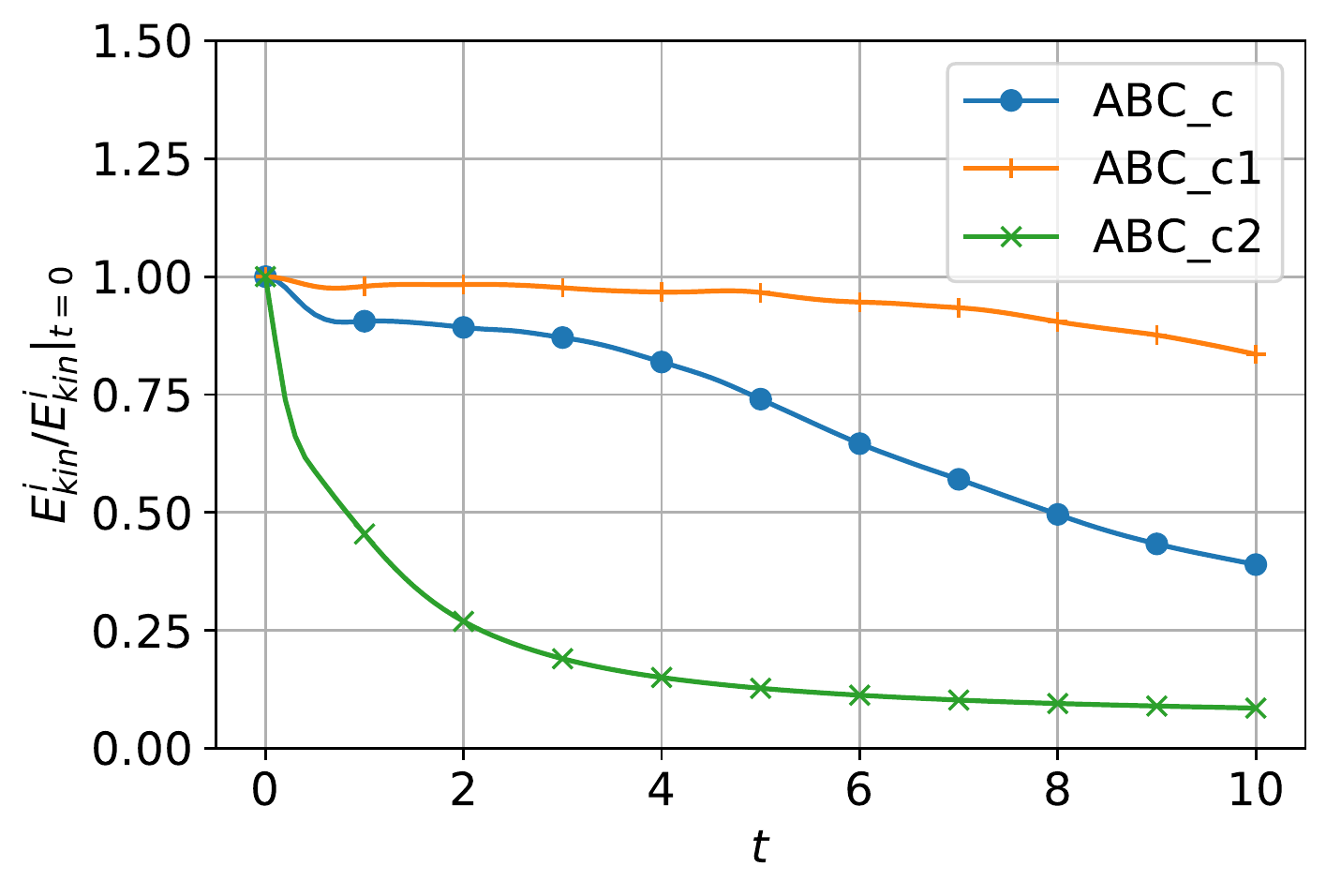}
		\end{minipage}\hfill
		\begin{minipage}{0.5\textwidth} 
			b)\\
			\includegraphics[width=\textwidth]{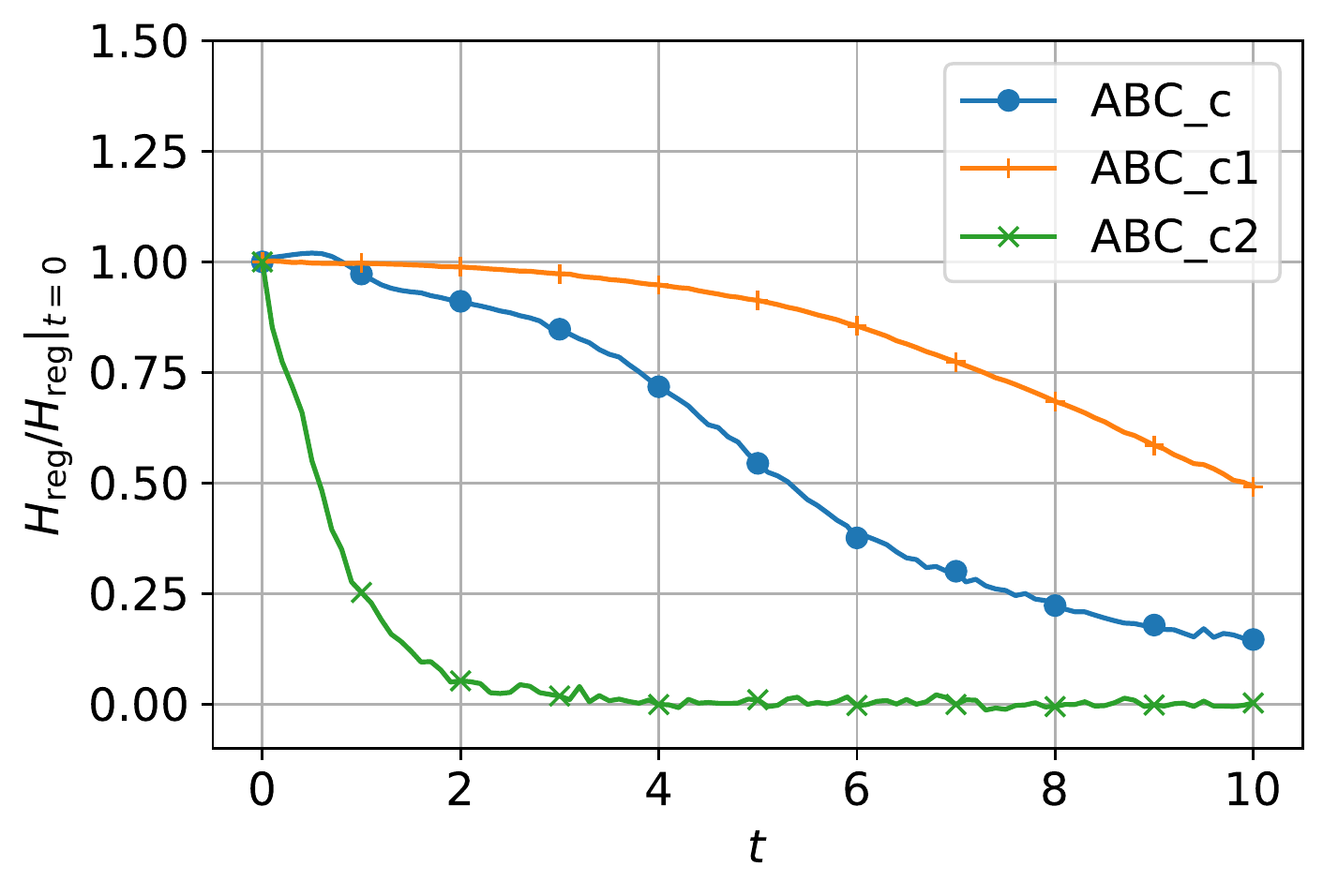}
		\end{minipage}
	\end{center}
	\caption{ABC-QT. Influence of the Mach number.  
		Time evolution of the incompressible kinetic energy $\Ekini$ (a) and  regularized helicity $H_{\rm reg}$  (b). To be compared with curves in Fig. \ref{fig:evol_energy_ABC}.
		}\label{fig:ABC_Mach_evol}
\end{figure}

\begin{figure}[h!]
							\begin{center}	
		\begin{minipage}{0.5\textwidth} 
			a)\\
			\includegraphics[width=\textwidth]{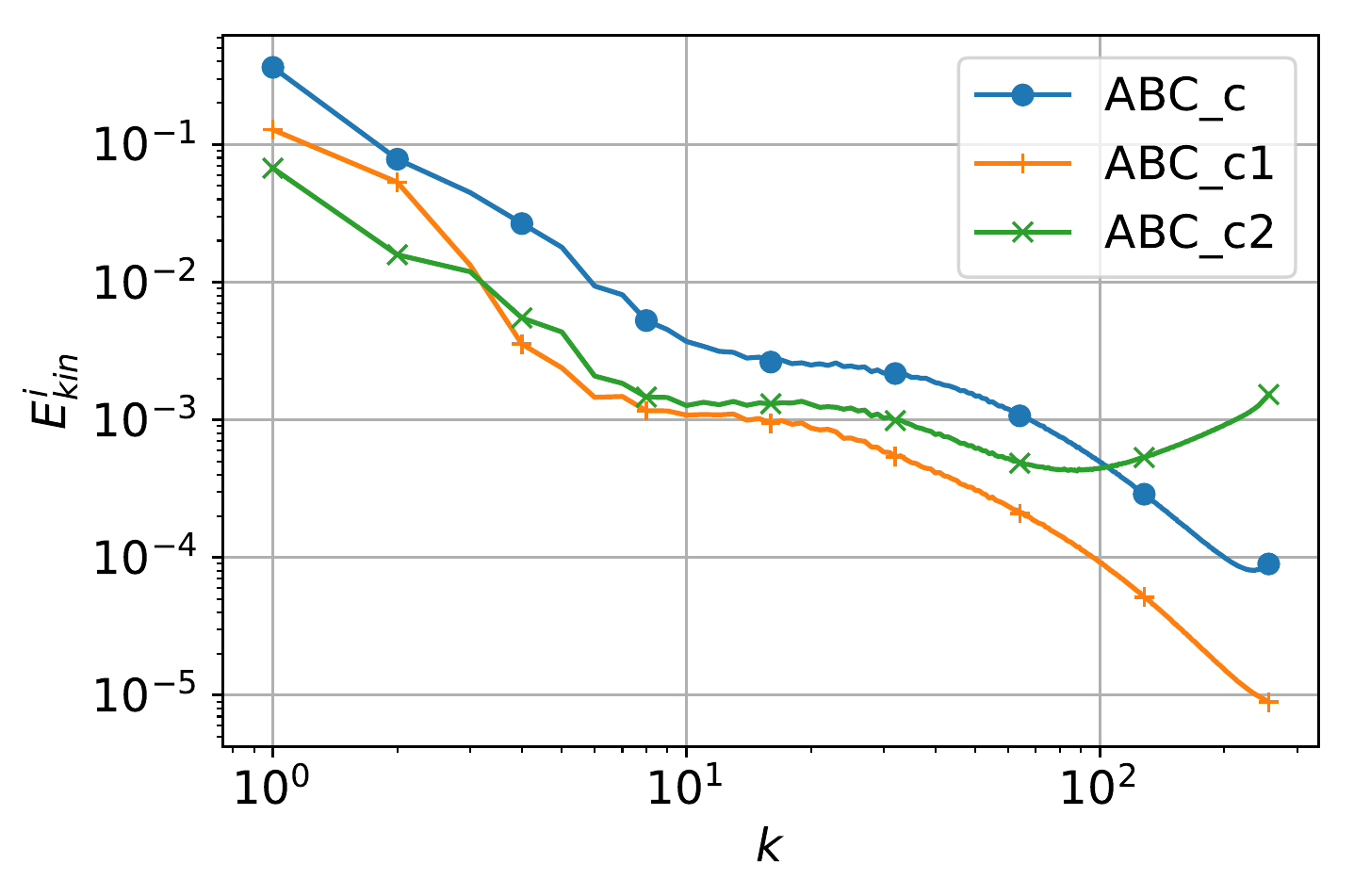}
		\end{minipage}\hfill
		\begin{minipage}{0.5\textwidth} 
			b)\\
			\includegraphics[width=\textwidth]{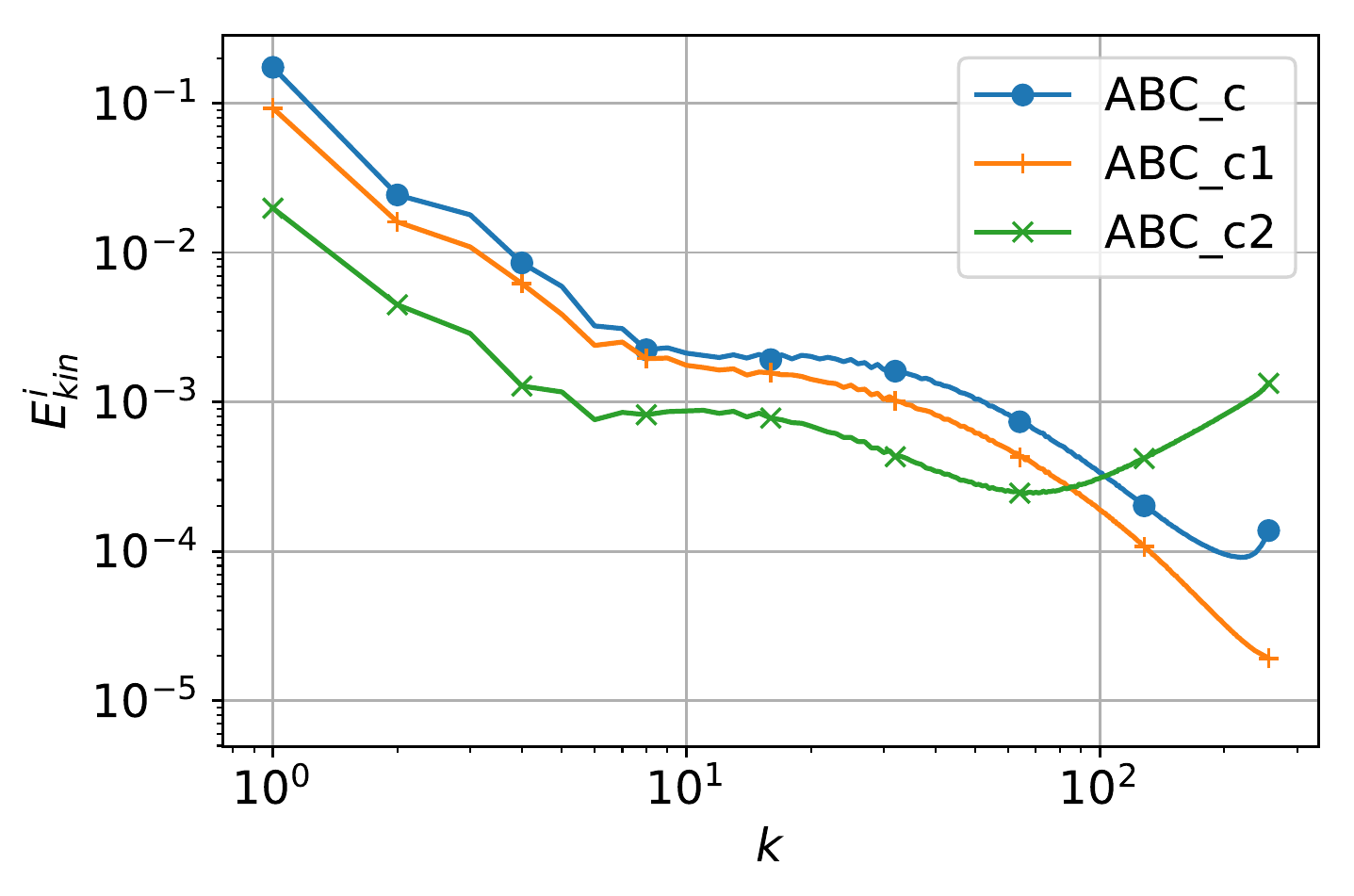}
		\end{minipage}
	\end{center}
	\caption{ABC-QT. Influence of the Mach number. Spectrum of the incompressible kinetic energy $\Ekini$ for case ABC\_c, ABC\_c1, ABC\_c2, at time instants $t=5$ (a) and $t=T_f=10$ (b).}\label{fig:ABC_Mach_spectra}
\end{figure}

We present in Fig. \ref{fig:ABC_Mach_evol} the time evolution of incompressible kinetic energy $\Ekini$ and regularized helicity $H_{\rm reg}$ for new cases with different Mach numbers. As expected from the analysis above, the energy and helicity associated with the classical flow $\bv_{ABC}$ are  divided (resp. multiplied) by 4 for case ABC\_c1 (resp. ABC\_c2). We note that the time evolution of these main quantities depends on the Mach number. To assess on the distribution of the incompressible kinetic energy among scales, we plotted in Fig. \ref{fig:ABC_Mach_spectra} spectra of $\Ekini$ at significant time instants, $t=5$ and final time $t=T_f=10$. The spectra for the three cases are quite similar showing that the obtained dynamics of the QT is equivalent when varying the Mach number of the flow.

\clearpage

\subsection{Benchmark \#3: Smoothed random phase quantum turbulence (SRP-QT)}

The SRP initial field was prepared as described in \S \ref{sec:init-srp}.  The advantage of this new initial condition is that the time-imaginary ARGLE simulation is no longer necessary in the preparation  of the initial field. We display in  Table \ref{tab:SRP-params}  the values of the time step $\delta t$ used in the GP solver (see \S\ref{sec:real-time}), the final time $T_f$ of each simulation, and the parameters $K$ (maximum amplitude of the phase) and $N_r$ (number of random values) of the method generating the phase field (see Fig. \ref{fig:SRP_psi}). We recall that  the characteristic velocity of the generated flow field results is ${v} = 2\alpha  (K N_r / \pi)$, and the corresponding theoretical Mach number $M=\sqrt{2 \alpha} K N_r / \pi \sqrt{\beta}$.
\begin{table}[h!]
	\centering
	{\small	
		\begin{tabular}{|c|c|c|c|c|c|} \hline
			Run &     $N_x$ & $\delta t$ &$T_f$& $K$& $N_r$ \\ \hline
			SRP\_a    & 128   &  $1/1024$   & 8 &$8 \pi$ &4\\ \hline
			SRP\_b    & 256   &  $1/2048$   & 8 &$16 \pi$ &4\\ \hline
			SRP\_c    & 512   &  $1/4096$   & 6.5 &$32 \pi$ &4\\ \hline
		\end{tabular}
	}	
	\caption{Runs for the SRP-QT case. For each space resolution $N_x$, the corresponding physical and numerical parameters are displayed in Table \ref{tab:ALL-params}. }\label{tab:SRP-params}
\end{table}

\begin{figure}[!h]
	\begin{center}
		\includegraphics[width=0.4\textwidth]{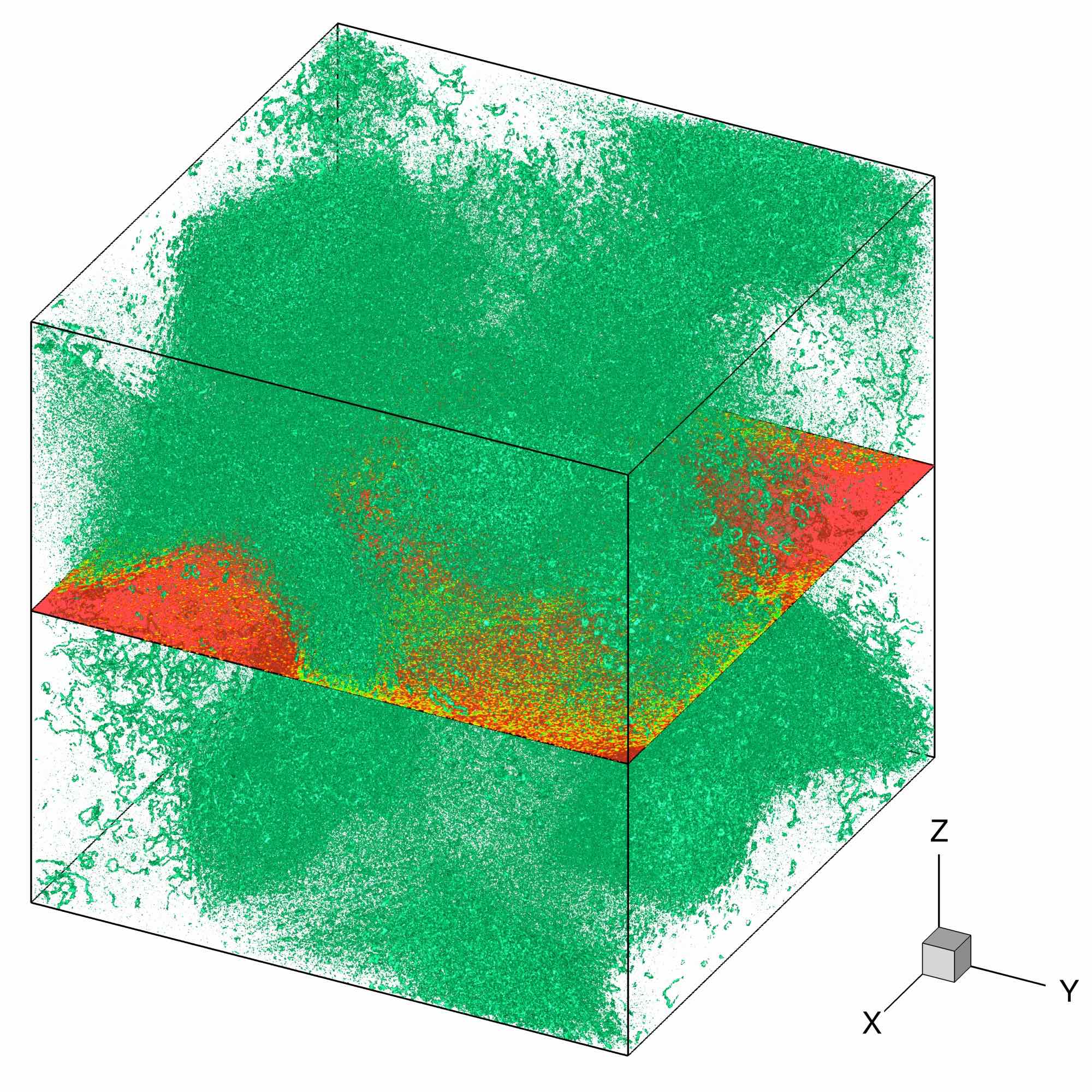}
	\end{center}
	\caption{SRP-QT. Instantaneous fields computed with the real-time GP solver, starting from the initial condition presented in Fig. \ref{fig:SRP_psi}. Vortex structures (iso-surfaces of low $\rho$) of the  wave function at final time $T_f$.
		Grid resolution $N_x=512$, corresponding to run SRP\_c in Table \ref{tab:SRP-params}). }
	\label{fig:SRP-at-Tf}
\end{figure}

Figure \ref{fig:SRP-at-Tf} illustrates the vortex structures in the QT super-flow generated with this method. Compared to TG and ABC cases, in the SRP case vortices nucleate progressively and do not display long vortex lines. A very fine grain structure of vortices is observed in all SRP runs. 

\begin{figure}[h!]
\begin{center}	
		\begin{minipage}{0.5\textwidth} 
			a)\\
			\includegraphics[width=\textwidth]{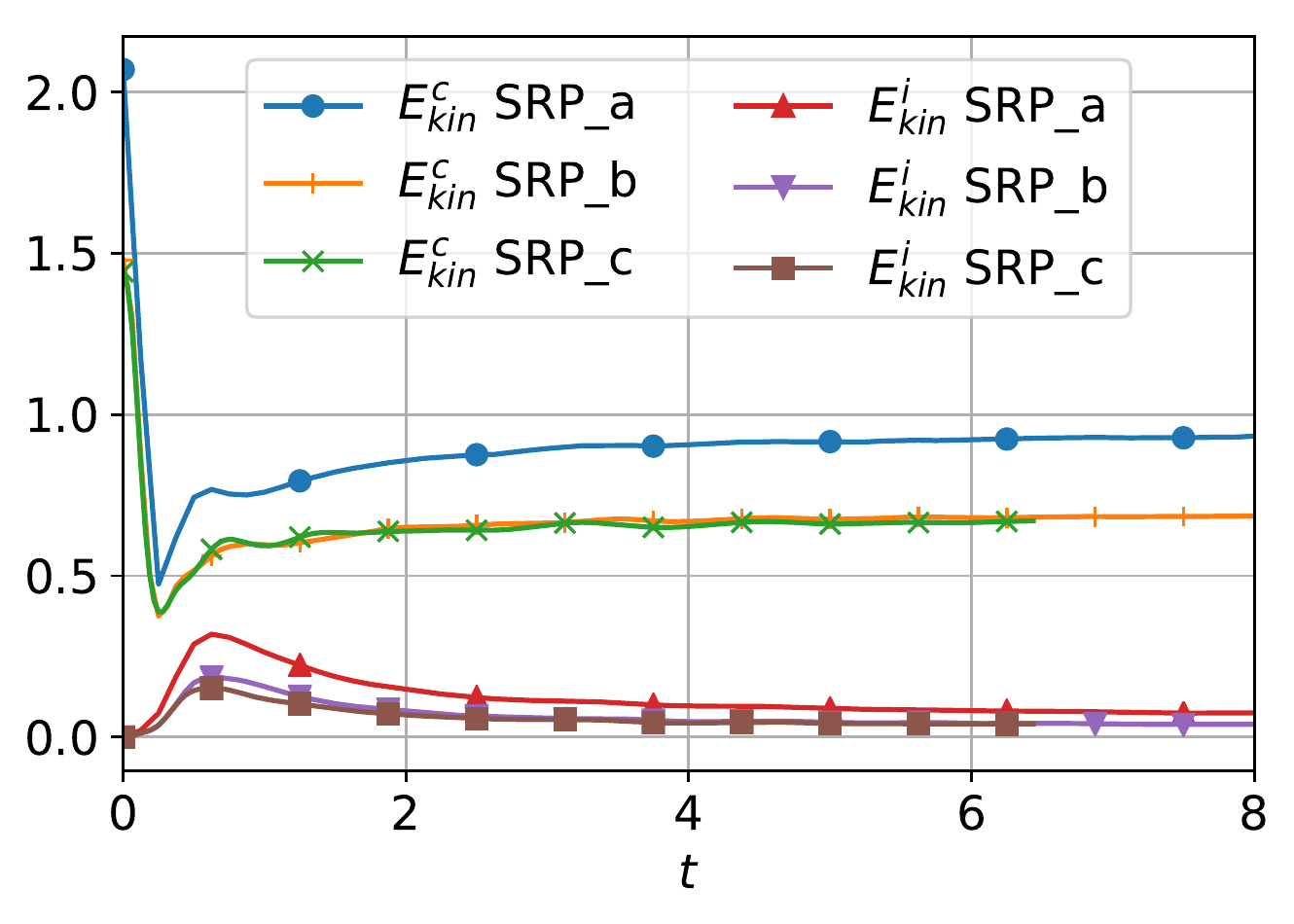}
		\end{minipage}\hfill
		\begin{minipage}{0.5\textwidth} 
			b)\\
			\includegraphics[width=\textwidth]{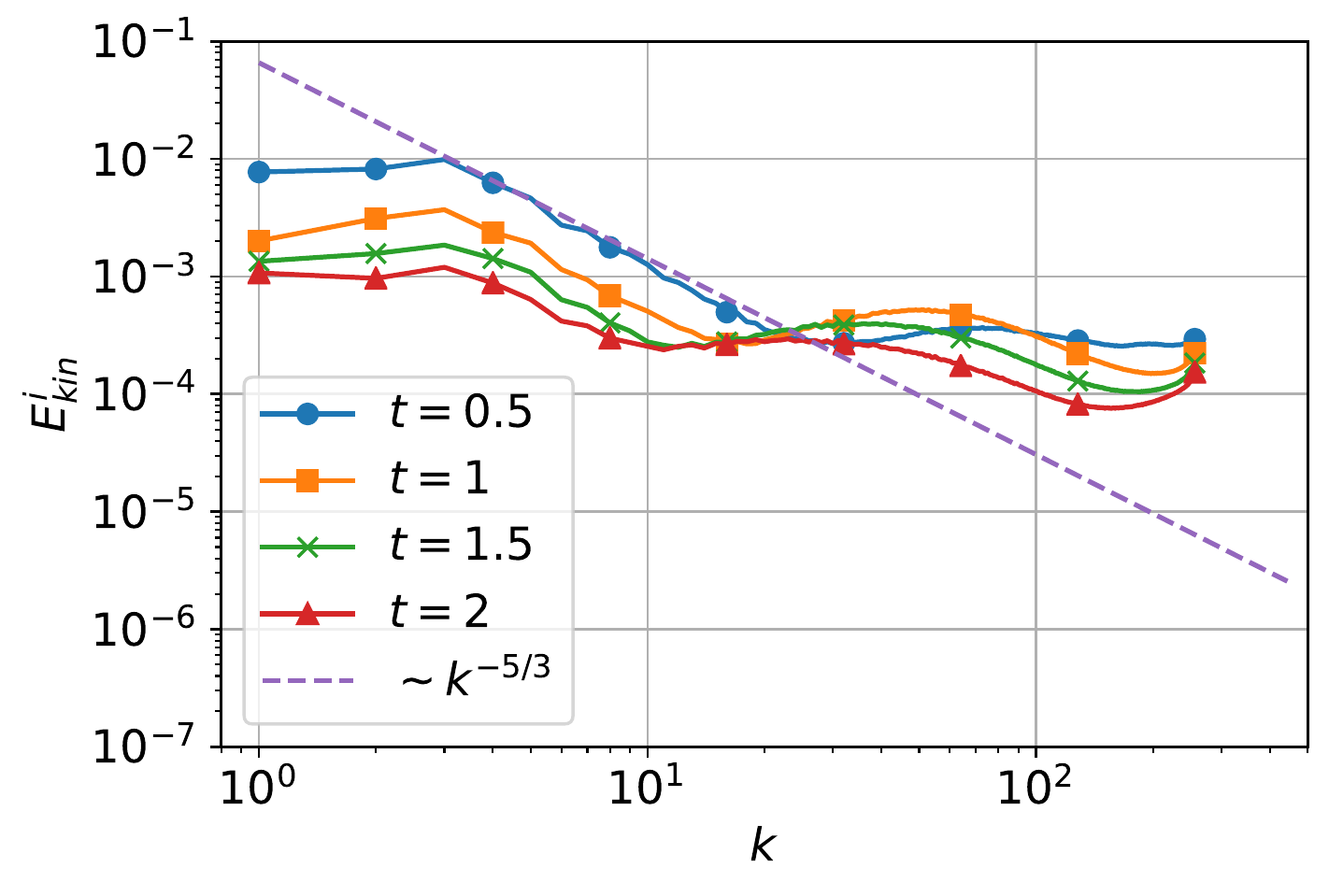}
		\end{minipage}
\end{center}	
	\caption{SRP-QT.  (a) Time evolution of the compressible $\Ekinc$  and incompressible $\Ekini$  kinetic energies.
		(b) Spectrum of  $\Ekini$ at different time instants. Case SRP\_c ($N_x=512$). 
In both panels, the results represent an ensemble average for 10 different (random) initial conditions.
}
\label{fig:SRP-energy}
\end{figure}

To analyze  the SRP-QT flow we plotted in Fig. \ref{fig:SRP-energy}a the time evolution of the compressible $\Ekinc$ and incompressible $\Ekini$ kinetic energies.
An ensemble average for 10 different (random) initial conditions was taken to display the results.
Since the initial filed (at $t=0$) does not contain vortices, the incompressible kinetic energy $\Ekinc$ is initially zero and subsequently increases due to vortex nucleations. After reaching the maximum value at $t \sim 0.5$, $\Ekinc$ gradually decreases to the end of the simulation ($t=T_f$).
During the entire time evolution, the dynamics of the flow is dominated by the compressible kinetic energy $\Ekinc$, which is always larger than $\Ekini$.
 Figure \ref{fig:SRP-energy}b shows the spectrum of $\Ekini$. As for TG and ABC cases, a Kolmogorov-like scaling is obtained, with a 
 $-5/3$ power-law at low wave numbers $k$. Hereof, the SRP-QT flow is statistically similar to the TG and ABC QT flows and can be used in a detailed parametric study of the decay of quantum turbulence (which is beyond the scope of this contribution).

\subsection{Benchmark \#4: Random vortex rings quantum turbulence (RVR-QT)}

The RVR initial field was prepared as described in \S \ref{sec:init-rvr}.  Like in the SRP case, building this new initial condition avoids the use of the
time-imaginary ARGLE computation. We display in  Table \ref{tab:RVR-params}  the values of the time step $\delta t$ used in the GP solver (see \S\ref{sec:real-time}), the final time $T_f$ of each simulation, the parameter $N_V$ representing the number of pairs of vortex rings seeded in the initial field, the radius $R$ of a vortex ring, and the distance $d$ between the vortex rings forming a pair (see Eq. \eqref{psiRVR}). 
\begin{table}[!h]
	\centering
	{\small	
		\begin{tabular}{|c|c|c|c|c|c|c|c|} \hline
			Run &     $N_x$ & $\delta t$ &$T_f$& $N_V$ &$R$ & $d$\\ \hline
			RVR\_a    & 128   &  $1/1024$   & 8 &200&$\pi/2$ & $\pi$\\ \hline
			RVR\_b    & 256   &  $1/2048$   & 8 &400&$\pi/2$ & $\pi$\\ \hline
			RVR\_c    & 512   &  $1/4096$   & 4.5 &800&$\pi/2$ & $\pi$\\ \hline
		\end{tabular}
	}	
	\caption{Runs for the RVR-QT case. For each space resolution $N_x$, the corresponding physical and numerical parameters are displayed in Table \ref{tab:ALL-params}. }\label{tab:RVR-params}
\end{table}

\begin{figure}[!h]
	\begin{center}
		\begin{minipage}{\textwidth}
			\includegraphics[width=0.33\textwidth]{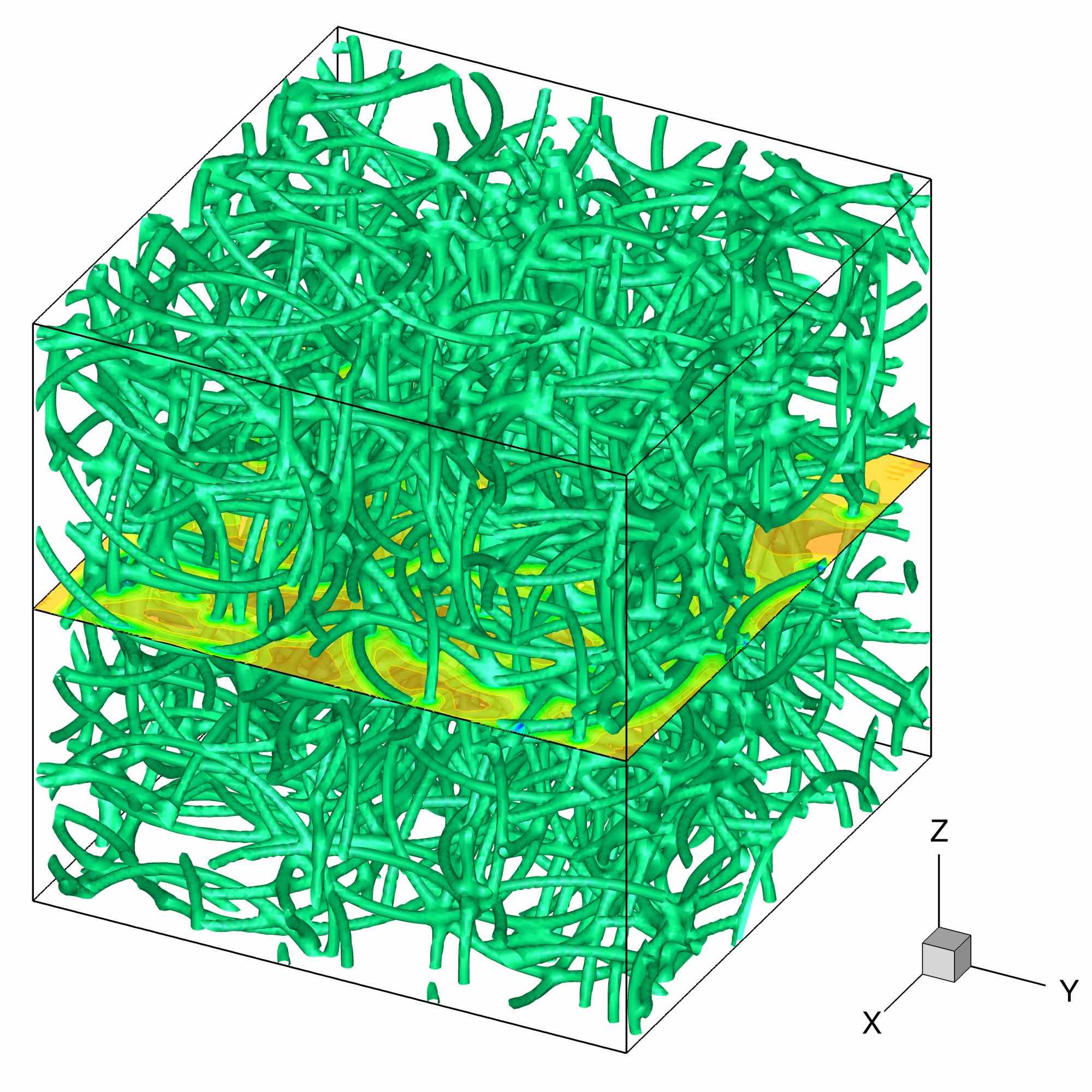}
			\hfill
			\includegraphics[width=0.33\textwidth]{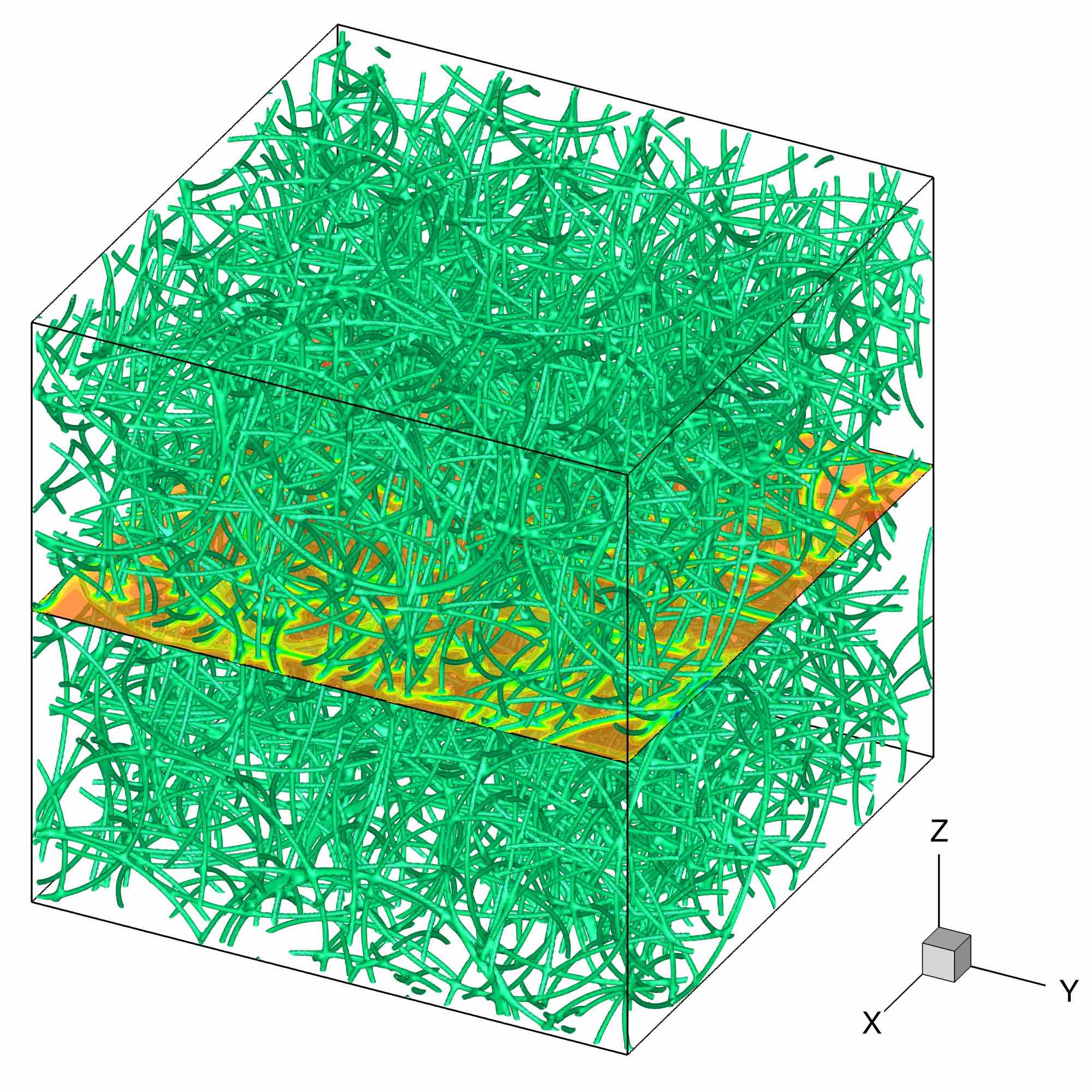}
			\hfill
			\includegraphics[width=0.33\textwidth]{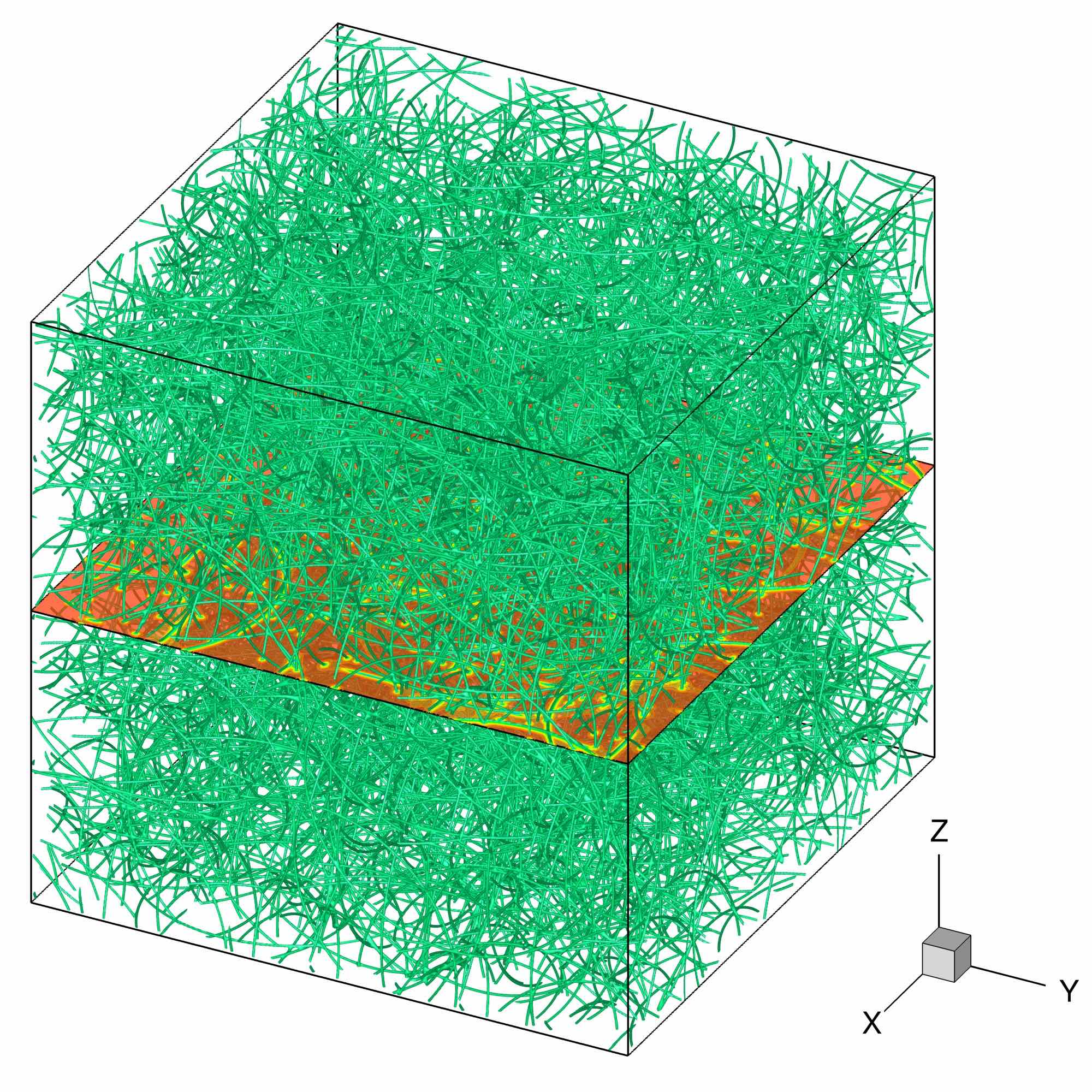}
		\end{minipage}
	\end{center}
	\caption{RVR-QT. Initial field containing $N_V$  randomly distributed vortex ring pairs. Vortex lines (iso-surfaces of low $\rho$) of the  wave function.
		From left to right: grid resolutions $N_x=$ 128, 256, 512 and $N_V=$ 200, 400, 800 (corresponding to runs RVR\_a, RVR\_b and RVR\_c in Table \ref{tab:RVR-params}).}
	\label{fig:RVR-at-t0}
\end{figure}

Note that in Fig. \ref{fig:RPR} we represented, to illustrate the method, a few number of vortex pairs ($N_V$=1, 20 and 50). In the GP calculations we used a much larger value for $N_V$, up to 800 for the case RVR\_c. The initial field for the three considered cases is displayed in Fig. \ref{fig:RVR-at-t0}. Like in the TG and ABC cases, when the grid resolution $N_x$ is increased,  $\xi$ diminishes and, consequently, thinner vortex rings are seeded in the initial field.

\begin{figure}[!h]	
	\begin{center}
		\begin{minipage}{\textwidth}
			\includegraphics[width=0.33\textwidth]{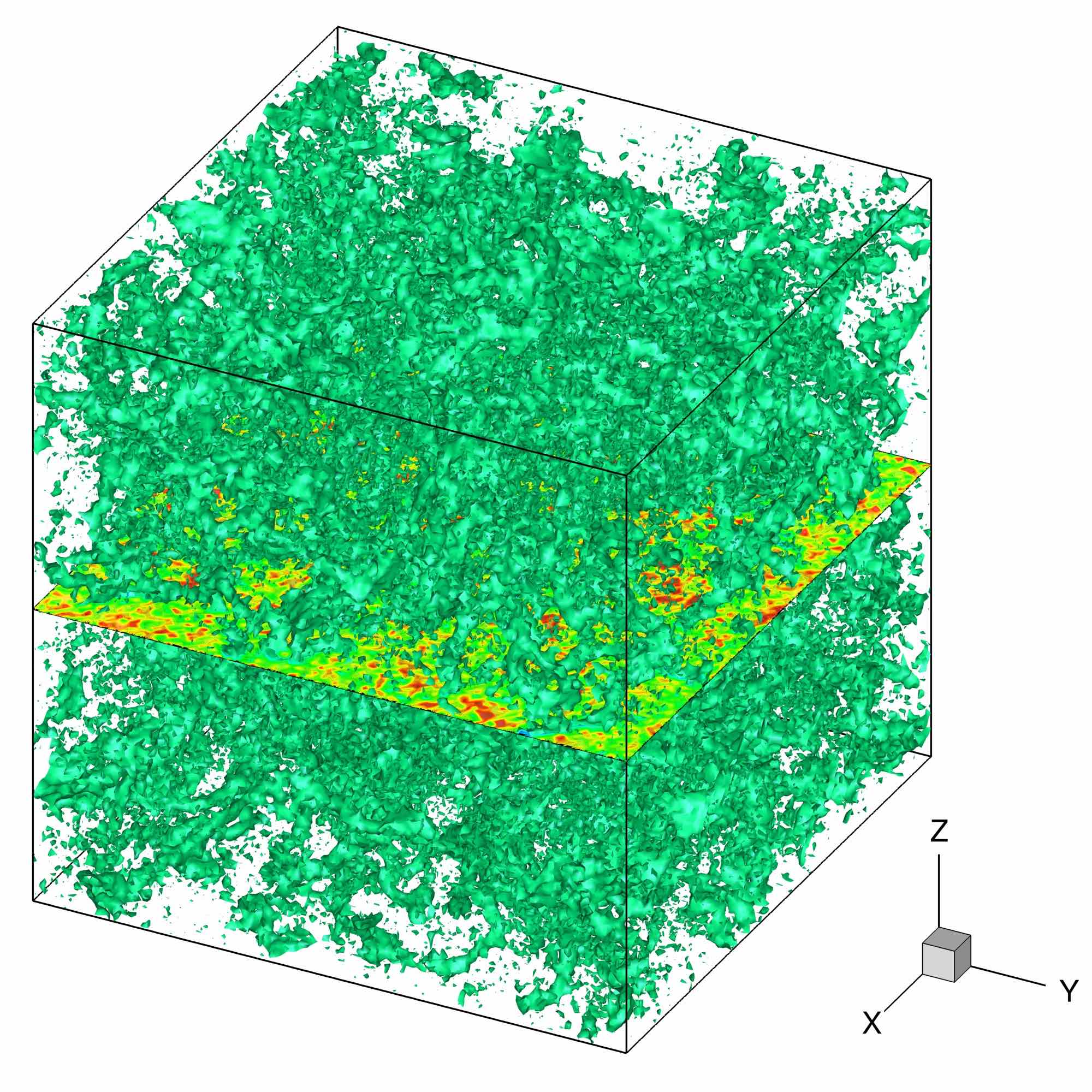}
			\includegraphics[width=0.33\textwidth]{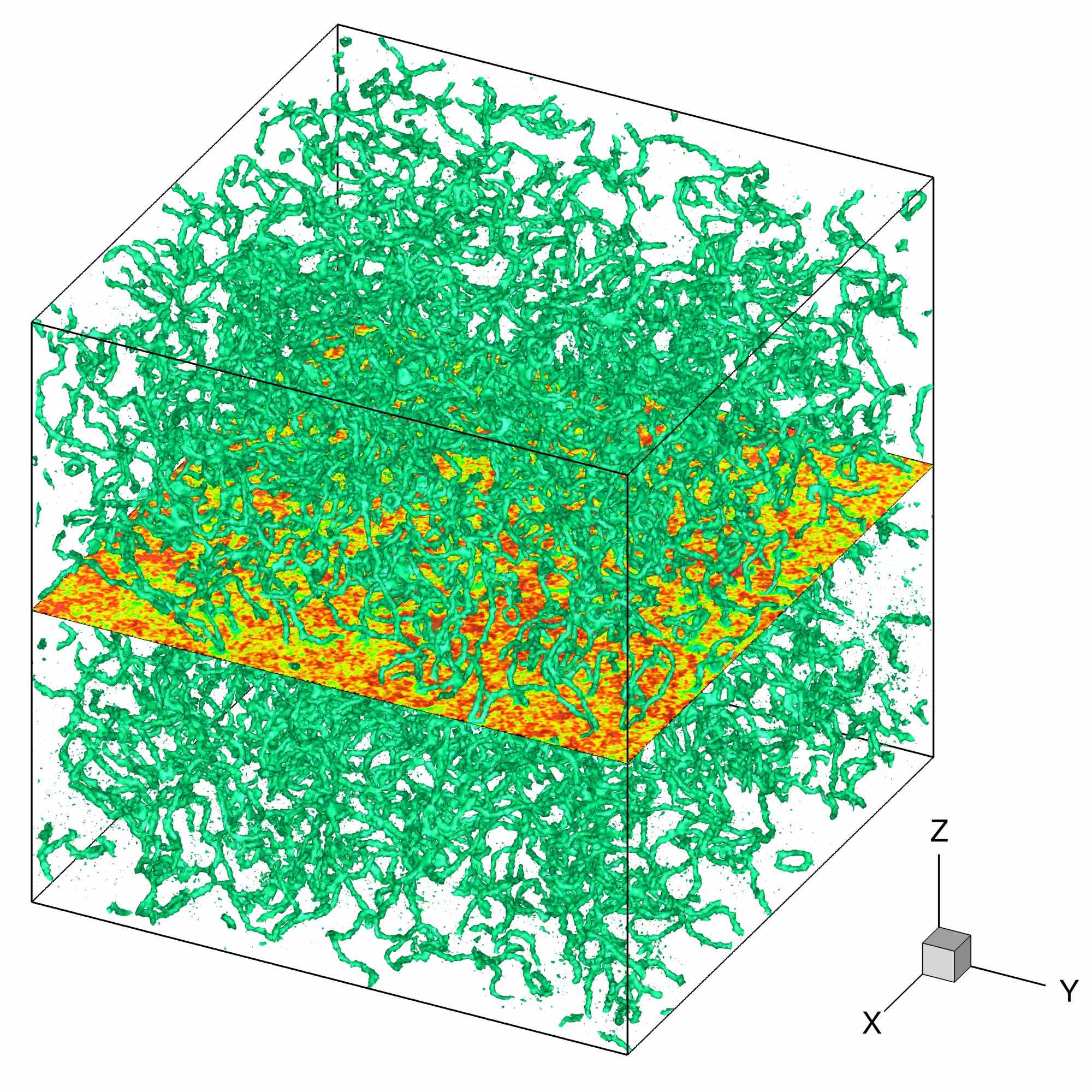}
			\includegraphics[width=0.33\textwidth]{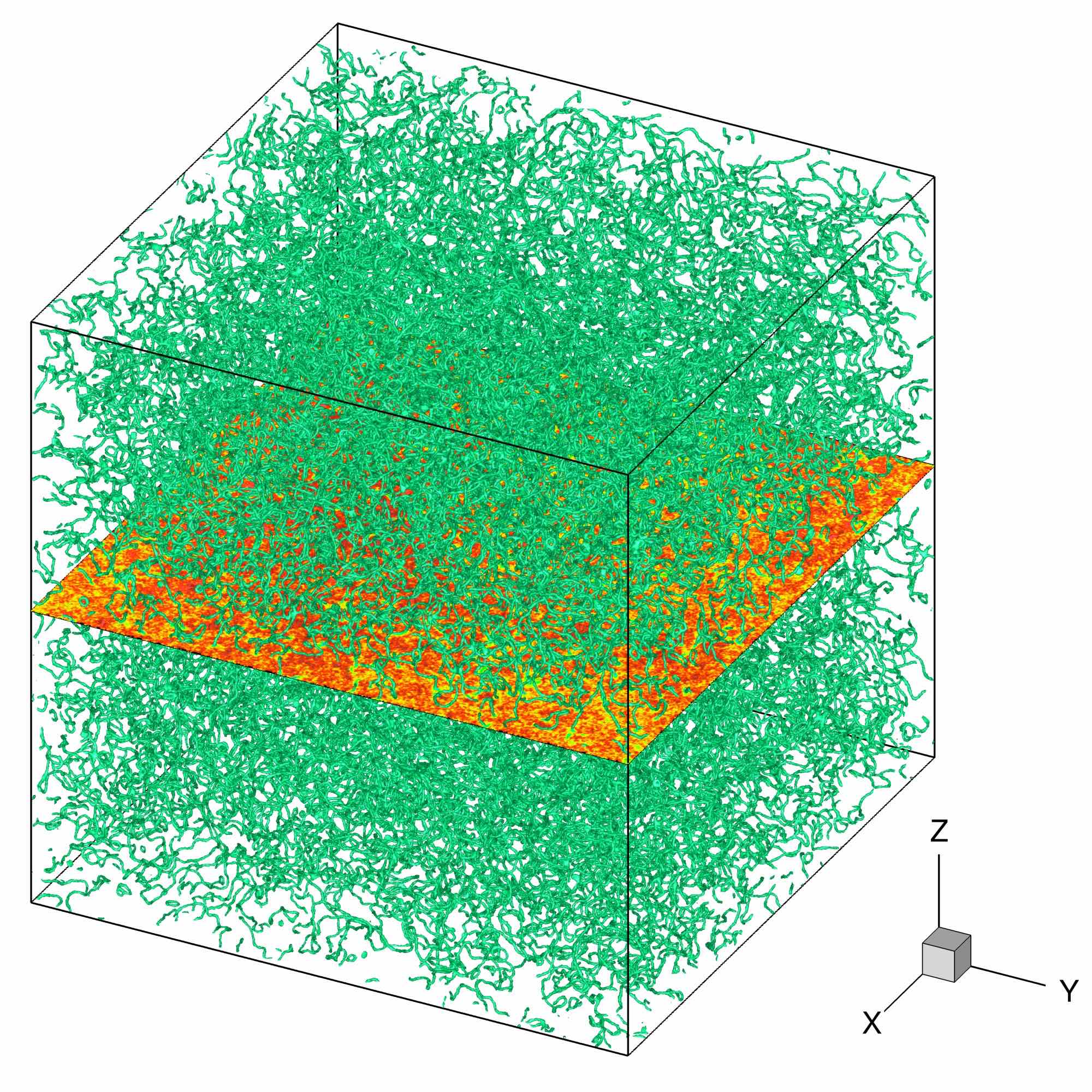}
		\end{minipage}
	\end{center}
	\caption{RVR-QT. Instantaneous fields computed with the real-time GP solver, starting from the initial condition presented in Fig. \ref{fig:RVR-at-t0}. Vortex lines (iso-surfaces of low $\rho$) of the  wave function at final time $T_f$.
		From left to right: grid resolutions $N_x=$ 128, 256, 512 (corresponding to runs RVR\_a, RVR\_b and RVR\_c in Table \ref{tab:RVR-params}).}
	\label{fig:RVR-at-Tf}
\end{figure}

The obtained RVR-QT flow is illustrated in Fig. \ref{fig:RVR-at-Tf}. Multiple vortex ring reconnections lead to a dense vortex distribution in the QT field, similar to that obtained for the ABC  flow (see Fig. \ref{fig:ABC-at-Tf}). 

\begin{figure}[h!]
	\begin{center}	
		\begin{minipage}{0.5\textwidth} 
			a)\\
			\includegraphics[width=\textwidth]{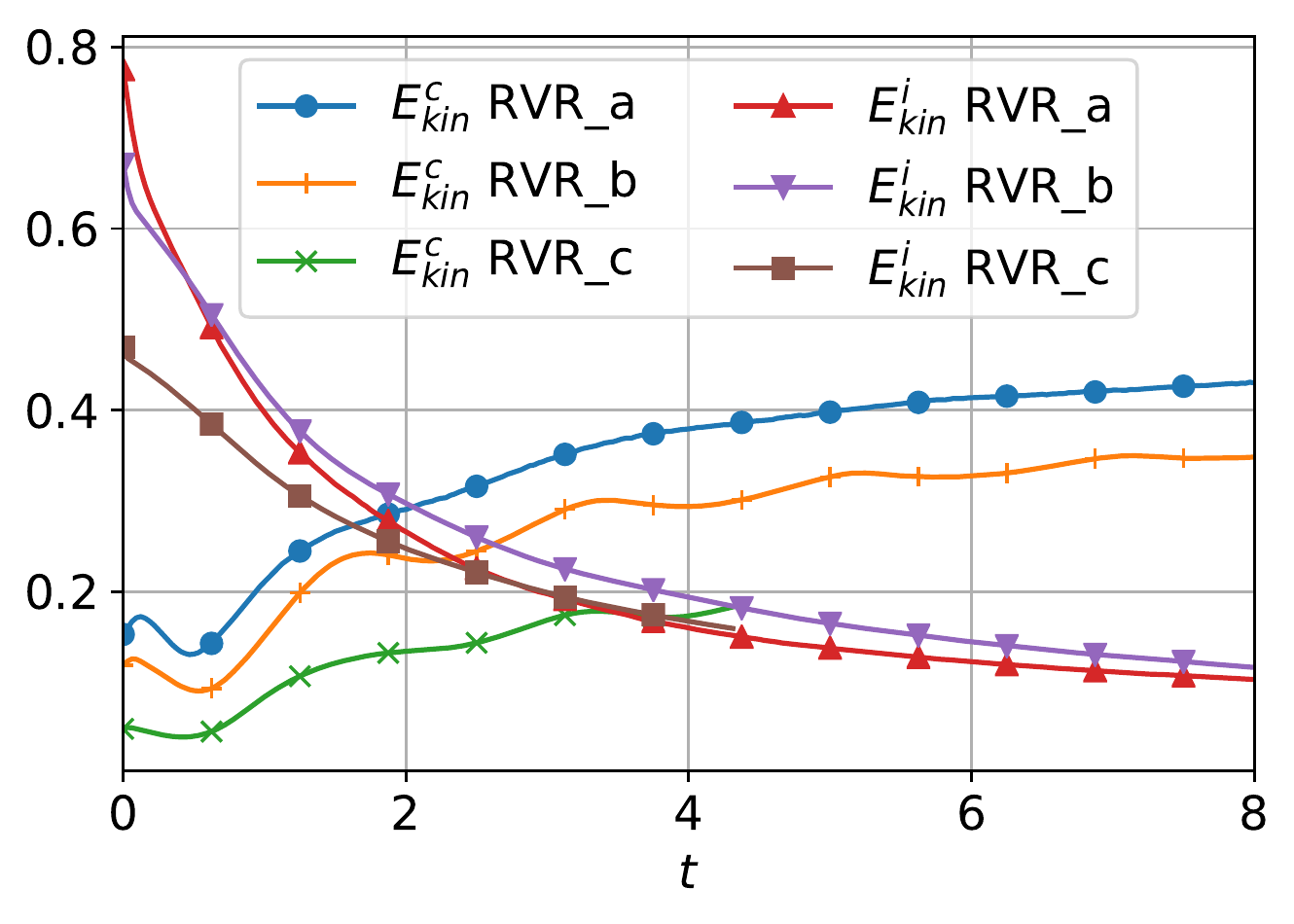}
		\end{minipage}\hfill
		\begin{minipage}{0.5\textwidth} 
			b)\\
			\includegraphics[width=\textwidth]{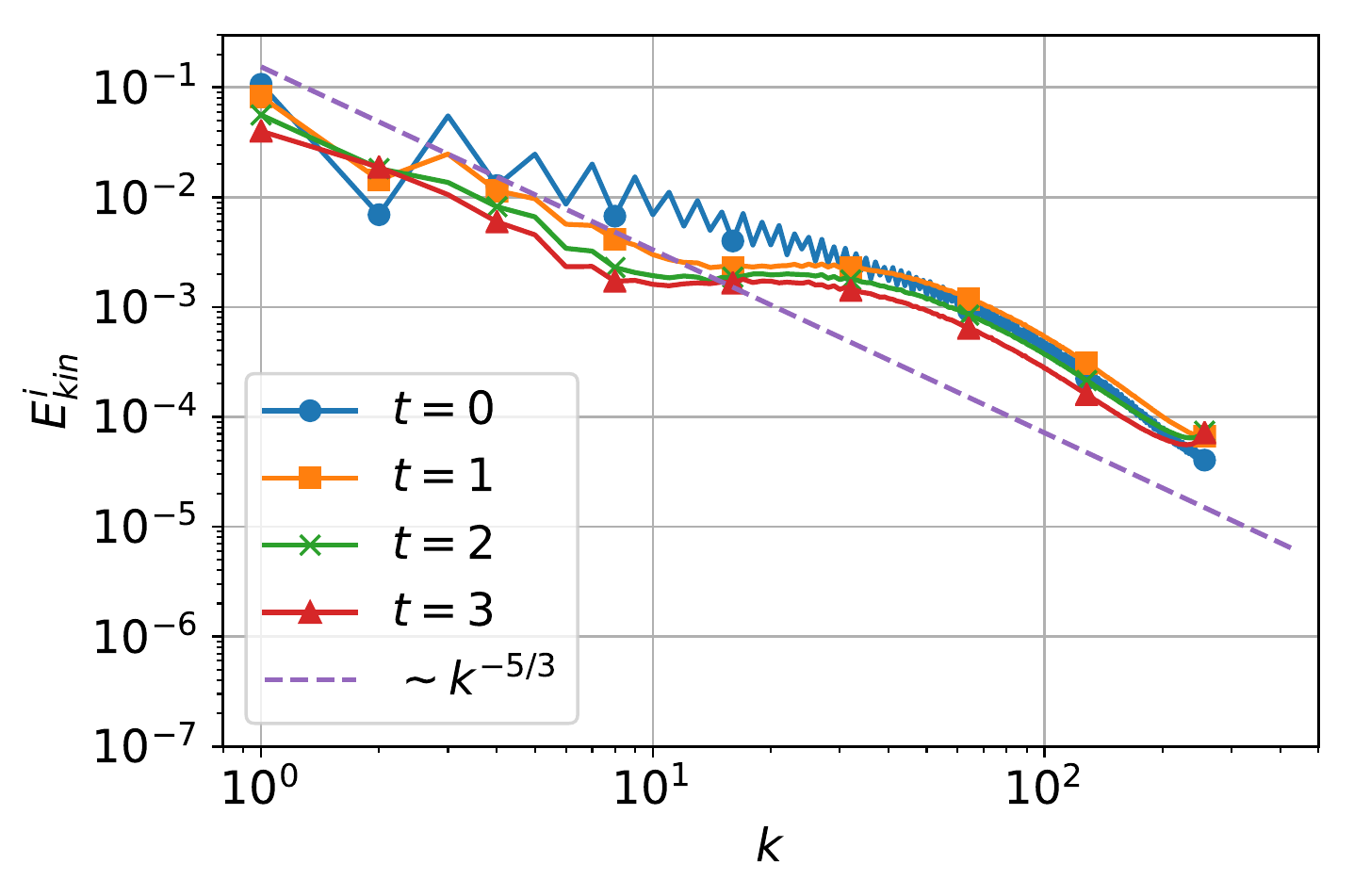}
		\end{minipage}
	\end{center}	
	\caption{RVR-QT.  (a) Time evolution of the compressible $\Ekinc$  and incompressible $\Ekini$  kinetic energies.
		(b) Spectrum of  $\Ekini$ at different time instants. Case RVR\_c ($N_x=512$). 
		In both panels, the results represent an ensemble average for 10 different runs, with random initial distribution of $N_V=800$ vortex ring pairs in the computational domain.}
\label{fig:RVR-energy}
\end{figure}

For the analysis of the RVR-QT flow we provide in Fig. \ref{fig:RVR-energy}(a) the time evolution of the compressible $\Ekinc$ and incompressible $\Ekini$ kinetic energies for the case RVR\_c. Since the initial distribution of vortex rings pairs is random in the computational box, we present the ensemble average results for 10 runs with random positions of  the same number of vortex ring pairs ($N_V=800$). In the early stages of the time evolution ($t < 1$), $\Ekini$ is dominant. The compressible kinetic energy $\Ekinc$ starts to increase at $t \sim 1$, due to sound emissions through vortex reconnections. This evolution is opposite to that observed for the SRP-QT cases.  Figure \ref{fig:RVR-energy}b shows the  spectrum of $\Ekini$. Like in the SRP cases (see Fig. \ref{fig:SRP-energy}), we note a Kolmogorov-like scaling of the spectrum, with a $-5/3$ power-law at low wave numbers $k$.

\section{Conclusion}\label{sec:Conclusion}

We simulated in this paper quantum turbulence superfluid flows described by the Gross-Pitaevskii equation.  Numerical simulations were performed using a parallel (MPI-OpenMP) code based on a pseudo-spectral spatial discretization and  second order splitting for the time integration. As expected from the theoretical numerical analysis, this approach ensured an accurate capture of the dynamics of the flow, with a perfect conservation of the number of particles and a negligible drift in time of the total energy. Several configurations of QT were simulated using four different initial conditions: Taylor-Green (TG) vortices, Arnold-Beltrami-Childress (ABC) flow, smoothed random phase (SRP) fields and  random vortex rings (RVR) pairs. Each of these case was described in detail by setting corresponding benchmarks that could be used to validate/calibrate new GP codes. Particular care was devoted in describing dimensionless equations, characteristic scales and optimal numerical parameters. We  presented values, spectra and structure functions of main quantities of interest (energy, helicity, etc.)  that are useful to describe the turbulent flow.  Some general features of QT  were identified, despite the variety of initial states: the spectrum of the incompressible kinetic energy exhibits a Kolmogorov-type $-5/3$  power-law scaling for the large scales,  the flow dynamics is characterized by a continuous transfer between incompressible and compressible energy, etc. 

The first two benchmarks (TG and ABC) are classical and inspired from classical turbulence. They start from defining a velocity field containing vortices and use an imaginary-time ARGLE procedure to reduce the acoustic emission of the initial field. The last two benchmarks (SRP and RVR) are new and based on the direct manipulation of the wave function. The new initial conditions have the advantage to be simple to implement and to avoid supplementary computations through the ARGLE procedure. The SRP initial condition has the particularity of being vortex free, with kinetic energy dominated at initial stages  by its compressible part.  The situation is reversed in the RVR initial condition, since at early stages the incompressible kinetic energy dominates. Therefore, the new initial conditions could be used as new QT settings to explore various physical phenomena, such as the interaction of particles with quantized vortices in QT \citep{QT-2019-Giuriato}. Another possible use of the new SRP and RVR initial conditions is for the simulation of QT in atomic Bose-Eintein condensates (BEC). GP-QT dynamics in  BECs  is generally triggered by directly manipulating the wave function field. \cite{QT-2002-berloff} used a randomly distributed initial wave function field,  \cite{QT-2005-parker} applied a simple rotation of the initial field, \cite{Kobayashi2007} used combined rotations around two axes, while \cite{QT-2010-White2010} suggested a random phase imprinting. The extension of our SRP and RVR models to BEC-QT will be reported in a forthcoming contribution.

Supplementary images and movies depicting the dynamics of QT-GP cases simulated  in this paper are provided as Supplemental Material at \\ \href{http://qute-hpc.math.cnrs.fr/2020_03_QT_GP.html}{http://qute-hpc.math.cnrs.fr/2020\_03\_QT\_GP.html}.

\section*{Acknowledgements}
%
The authors acknowledge financial support from the French ANR grant ANR-18-CE46-0013 QUTE-HPC. Part of this work used computational resources provided by IDRIS (Institut du d{\'e}veloppement et des ressources en informatique scientifique) and CRIANN (Centre R{\'e}gional Informatique et d'Applications Num{\'e}riques de Normandie).

\appendix

\section{Parallel performance of the code}

\subsection{Execution time}
Execution times for runs ABC\_a to ABC\_c and ABC\_aIT to ABC\_cIT are reported in Table \ref{tab:ExecTime}.

\begin{table}[!h]
	\centering%
	\begin{tabular}{|c|c|r|r|r|r|}\hline
		Case & $N_x$ & Iterations & MPI proc. & Execution time (s) & ratio \\ \hline
		ABC\_aIT & 128 & 7500 & 56 & 539.126        &   0.000257075 \\
		ABC\_bIT & 256 & 15000 & 112 & 5940.076   &  0.000354056 \\
		ABC\_cIT & 512 & 30000 & 224 & 53066.445 &  0.000395375 \\
		ABC\_a & 128 & 12500 & 56 & 700.524          &  0.000334036 \\
		ABC\_b & 256 & 25000 & 112 & 7364.754      &  0.000438973 \\
		ABC\_c & 512 & 50000 & 224 & 63397.341    &  0.000472346 \\ \hline
	\end{tabular}
	\caption{Execution time for ABC runs. The last column reports the execution time divided by the number of degrees of freedom ($N_x^3$).}\label{tab:ExecTime}
\end{table}

When switching from one case to the next one, we doubled the total number of iterations and also the number of processes. We expected a small variation of the value of the execution time divided by the grid resolution. For the ARGLE procedure, we monitored an efficiency of 65\% from case ABC\_aIT to case ABC\_cIT. For the time-dependent GP simulation, we obtained an efficiency of 70\% from case ABC\_a to case ABC\_c. Note that the measured time is the total time for the execution of the program, not solely the computational part of the code.

\subsection{Strong scalability of GPS}

\begin{figure}[h]
\begin{center}
	\includegraphics[width=0.7\textwidth]{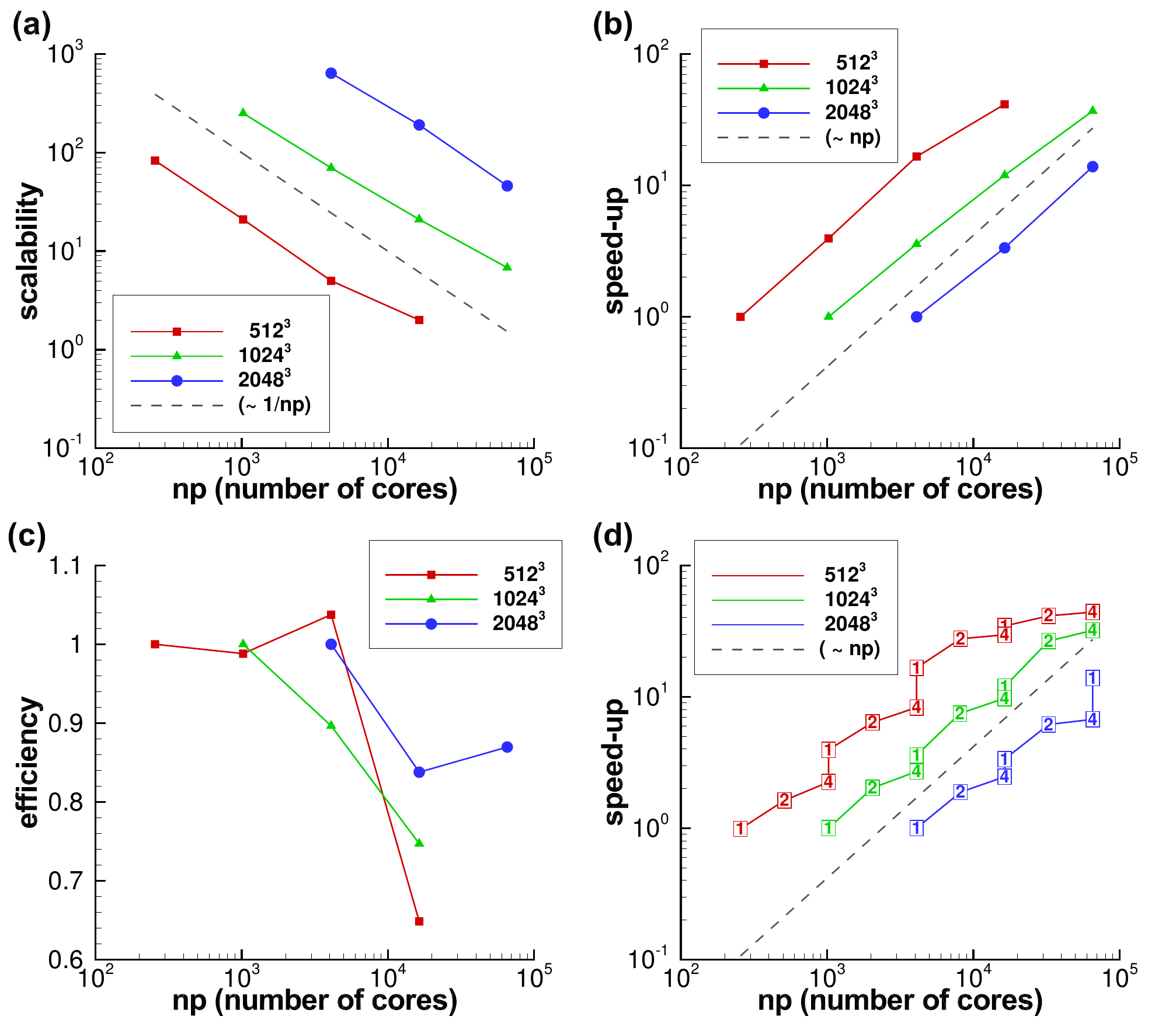}
\end{center}
	\caption{Parallel performance of the GPS code when computing 3D cases with $512^3$,  $1024^3$ and $2048^3$  grid points. 
		(a-c) Strong scalability test using only MPI (from $256$ to $64 536$) processes: scalability, speed-up and efficiency. (d) Strong scalability test of the hybrid code MPI-OpenMP using MPI processes as in the previous test and $1$, $2$ or $4$ OpenMP threads. Dashed lines represent ideal scaling.}
	\label{fig:StrongScalability}
\end{figure}

Strong scalability results of the GPS code are presented in figure \ref{fig:StrongScalability}.  A 3D test case (with grid resolutions up to $2048^3$) was performed using a  different number of processes (up to $64 536$), and the execution time was monitored. Strong scalability tests using only MPI  (Fig. \ref{fig:StrongScalability}a) or the hybrid code MPI-OpenMP  (Fig. \ref{fig:StrongScalability}b)  show scalability and speed-up close to ideal performances. 

\end{document}